\documentclass[oldversion,longauth]{aa} 

\usepackage{txfonts,graphicx}

\begin{document}

   	\title{Galactic kinematics  with RAVE data}  
	\subtitle{I.   The distribution  of  stars  
	towards the Galactic poles}
	\titlerunning{Galactic kinematics with RAVE data}

   \author{L. Veltz\inst{1,2,13} \and O. Bienaym\'e\inst{1} \and K.~C. 
Freeman\inst{2} \and
J. Binney\inst{3} \and
J. Bland-Hawthorn\inst{4} \and
B.~K. Gibson\inst{5} \and
G. Gilmore\inst{6} \and
E.~K. Grebel\inst{7,8} \and
A. Helmi\inst{9} \and
U. Munari\inst{10} \and
J.~F. Navarro\inst{11} \and
Q.~A. Parker\inst{4, 12} \and
G.~M. Seabroke\inst{6} \and
A. Siebert \inst{1,13} \and
M. Steinmetz\inst{13} \and
F.~G. Watson\inst{4} \and
M. Williams\inst{2,13}\and
R.~F.~G. Wyse\inst{14} \and
T. Zwitter\inst{15}
}
   \authorrunning{L. Veltz et al.}

   \offprints{veltz@astro.u-strasbg.fr}

   \institute{
   Observatoire Astronomique de Strasbourg, Strasbourg, France
\and   RSAA, Mount Stromlo Observatory, Canberra, Australia
\and   Rudolf Peierls Centre for Theoretical Physics, University of Oxford, UK
\and   Anglo Australian Observatory, Sydney, Australia
\and   University of Central Lancashire, Preston, UK
\and   Institute of Astronomy, University of Cambridge, UK
\and   Astronomisches Rechen-Institut, Zentrum f\"ur Astronomie der Universit\"at Heidelberg,  Heidelberg, Germany.
\and   Astronomical Institute of the University of Basel, Basel, Switzerland
\and   Kapteyn Astronomical Institute, University of Groningen, Groningen, The 
Netherlands
\and   INAF Osservatorio Astronomico di Padova, Asiago, Italy
\and   University of Victoria, Victoria, Canada
\and   Macquarie University, Sydney, Australia
\and   Astrophysikalisches Institut Potsdam, Potsdam, Germany
\and   Johns Hopkins University, Baltimore MD, USA
\and   University of Ljubljana, Department of Physics, Ljubljana, Slovenia
}

   \date{Received \today }
   
 \abstract{We analyze the  distribution of G and K  type stars towards
the  Galactic poles  using RAVE  and ELODIE  radial  velocities, 2MASS
photometric star counts, and UCAC2 proper motions.  The combination of
photometric  and  3D  kinematic  data  allows us  to  disentangle  and
describe the vertical distribution of dwarfs, sub-giants and giants and
their kinematics.

We identify  discontinuities within the kinematics and magnitude counts
 that separate the  thin disk,  thick disk and  a hotter  component.  The
 respective scale heights of the thin  disk and thick disk are 225$\pm$10\,pc 
 and 1048$\pm$36\,pc. We also constrain the luminosity  
 function  and  the kinematic distribution  function. The existence of 
 a  kinematic gap between the thin and thick  disks is incompatible with 
 the thick disk having formed  from the  thin disk by  a continuous 
 process,  such as scattering  of  stars by  spiral  arms  or  molecular 
 clouds. Other mechanisms of formation of the thick disk such as `created on 
 the spot' or smoothly `accreted' remain compatible with our findings. 

}

   \maketitle

\section{Introduction}

It  is  now widely  accepted  that  the  stellar density  distribution
perpendicular  to  the  Galactic  disk  traces at  least  two  stellar
components,  the thin and  the thick  disks.  The  change of  slope 
in the logarithm of the vertical density distributions at
$\sim$ 700\,pc (Cabrera-Lavers et  al. \cite{cab05}) or $\sim$ 1500 pc
(Gilmore  \&  Reid  \cite{gil83})  above  the Galactic  plane  
is  usually explained as  the signature of  a transition 
between these two  distinct components: the thin and  the thick disks.
The thick disk is an  intermediate stellar population between the thin
disk and  the stellar halo, and  was initially defined  with the other
stellar  populations  by combining  spatial,  kinematic and  abundance
properties (see a summary of  the Vatican conference of 1957 by Blaauw
\cite{bla95} and Gilmore \& Wyse \cite{gil89}).  Its properties
are described  in a long  series of publications with  often diverging
characteristics (see the analysis  by Gilmore \cite{gil85}, Ojha
\cite{ojh01}, Robin et al.  \cite{rob03} and also by Cabrera-Lavers et
al.   \cite{cab05}, that  give  an overview  of recent  improvements).
Majewski (\cite{maj93}) compared a nearly exhaustive list of scenarios
that describe many possible formation mechanisms for the thick disk.

In this  paper, we attempt to  give an answer  to the simple
but  still open questions:  are the  thin and  thick disks  really two
distinct components?  Is there any continuous transition between them?
These questions were  not fully settled by analysis  of star counts by
Gilmore \& Reid (\cite{gil83}) and later workers.
 Other  important   signatures  of   the  thick  disk   followed  from
kinematics:  the  age--velocity   dispersion  relation  and  also  the
metallicity--velocity dispersion relation.  However the identification
of  a thin--thick  discontinuity depends  on the  authors, due  to the
serious difficulty of assigning accurate ages to stars (see Edvardsson
et  al.   \cite{edv93} and Nordstr\"om et al. \cite{nor04}).   More  
recently   it  was  found   that  the
[$\alpha$/Fe] versus [Fe/H] distribution  is related to the kinematics
(Fuhrmann  \cite{fur98}; Feltzing  et al.   \cite{fel03};  Soubiran \&
Girard \cite{sou05};  Brewer \& Carney \cite{bre06}; Reddy et al. 
\cite{red06})  and provides an
effective  way  to  separate  stars  from  the  thin  and  thick  disk
components.  Ages and abundances are important to describe the various
disk components  and to depict  the mechanisms of their  formation.  A
further complication comes from the recent indications of the presence
of  at  least  two   thick  disk  components  with  different  density
distributions,   kinematics    and   abundances   (Gilmore    et   al.
\cite{gil02};    Soubiran    et    al.     \cite{sou03};    Wyse    et
al. \cite{wys06}). 

Many of the  recent works favor the presently  prevailing scenarios of
thick disk formation by the  accretion of small satellites, puffing up
the early stellar Galactic disk or tidally disrupting the stellar
disk (see for example Steinmetz  \& Navarro \cite{ste02}; Abadi et al.
\cite{aba03};  Brook  et al.   \cite{bro04}).   We  note however  that
chemodynamical  models of  secular Galactic  formation including
extended ingredients  of stellar formation  and gas dynamics  can also
explain  the formation of  a thick  disk distinct  from the  thin disk
(Samland \& Gerhard \cite{sam03}; Samland \cite{sam04}).

In this paper,  we use the recent RAVE  observations of stellar radial
velocities, combined  with star counts and proper  motions, to recover
and  model the  full 3D  distributions of  kinematics and  densities for
nearby   stellar   populations.  In  a   forthcoming   study,
metallicities  measured from  RAVE  observations will  be included  to
describe  the galactic  stellar populations  and their  history.  The
description of data is given in Sect.~2, the model in Sect.~3, and the
interpretation  and  results  in  Sect.~4.  Among  these  results,  we
identify discontinuities visible  both within the density distributions
and the kinematic  distributions.  They  allow to  define more
precisely the  transition between the thin and  thick stellar Galactic
disks.

\section{Observational data}
\label{data}

Three types of data are used to constrain our Galactic model for the
stellar kinematics and star counts (the model description is given in 
Sect.~3): the Two-Micron
All-Sky Survey (2MASS PSC; Cutri et al. \cite{cut03}) magnitudes, the
RAVE  (Steinmetz  et  al.  \cite{ste06})  and
ELODIE    radial   velocities,   and    the   UCAC2    (Zacharias   et
al. \cite{zac04}) proper motions. Each sample of stars is selected independently 
of the other, with its own magnitude limit and coverage of sky due to the 
different source (catalogue) characteristics.

(1) We select 22\,050 2MASS stars within an 8-degree radius of the South and
    North  Galactic Poles, with  $m_{\rm K}$ magnitudes between 5-15.4.  Star
    count  histograms for both  Galactic poles are  used to
    constrain the Galactic model.

(2) We select 105\,170 UCAC2 stars within a  radius of 16  degrees of the
    Galactic poles, with $m_{\rm  K}$  2MASS magnitudes between 
    6-14.  We adjust  the model to fit histograms of
    the  $\mu_U$  and $\mu_V$  proper  motion marginal  distributions;
    the histograms combine  stars in  1.0 magnitude intervals  for $m_{\rm
    K}$=6 to 9 and 0.2 magnitudes intervals for $m_{\rm K}$=9 to 14.

(3) We select 543 RAVE stars ( with $m_{\rm K}$ 2MASS magnitudes from 
 8.5 to 11.5) within a radius of 15 degrees of the
    SGP.  We group  them in three histograms according  to $m_{\rm K}$
    magnitudes.   We   complete  this  radial   velocity  sample  with
    392 other similar stars: TYCHO-II stars selected towards the
    NGP  within an area of 720  square degrees,  with B-V  colors
    between 0.9-1.1.   Their  magnitudes  are brighter  than  $m_{\rm
    K}$=8.5, they were observed  with the ELODIE spectrograph and were
    initially   used  to   probe  the   vertical   Galactic  potential
    (Bienaym\'e  et  al.  \cite{bie06}).   All  these radial  velocity
    samples  play  a  key  role  in constraining  the  vertical  velocity
    distributions of stars and the shape of the velocity ellipsoid.

\begin{figure}[!htbp]
\resizebox{8cm}{!}{
\includegraphics{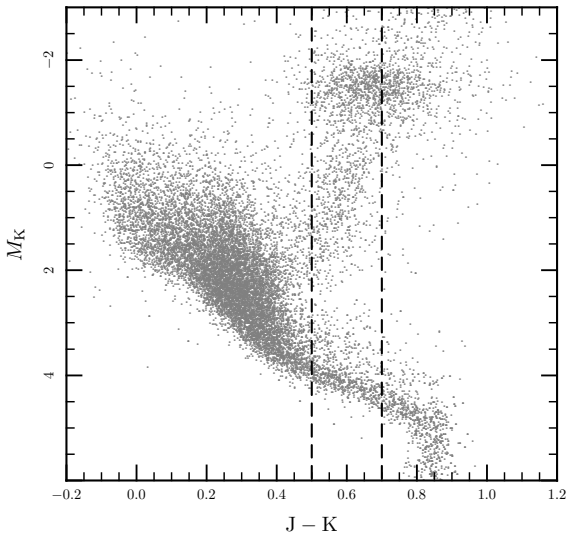}
}
\caption{ $\rm M_K / J-K $ HR Diagram from Hipparcos stars with $\sigma_\pi / \pi \le 0.1$ cross-matched with the 2MASS catalogue. Vertical dashed lines represent our color selection
 J$-$K=[0.5-0.7]}
\label{f:HR}
\end{figure}

\subsection{Data selection}
\label{s:selection}

 In this paper, we restrict our analysis to  stars near the Galactic poles with J$-$K
colors  between  0.5-0.7 (see fig.\ref{f:HR}).  This  allows  us  to  recover some  Galactic
properties, avoiding the coupling with other Galactic parameters that  
occurs in other  Galactic directions (density and kinematic scale
lengths, Oort's constants, $R_0$, $V_0$...).

The  selected  J$-$K=[0.5-0.7]  color  interval corresponds  to  K3-K7
dwarfs   and   G3-K1  giants   (Koornneef   \cite{koo83};  Ducati   et
al. \cite{duc01}).   They may be  G or K  giants within the  red clump
region  (the part  of the  HR  diagram populated  by high  metallicity
He-burning core  stars).  The absolute  magnitudes of red  clump stars
are well  defined: nearby HIPPARCOS  clump stars have a  mean absolute
magnitude $M_{\rm  K}=-1.61$ with a  dispersion of $\sim  0.22$ (Alves
\cite{alv00}, see  Cannon \cite{can70} for the first  proposed use of 
clump  stars  as  distance  indicators,  see also  Salaris  \&  Girardi
\cite{sal02};  Girardi et  al.  \cite{gir98}  and other  references in
Cabrera-Lavers  et al.  \cite{cab05}).   This mean  absolute magnitude
does  not  vary  significantly  with  [Fe/H] in  the  abundance  range
$[-0.6,0]$  (Alves  \cite{alv00}). Studying nearby stars in 13 open 
clusters and 2 globular  clusters,   Grocholski  \&
Sarajedini \cite{gro02} find  that the mean absolute  magnitude of
clump stars is not dependent  on metallicity when the [Fe/H] abundance
remains within  the interval $[-0.5,0.1]$.  Sarajedini \cite{sar04}  finds that,  
at metallicity [Fe/H]=$-0.76$,  the mean absolute magnitude  of red 
clump stars  drops to  $M_{\rm K}=-1.28$,  a shift  of 0.33  mag.  Most of the giants with metallicity [Fe/H] lower than -0.8 dex are excluded 
by our color selection from our sample. Hence, we did not model giants of
the metal-weak thick disk, first identified by Norris \cite{nor85} 
(see also, Morrison, Flynn \& Freeman \cite{mor90}). 
This represents however only a minor component of the thick disk. 
Although, Chiba \& Beers (\cite{chi00}) find that $\sim$ 30 \% 
of the stars with $-1 >{\rm [Fe/H]}> -1.7$ are thick disk stars,  but 
stars with ${\rm [Fe/H]} < -1$ represent only 1 per cent of the local 
thick disk stars (Martin \& Morrison \cite{mar98}).

K  dwarfs within the  J$-$K=[0.5-0.7] color  interval also  have well
defined absolute  magnitudes that  depend slightly on  metallicity and
color.  We determine their mean absolute magnitude, $M_{\rm K}$=4.15,
from nearby  HIPPARCOS stars using  color magnitude data  provided by
Reid  (see http://www-int.stsci.edu/$\sim$inr/cmd.html).   From Padova
isochrones (Girardi  et al. \cite{gir02}),  we find that  the absolute
magnitude varies by 0.4 magnitudes  when J$-$K changes from 0.5 to 0.7.
A  change  of  metallicity  of  $\Delta$[Fe/H]=0.6  also  changes  the
magnitude  by  about  0.3,  in  qualitative  agreement  with  observed
properties of K dwarfs  (Reid \cite{rei98}, Kotoneva et al  \cite{kot02}).  Thus, we
estimate that  the dispersion of  absolute magnitude of dwarfs  in our
Galactic pole sample is $\sim$0.2-0.4.

Another important motivation for selecting the J$-$K=[0.5-0.7] color
interval is the absolute magnitude  step of 6 magnitudes between
dwarfs  and  giants.   This  separation is the  reason the  magnitude
distributions for these two kinds  of stars are very different towards
the Galactic  poles. If giants and dwarfs have the same density distribution in the disk, in the apparent magnitude count, giants will appear before and well separated from dwarfs.  Finally we  mention  a convenient property of  the Galactic pole  directions: there, the  kinematic data
are simply  related to  the cardinal velocities  relative to  the local standard of rest (LSR).
UCAC2  proper  motions  are  nearly   parallel  to  the  $U$  and  $V$
velocities, and RAVE  radial velocities are close to  the vertical $W$
velocity component.

\subsection{How accurate is the available data?}

 The star magnitudes are taken from the 2MASS survey which is presently the most accurate  photometric all sky survey for  probing the Galactic stellar populations. Nevertheless, since our color rang is narrow, we have to take care that the photometric errors on J and K do not bias our analysis.  
 
\begin{figure}[!htbp]
\resizebox{8cm}{!}{
\includegraphics{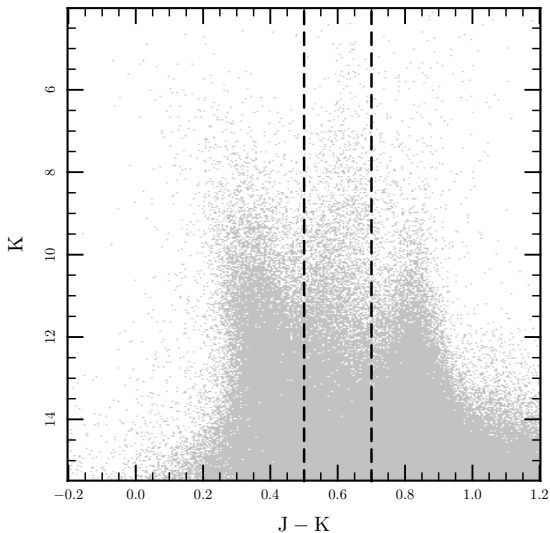}
}
\caption{ K / J$-$K Color Magnitude Diagram obtained with 2MASS stars within a 8 degrees radius around the North Galactic pole. Dashed lines represent the  limit of our color selection: $\rm J-K = [0.5,0.7]$.  }
\label{f:cmd}
\end{figure}

The mean  photometric accuracy ranges from 0.02 in K and J at magnitudes  $m_{\rm K}$=5.0, to 0.15 in K and 0.08 in J at magnitudes  $m_{\rm K}$=15.4. The error in J$-$K is  not small considering the size ($\Delta$(J--K) = 0.2) of the analyzed  J$-$K interval, 0.5 to 0.7. We do not expect, however, that it  substantially biases  our  analysis.  For $m_{\rm K}$ brighter  than 10, the  peak of  giants is  clearly  identified in  the J$-$K  distribution within  the  J$-$K= [0.5-0.7]  interval  (see Fig.\ref{f:cmd} or Figure~6  from Cabrera-Lavers et  al., \cite{cab05}). This peak  vanishes only beyond $m_{\rm K}$=11.  At fainter 
K magnitudes, the dwarfs dominate and the J$-$K histogram of colors has a constant slope.  This implies that the error in color  at faint magnitudes does not  affect to first order the star counts.

We find from the shape of the count histograms that, in the direction of the Galactic Pole and with our color selection J$-$K = [0.5-0.7] the limit of completeness is $m_{\rm K}\sim$15.5-15.6. Moreover, the contamination by galaxies must be low  within the 2MASS PSC.  It is  also unlikely that compact or unresolved galaxies are present:  according to recent deep J and K photometric  counts (see Figure 15 of  Iovino et al. \cite{iov05}),  with our color  selection,  galaxies  contribute  only  beyond $m_{\rm K}\sim$16. We conclude that we have a complete sample of stars for magnitudes from 5.0 to 15.4 in K, towards the Galactic poles. 

The UCAC2 and RAVE catalogues however are not complete. Making it necessary to scale the proper motions and radial velocities distributions predicted by our model for complete samples. The total number of stars given by the model for the distribution of proper motions (or radial velocity) in a magnitude interval is multiplied by the ratio between the number of stars observed in UCAC2 (or RAVE) divided by the number of stars observed in 2MASS.

Towards the North Galactic Pole (NGP), the error  on the UCAC2  proper motions used  in our
analysis  varies  from 1\,mas\,yr$^{-1}$  for  the  brightest stars  to
6\,mas\,yr$^{-1}$ at $m_{\rm K}$=14.   Towards the South Galactic Pole (SGP),
the  error distribution  looks similar,  with the exception  of a  small
fraction of stars with $m_{\rm K}$  from 11 to 14 having errors around
8  or 13  mas\,yr$^{-1}$.   The only  noticeable difference between
the histograms at the NGP and SGP is that the peak of the proper
motion distribution is slightly  more  flattened  at the  SGP,
for magnitudes $m_{\rm  K}>$13 (see fig. \ref{f:vitesse2} ). 
This difference is related  to the  different  error distributions  towards the NGP and SGP.

The analyzed stars are located  at distances from 200\,pc to 1\,kpc for  dwarfs and  to 1.5\,kpc for  giants.   A 2\,mas\,yr$^{-1}$  error represents 10\,km\,s$^{-1}$ at  1\,kpc, and 6\,mas\,yr$^{-1}$, an error
of 30\,km\,s$^{-1}$.  This  can be  compared to  the  $\sigma_U$ values 
 for the  isothermal components, for  instance $\sim$\,60\,km\,s$^{-1}$ 
 for  the thick disk that  is the  dominant  stellar population  1.5\,kpc  
 from the  plane. Adding the errors in quadrature to the velocity dispersion would modify
a  real proper motion  dispersion of  60\,km\,s$^{-1}$ to  an apparent
dispersion of 67\,km\,s$^{-1}$.  The apparent dispersion would be only
60.8\,km\,s$^{-1}$ if the stars have a 2\,mas\,yr$^{-1}$ accuracy. 
Therefore, we overestimate  the $\sigma_U$  dispersion of  the 
thick disk by 5 to 10 percent.  This effect is lower for the thin disk
components  (the stars  are closer  and their  apparent  proper 
motion distributions are broader).  We have not  yet included the 
effect  of proper motion errors within our model.  This error has just an impact of the determination 
on the velocity dispersions $\sigma_U$ and $\sigma_V$ and 
on the ellipsoid axis ratio $\sigma_U/\sigma_W$ of each stellar 
disk component, but does not change the determination of 
vertical velocity dispersions $\sigma_W$ which are mainly 
constrained by the magnitude star count and the radial velocities. Hence, 
it is not significant in our kinematic decomposition of the 
Galactic disk.

 The accuracy of proper motions  can also be gauged from the stability
of the peaks of proper motion distributions: comparing 112 $\mu_U$ and
$\mu_V$  histograms  for different  magnitude intervals,  we  find  no
fluctuations larger than 3-5 mas\,yr$^{-1}$ .

A  more complete  test  is  performed by  comparing  the UCAC2  proper
motions (with our J$-$K color selection) to the recent PM2000 catalogue
(Ducourant  et al.  \cite{duc06})  in an  area of  8$\times$16 degrees
around  $\alpha_{2000}$=12h50m, $\delta_{2000}=14\deg$  close  to the
NGP. PM2000 proper motions are more  accurate, with
errors from 1 to 4  mas\,yr$^{-1}$.  The mean differences between proper
motions  from   both  catalogues  versus   magnitudes  and  equatorial
coordinates do not show significant shifts, just  fluctuations of the
order of  $\sim$0.2 mas\,yr$^{-1}$.  We also find  that the dispersions
of  proper motion  differences are  $\sim$2\,mas\,yr$^{-1}$  for $m_{\rm
K}<$10, 4\,mas\,yr$^{-1}$ with  $m_{\rm K}$=10-13, and 6\,mas\,yr$^{-1}$
with $m_{\rm K}$=13-14.  These  dispersions are dominated by the UCAC2
errors.

From the internal and  external error analysis, RAVE radial velocities
  show  a   mean  accuracy   of  2.3\,km\,s$^{-1}$  
(Steinmetz et  al. \cite{ste06}).   Radial  velocities of 
stars observed  with the ELODIE \'echelle spectrograph  are an order 
of magnitude more accurate.  These errors have no impact on the 
determination  of  the   vertical  velocity  dispersion  of  stellar
  components that ranges from  10 to 50\,km\,s$^{-1}$, but the reduced
  size of  our radial velocity  samples towards the poles  (about 1000
  stars)  limits  the  accuracy  achieved  in modeling  the  vertical
  velocity dispersions.

\section{Model of the stellar Galactic disks}

The basic ingredients of our Galactic model are taken from traditional
works on star count and  kinematic modeling, for instance see Pritchet
(\cite{pri83});    Bahcall   (\cite{bah84});   Robin    \&   Cr\'ez\'e
(\cite{rob86}).   It is  also similar  to the  recent  developments by
Girardi et al.  (\cite{gir05}) or by Vallenari et al. (\cite{val06}).

The  kinematic  modeling is  entirely  taken  from  Ratnatunga et  al.
(\cite{rat89}) and  is also  similar to Gould's (\cite{gou03})  analysis.
Both propose closed-form expressions for velocity projections; the
dynamical consistency  is similar to Bienaym\'e  et al. (\cite{bie87})
and Robin et al. (\cite{rob03}, \cite{rob04}).

Our analysis, limited to the Galactic poles, is based on a set of
20  stellar  disk  components.   The  distribution  function  of  each
component  or stellar disk  is built  from three  elementary functions
describing the vertical  density $\rho_i$ (dynamically self consistent
with the vertical gravitational potential), the kinematic distribution
$f_{i}$ (3D-gaussians) and the luminosity function $\phi_{ik}$.

We  define $\mathcal{N}(z,  V_R, V_\phi, V_z;  M)$ to be the  density
of stars in the Galactic position-velocity-(absolute magnitude) space
$$\mathcal{N}  =\sum_{ik}\, \rho_i(z) f_{i}( V_R, V_\phi,  V_z)  \phi_{ik}(M)$$
\noindent 
the index $i$ differentiates the stellar disk components and the
index $k$ the absolute magnitudes used to model the luminosity function.

From  this  model,  we  apply  the  generalized  equation  of  stellar
statistics:
$$ A(m,\mu_l,\mu_b, V_r)= \int N(z,  V_R, V_\phi, V_z; M)\, z^2 \, \omega \, dz $$
to determine the  $A(m)$ apparent  magnitude star  count  equation as 
 well as  the
marginal  distributions  of both  components  $\mu_l$  and $\mu_b$  of
proper  motions and  the distributions  of radial  velocities  for any
direction and apparent magnitudes.   For the Galactic poles, we define
$\mu_U$  and $\mu_V$ as the  proper motion  components  parallel to  the
cardinal directions of $U$ and $V$ velocities. For a more general 
inverse method of the equation of the stellar statistic, see Pichon et al. (\cite{pic02}).

\subsection{The vertical density}

 Each   stellar  disk   is  modeled   with  an   isothermal  velocity
distribution,   assuming  that   the  vertical   density  distribution
(normalized at $z$=0) is given by the relation:
\begin{equation}
\phantom{exponentiel egual }
\rho_i(z)=\exp   \left(-\Phi(z)/\sigma^2_{zz,i}\right)
\end{equation}
where $\Phi(z)$  is the vertical gravitational potential  at the solar
Galactic  position  and   $\sigma_{zz,i}$  is  the  vertical  velocity
dispersion of  the considered stellar component $i$. The Sun's
position $z_\odot$  above the Galactic plane  is also used  as a model
parameter.  Such  expressions were introduced  by Oort (\cite{oor22}),
assuming the  stationarity of the density  distributions.  They ensure
the   consistency   between   the   vertical  velocity   and   density
distributions.  For  the vertical  gravitational potential we  use the
recent  determination obtained  by Bienaym\'e  et  al.  (\cite{bie06})
based  on the  analysis of  HIPPARCOS and  TYCHO-II red  clump giants.
 The vertical potential is defined at the solar position by:
$$\Phi(z)=  4\pi  G \left(  \Sigma_0  \left( \sqrt{z^2+D^2}-D  \right)
+\rho_{\rm eff}\,  z^2 \right)$$ 
with $\Sigma_0=48\,{\rm M}_{\odot}\,{\rm pc}^{-2}$,
$D=800\,{\rm pc}$ and $\rho_{\rm eff}=0.07\,{\rm M}_{\odot}\,{\rm pc}^{-3}$.

It is  quite similar   to  the  potential  determined  by Kuijken \& Gilmore 
(\cite{kui89}) and  Holmberg  \&  Flynn (\cite{hol04}).

\subsection{The kinematic distributions}

The  kinematical  model  is  given  by shifted  3D  gaussian  velocity
ellipsoids.   The three  components of  mean streaming  motion 
($\langle U \rangle$, $\langle V \rangle$,  $\langle W \rangle$) and 
 velocity  dispersions ($\sigma_{RR}$,  $\sigma_{\phi\phi}$,
$\sigma_{zz}$),  referred  to  the  cardinal directions  of  the  Galactic
coordinate frame, provide a  set of six kinematic quantities. The
mean stream motion is relative to the LSR.
The   Sun's  velocity  $U_\odot$ and $W_\odot$   are  model parameters. 
We define the $\langle V \rangle$ stream motion as:
$\langle V \rangle = -  V_\odot - V_{\rm lag}$. We adopt an asymmetric drift
proportional to the square of $\sigma_{RR}$:  $V_{\rm lag}= \sigma_{RR}^2 /k_{a}$,  
where the coefficient  $k_{a}$ is also  a model parameter.  
 We assume  null  stream motions  for  the
other  velocity components, thus $\langle U \rangle=-U_\odot$ and 
$\langle W \rangle=-W_\odot$.

For      simplicity,      we      have      assumed      that      the
$\sigma_{RR}/\sigma_{\phi\phi}$  ratio   is  the  same   for  all  the
components. It is well
known      that      the      assumptions      of      a      constant
$\sigma_{RR}/\sigma_{\phi\phi}$  ratio, of  a linear  asymmetric drift
and of 2D  gaussian U and V velocity distributions  hold only for cold
stellar  populations   (see  for  instance   Bienaym\'e  \&  S\'echaud
\cite{bie97}).   These simple  assumptions allow  a  direct comparison
with similar  studies.  It allows  also an exact integration  of count
equations along the line of  sight. Thus the convergence of parameters
for any single  model is achieved in a reasonable  amount of time (one
week). The model includes 20 isothermal components  with $\sigma_{zz}$  from  3.5 to  70\,km\,s$^{-1}$. We choose a step of 3.5\,km\,s$^{-1}$ which is sufficient to give a realistic kinematic decomposition and permit calculation in a reasonable time. The  two first components  $\sigma_{zz}$= 3.5 and  7\,km\,s$^{-1}$ were suppressed since they do not contribute significantly to counts for $m_{\rm K}$$>$6 and are not constrained by our adjustments. The components between 10 and 60 \,km\,s$^{-1}$ are constrained by star counts, proper motions histograms up to magnitude 14 in K and radial velocity histograms for magnitudes  $m_{\rm K}$=[5.5-11.5]). The model includes isothermal components from 60 to 70\,km\,s$^{-1}$ to properly fit the star counts at the faintest apparent magnitudes $m_{\rm K} > 15.0$. All the values of the kinematic components depend on the adopted galactic potential.

The  velocity  ellipsoids are  inclined  along  the Galactic  meridian
plane.   The main  axis of  velocity  ellipsoids are  set parallel  to
con-focal hyperboloids as in St\"ackel  potentials. We set the focus at
$z_{hyp}$=6\,kpc on the main  axis giving them realistic orientations
(see Bienaym\'e \cite{bie99}).  The  non-zero inclination implies that
the vertical  density distributions of each  isothermal component is
not  fully  dynamically  consistent  with the  potential.   Since  the
$z$-distances  are  below 1.5\,kpc  for  the  majority  of stars  with
kinematic data,  and since  the main  topic of this  paper is  not the
determination  of the  Galactic potential,  we do  not develop  a more
consistent dynamical model.

\subsection{The luminosity functions}

The luminosity function of each stellar disk component is modeled with $n$ different kinds of stars according to their absolute magnitude:
$$\phi_{i}(M)=              \sum_{k=1,n}              \,\phi_{ik}(M)=
\frac{1}{\sqrt{2\pi}\sigma_{M}}\,\sum_{k=1,n}\,               c_{ik}\,
e^{-\frac{1}{2}   \left(  \frac{M-M_k}{\sigma_M}  \right)^2   }$$  
where $c_{ik}$ is the density for each type of star (index $k$) of each stellar disk component (index $i$). 

\begin{figure}[!htbp]
\resizebox{\linewidth}{!}{
\includegraphics{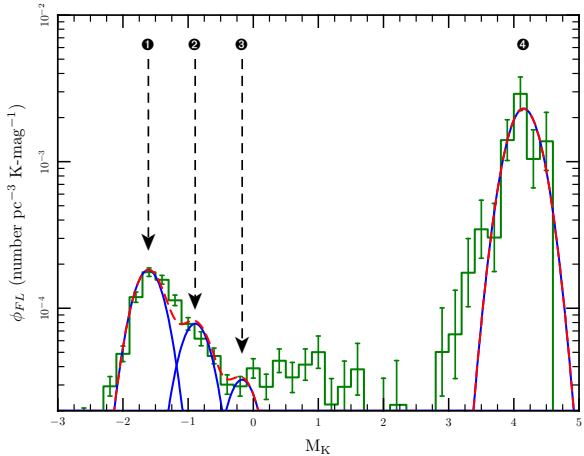}
}
\caption{ Local luminosity function: The histogram is our determination of the local luminosity function for nearby stars with error bars. The red (or dark grey) dashed line is a fit of the luminosity function with four gaussians (blue or light grey line) corresponding to the dwarfs, the giants and the two types of sub-giants. }
\label{f:LLF}
\end{figure}

We use four types of stars to model the local luminosity function (see Fig. \ref{f:LLF}). More details on the way that we have determined it is given in section \ref{lf}. Stars with a  mean  absolute  magnitude  $M_{\rm  K}=-1.61$ are identified to be the red clump giants ($k=1$) that we will call `giants',  with $M_{\rm  K}=-0.89$ and  $M_{\rm K}=-0.17$ for first ascent giants that we categorize as `sub-giants' ($k=2-3$) and  $M_{\rm K}=4.15$ are labelled dwarfs ($k = 4$) (see fig. \ref{f:HR}). We neglected 'sub-giant' populations having absolute magnitude $M_{\rm K}$ between 0.2 and 2. Their presences marginally change the ratio of giants to dwarfs, since their magnitudes are lower, and their total number in the magnitude counts appears significantly smaller than the other components. In fact, we initially tried to introduce 10 types of stars (spaced by 0.7 absolute magnitude intervals). This still improves the fit to the  data.  However due to the small contribution of the `sub-giants' components with $M_{\rm K}=[0.2-2]$, they were not determined with a useful accuracy. We adopt  $\sigma_M=0.25$, justified  by the narrow range of absolute magnitudes  both for red clump giants and for dwarfs on the luminosity function. 

The  4x20 coefficients $c_{ik}$  are parameters of the model.  In  order to obtain a realistic luminosity function,  we have added constraints to the minimization procedure.  For {\it each} kinematic  component $i$, we impose conditions on the proportion of dwarfs, giants and sub-giants following the local luminosity function. We have modeled our determination of the local luminosity function of nearby stars (see Fig. \ref{f:LLF}).  We obtained :
 \begin{itemize}
 \item a ratio of  the density of dwarfs ($k$=4) to the density of  giants ($k$=1) of 12.0, so we impose: $ \frac{c_{i,4}}{c_{i,1}} > 10 $
 \item a ratio of  the density of  giants ($k$=1) to the density of  sub-giants($k$=2) of 2.3, so we impose: $ \frac{c_{i,1}}{c_{i,2}} > 2 $
  \item and the density of sub-giants ($k$=2) is greater than the density of  sub-giants ($k$=3), so we impose: $c_{i,1}\,>\,c_{i,2}$.
\end{itemize} 
 If we do not include these constraints, the various components are populated either only with dwarfs 
 or only with giants.

\section{Results and discussion}

\begin{figure*}[!htbp]
\resizebox{\hsize}{!}{
\includegraphics[angle=270]{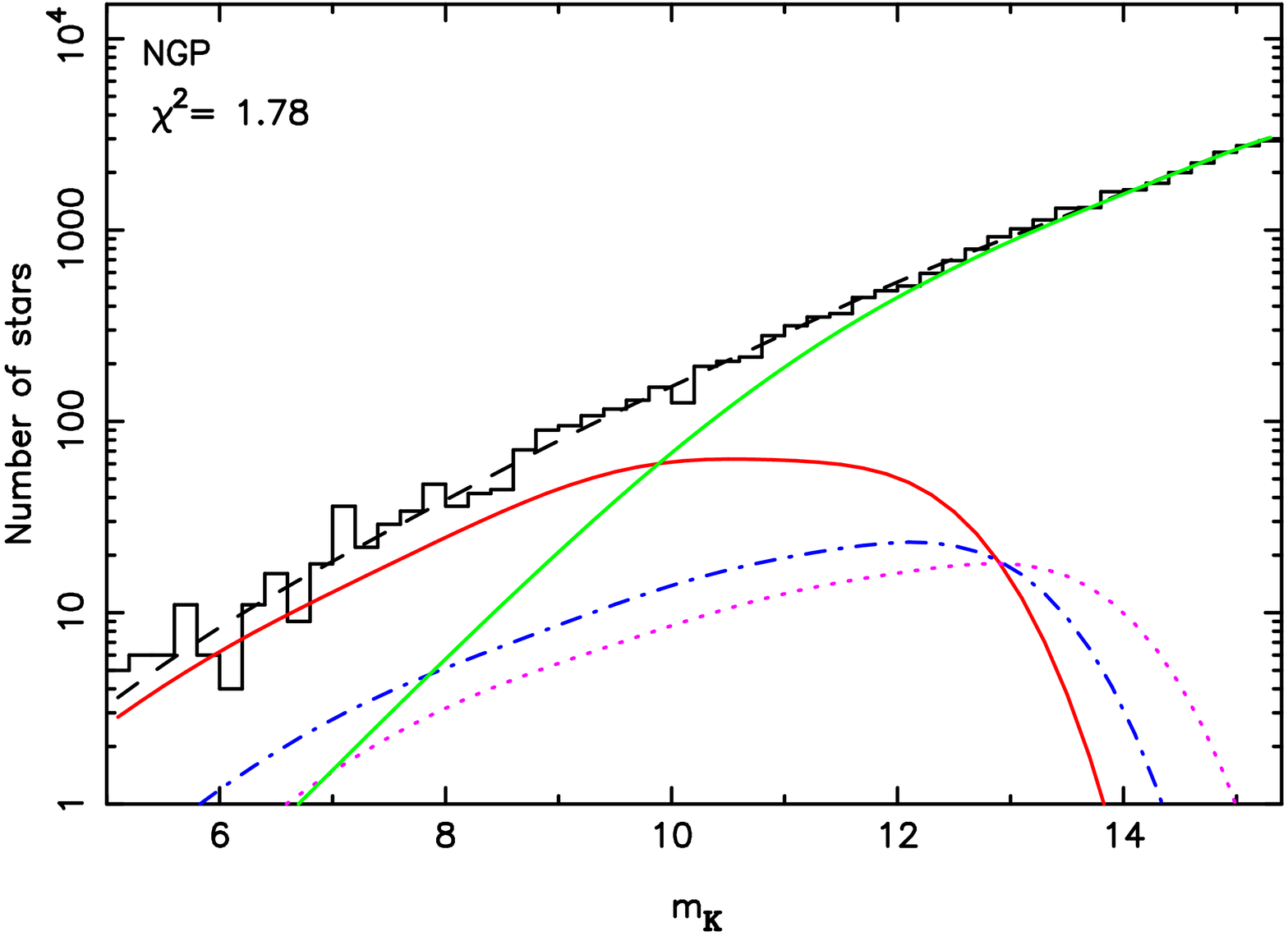}
\includegraphics[angle=270]{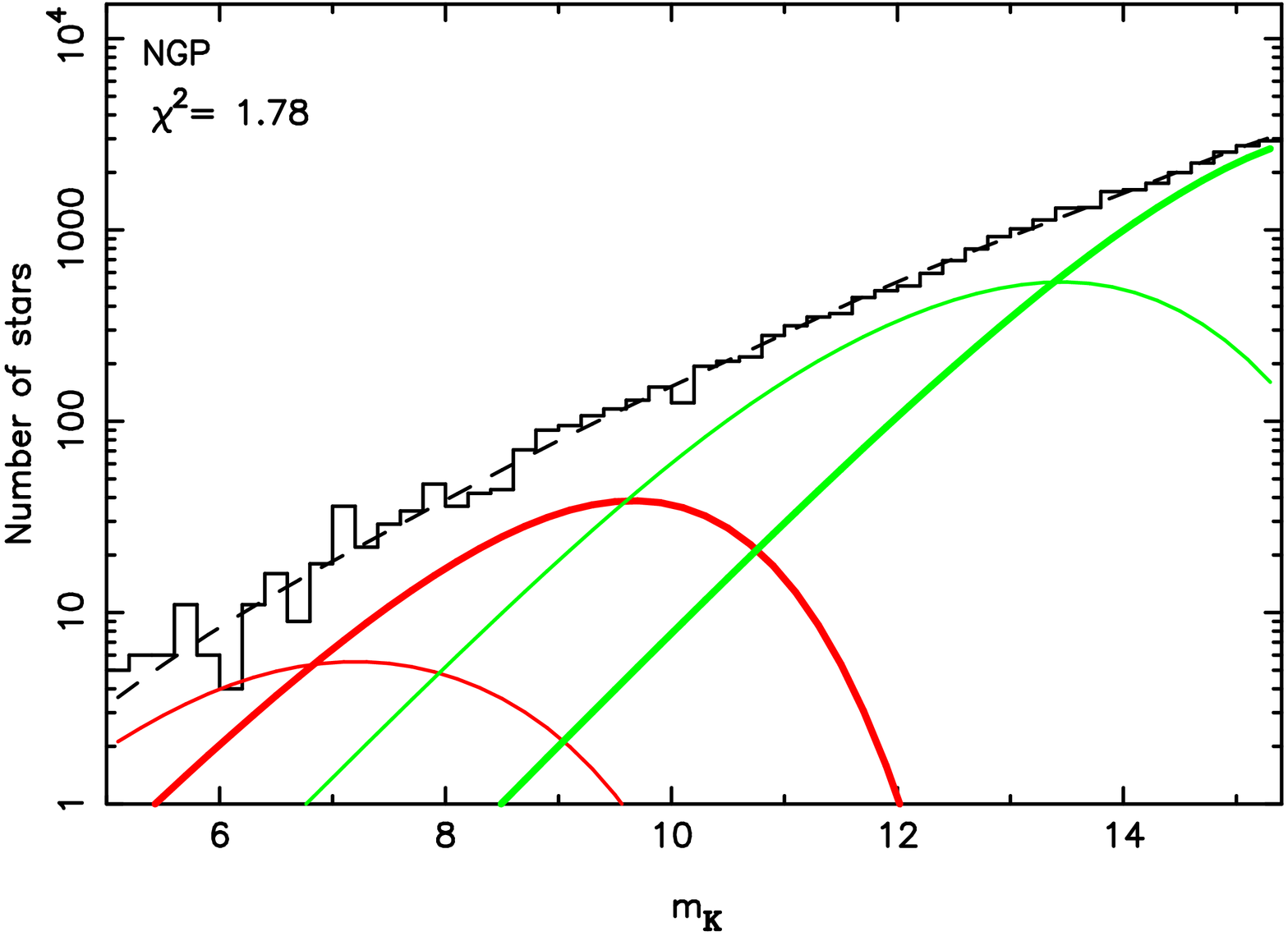}
}
\caption{Magnitude  count histogram towards  the North  Galactic Pole.
         Left: model prediction (dashed line) is split according to
         star types:  giants (red or black  line), sub-giants (dot-dashed
         and dotted) and dwarfs (green or grey line).  The right figure
         highlights   the  contributions  of   thin  and   thick  disks
         (respectively  thin and  thick lines),  for dwarfs  (green or
         grey) and giants (red or black).}
\label{f:count}
\end{figure*}

\begin{figure*}
\begin{minipage}{.48\textwidth}
\scalebox{0.17}[0.21]{\includegraphics[angle=270]{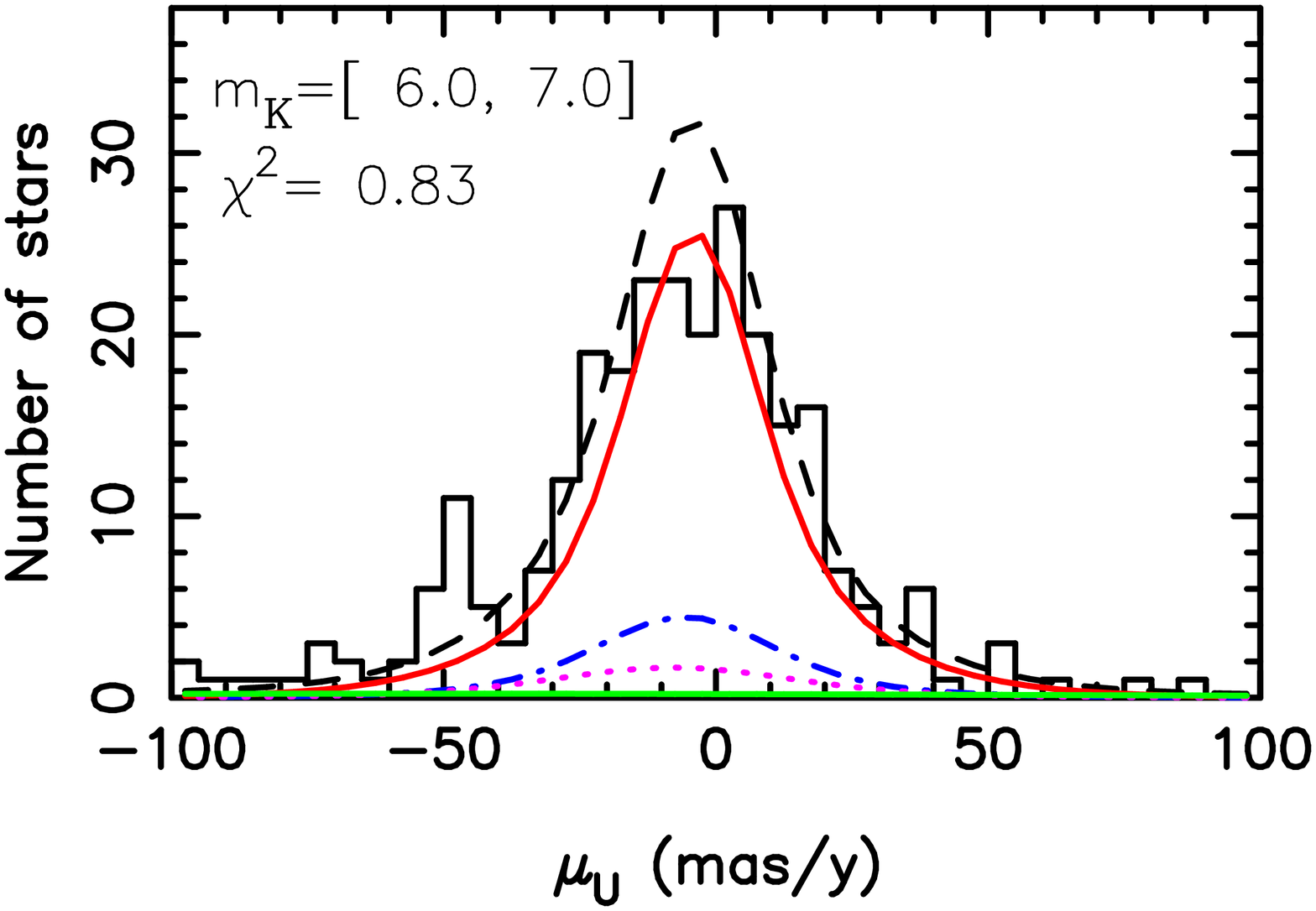}
	                      \includegraphics[angle=270]{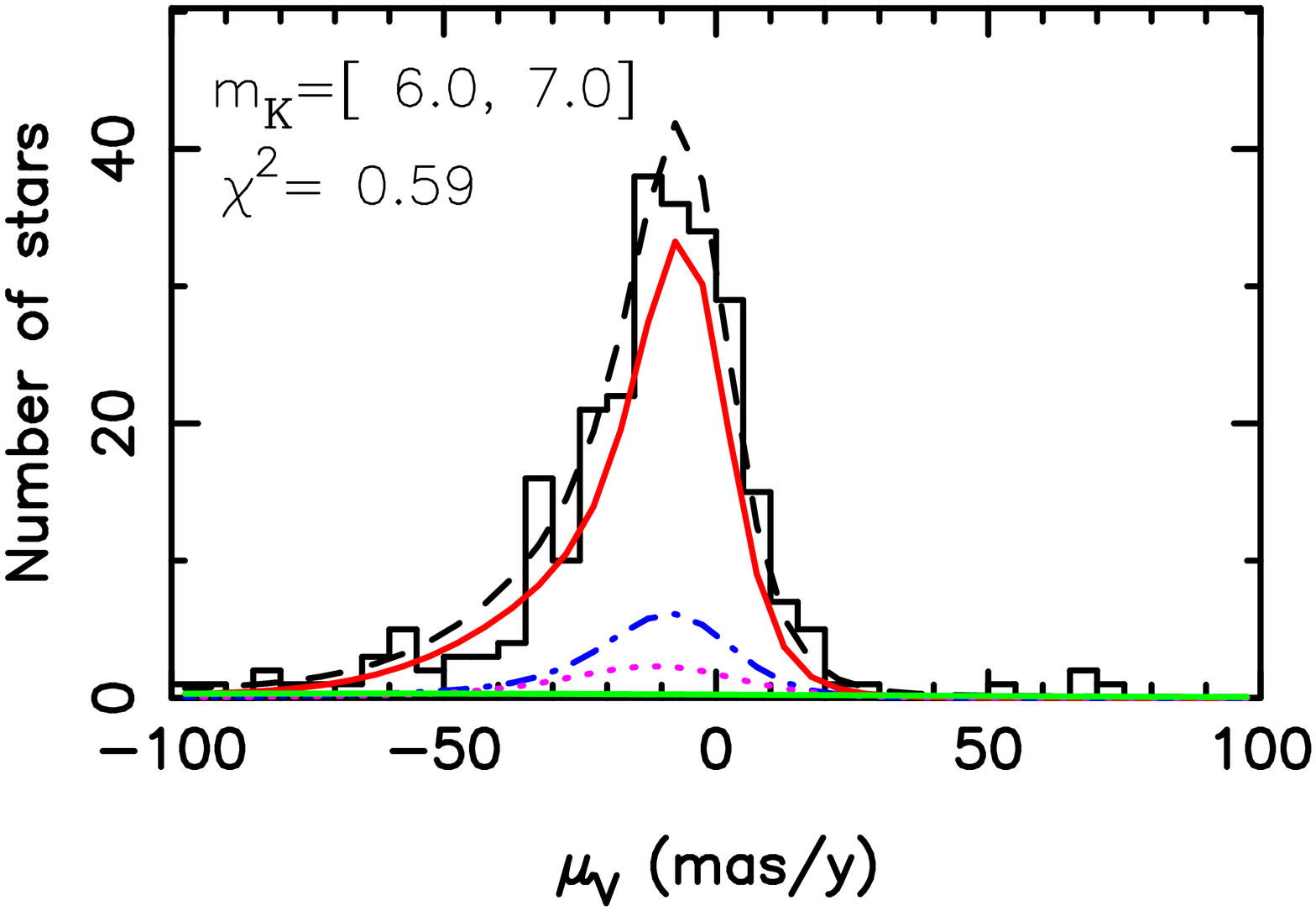}}
\scalebox{0.17}[0.21]{\includegraphics[angle=270]{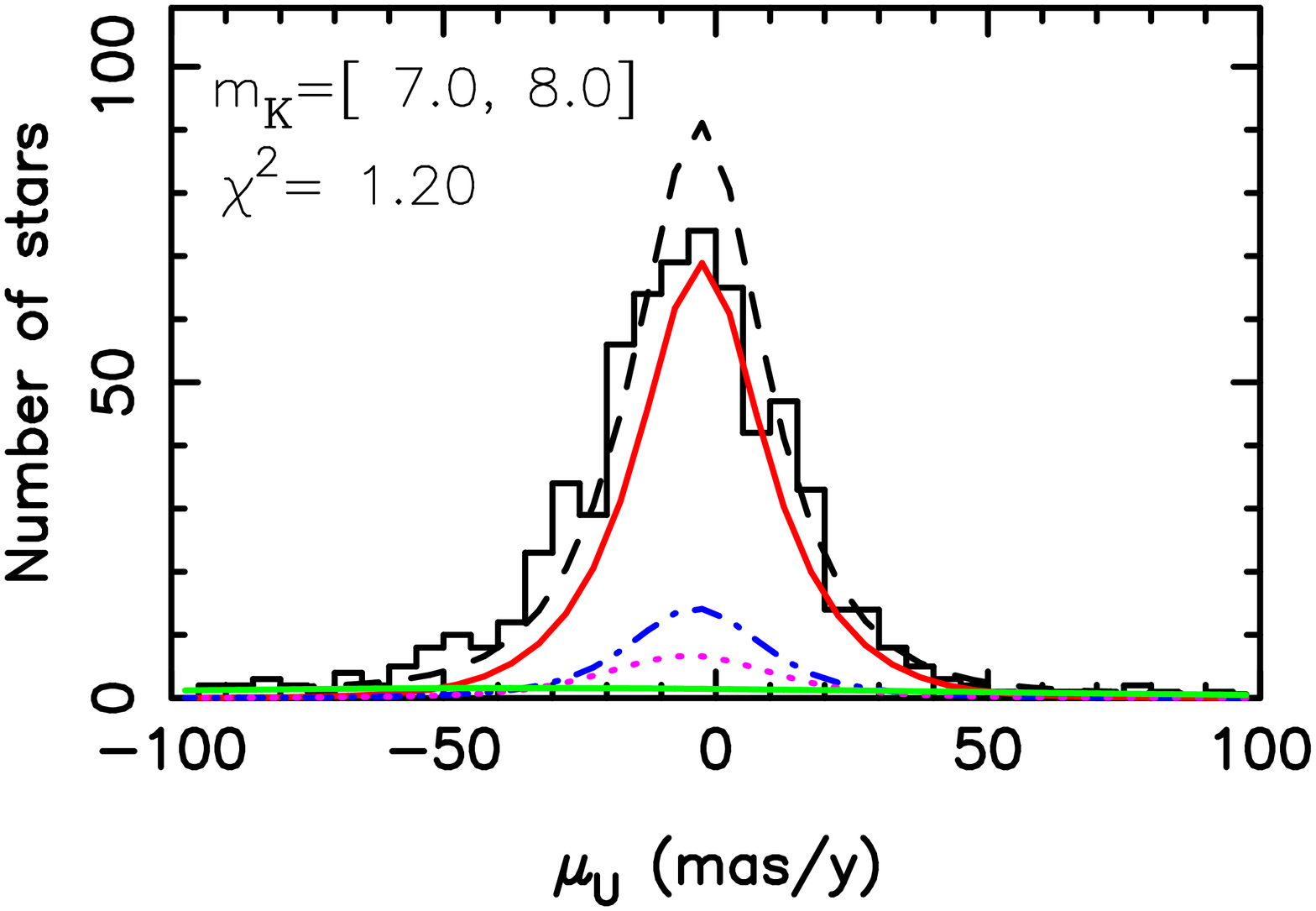}
	                      \includegraphics[angle=270]{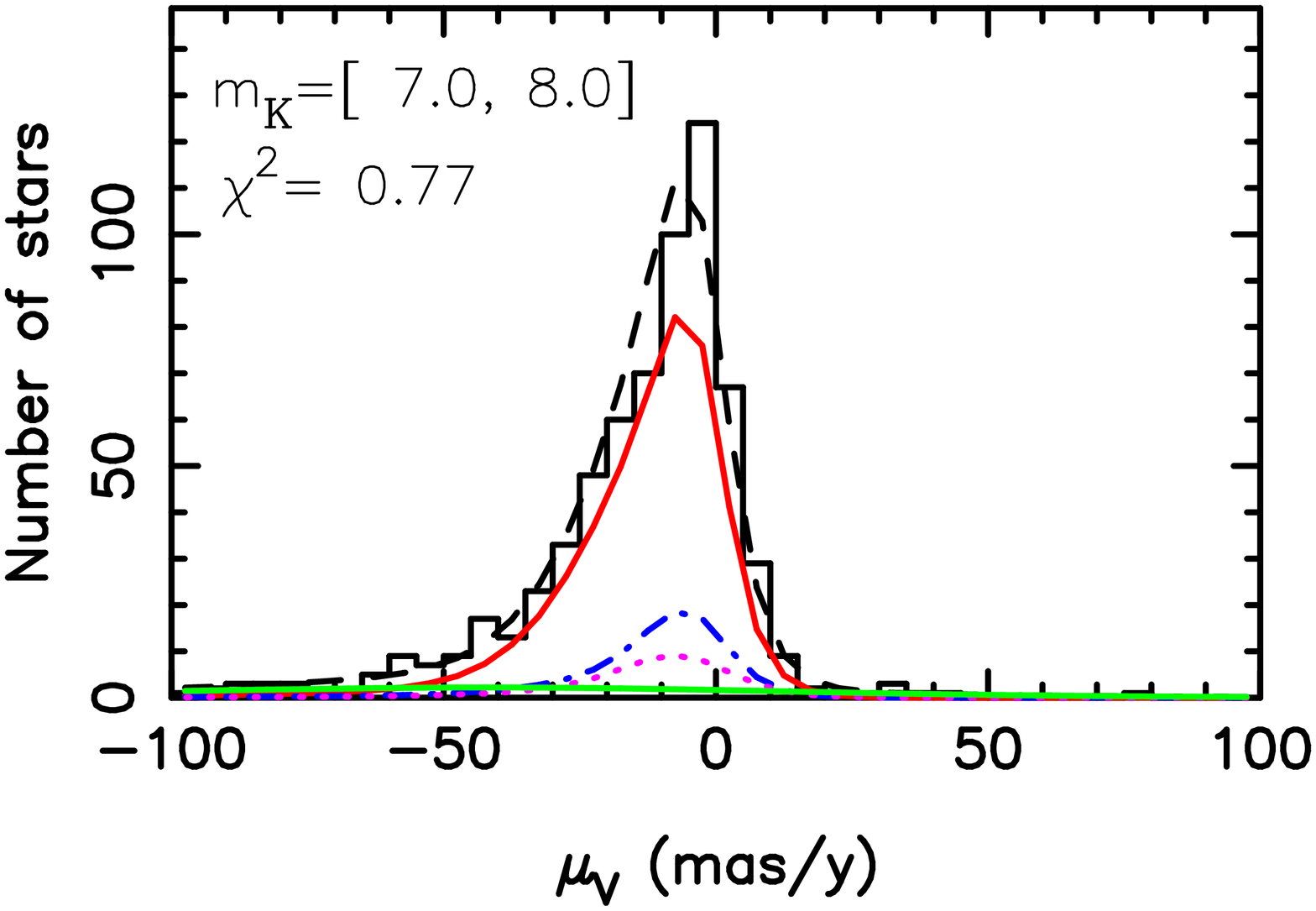}}
\scalebox{0.17}[0.21]{\includegraphics[angle=270]{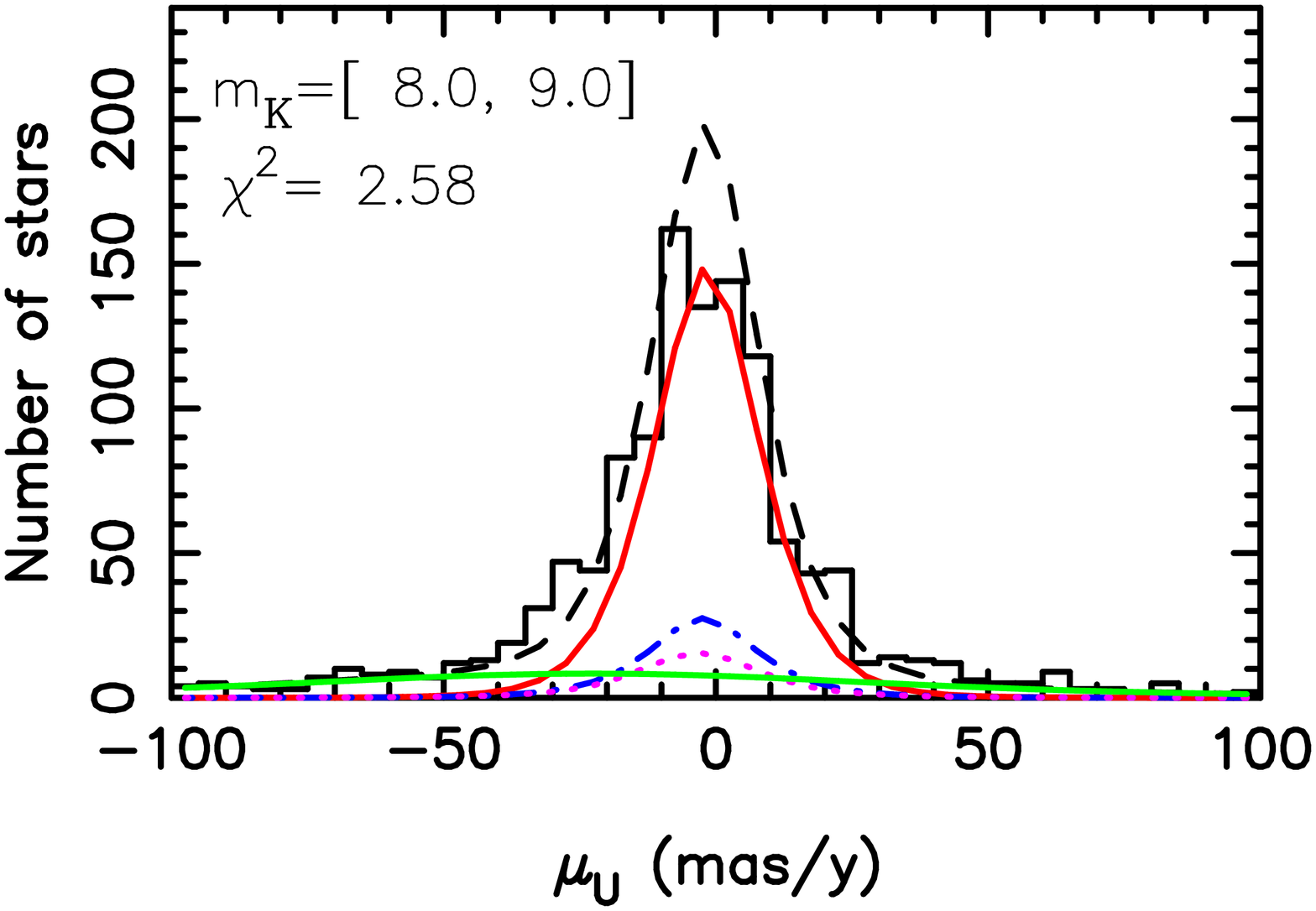}
	                      \includegraphics[angle=270]{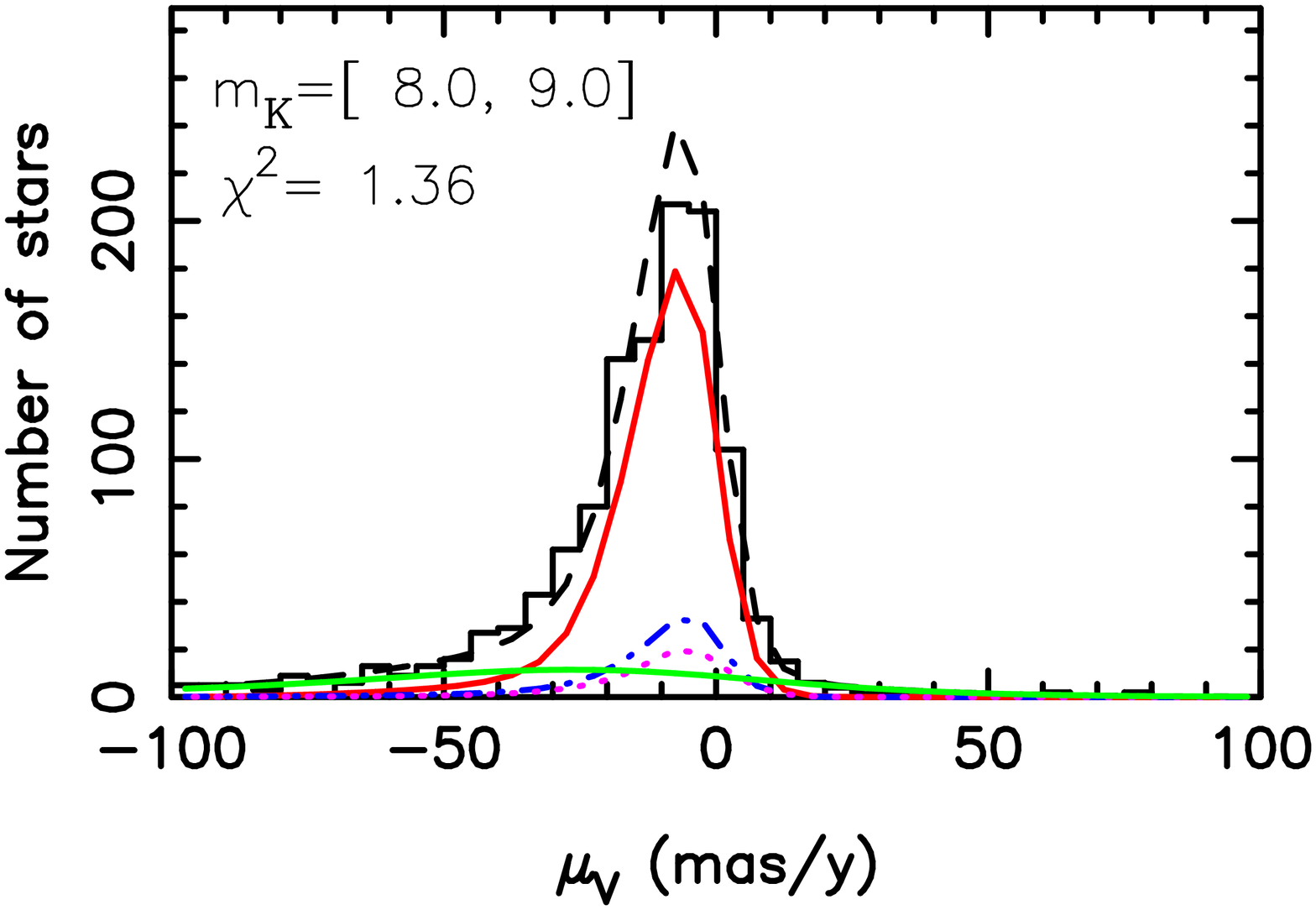}}
\scalebox{0.17}[0.21]{\includegraphics[angle=270]{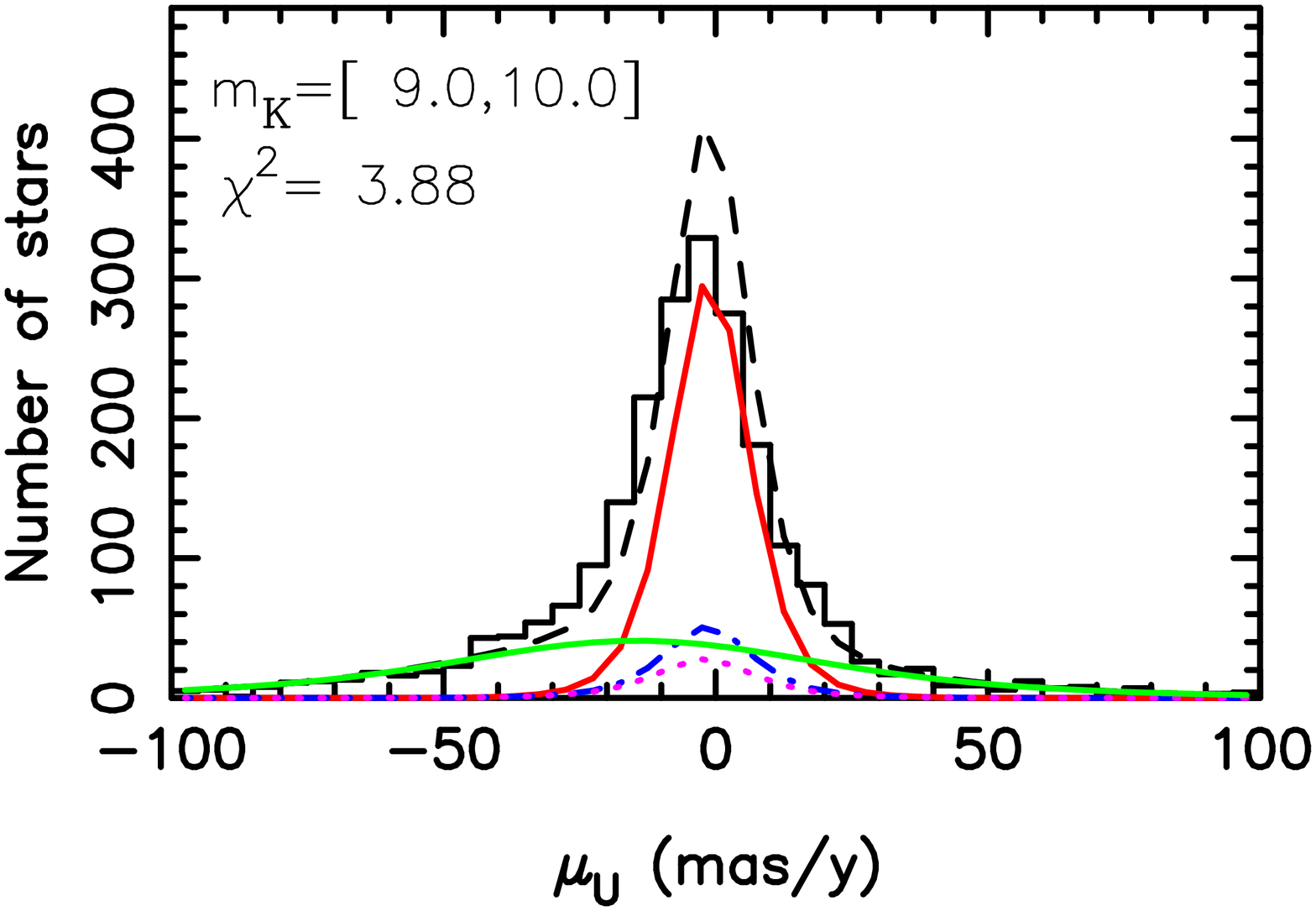}
	                      \includegraphics[angle=270]{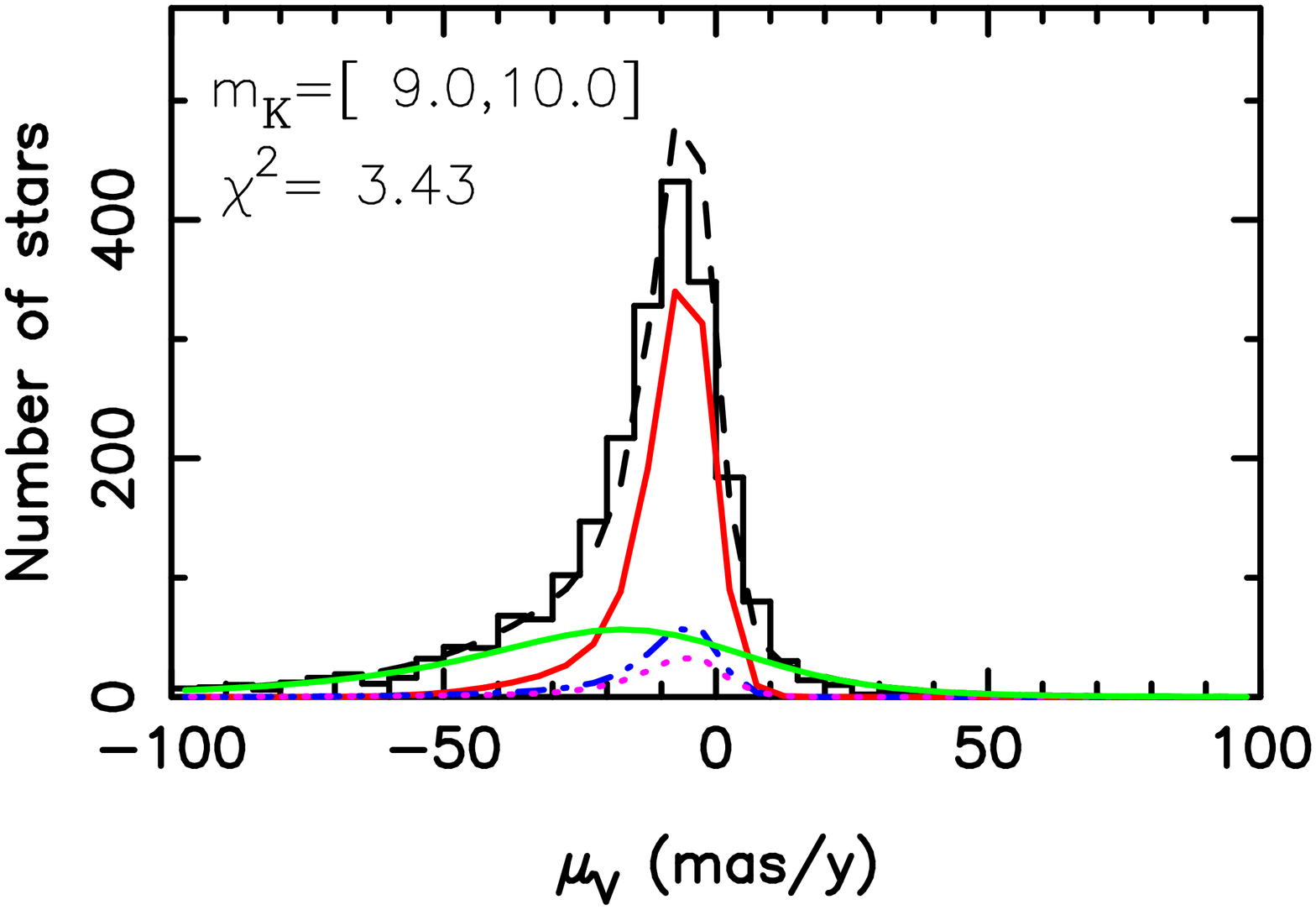}}
\end{minipage}
\hfill
\begin{minipage}{0.48\textwidth}
\scalebox{0.17}[0.21]{\includegraphics[angle=270]{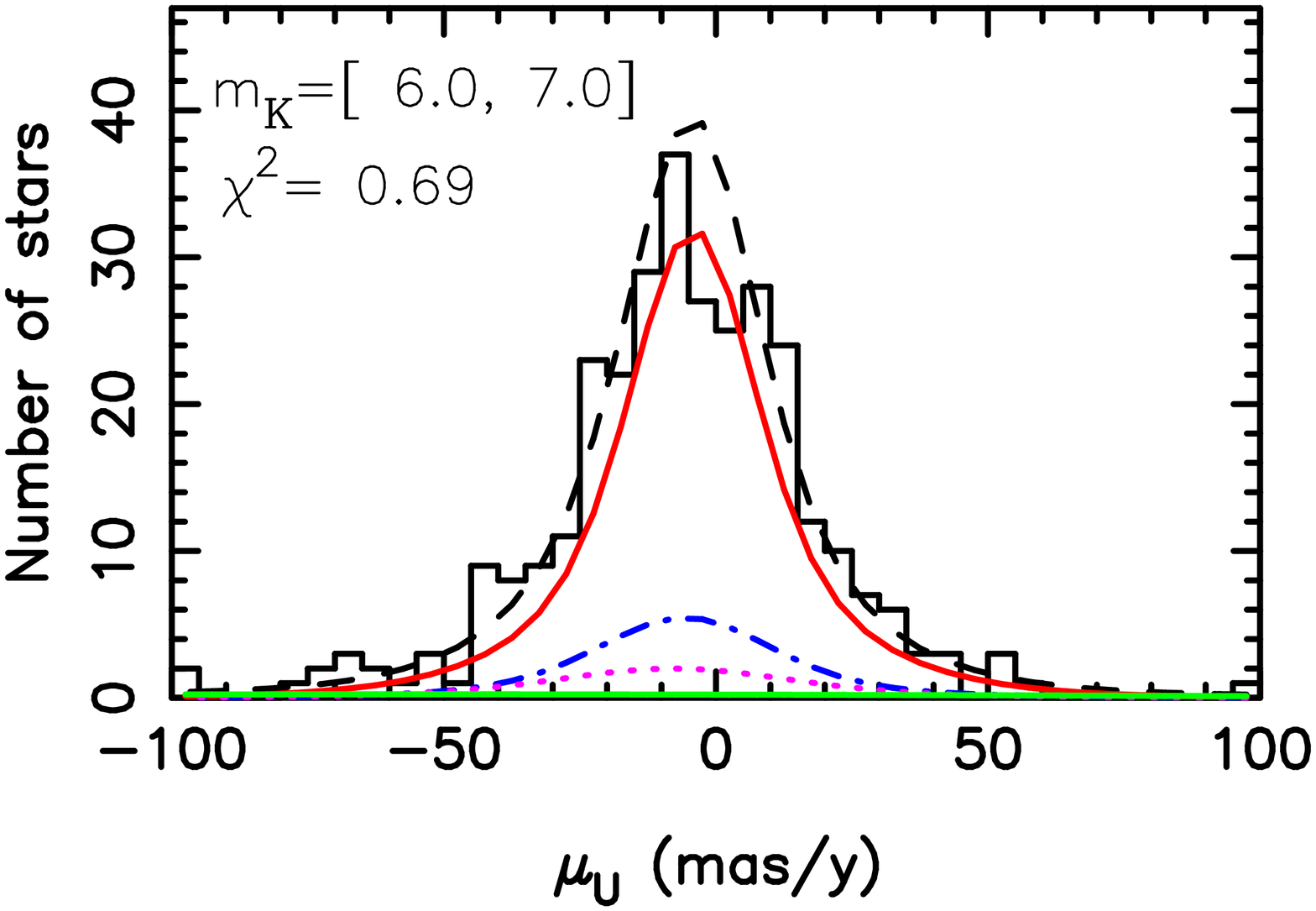}
	                      \includegraphics[angle=270]{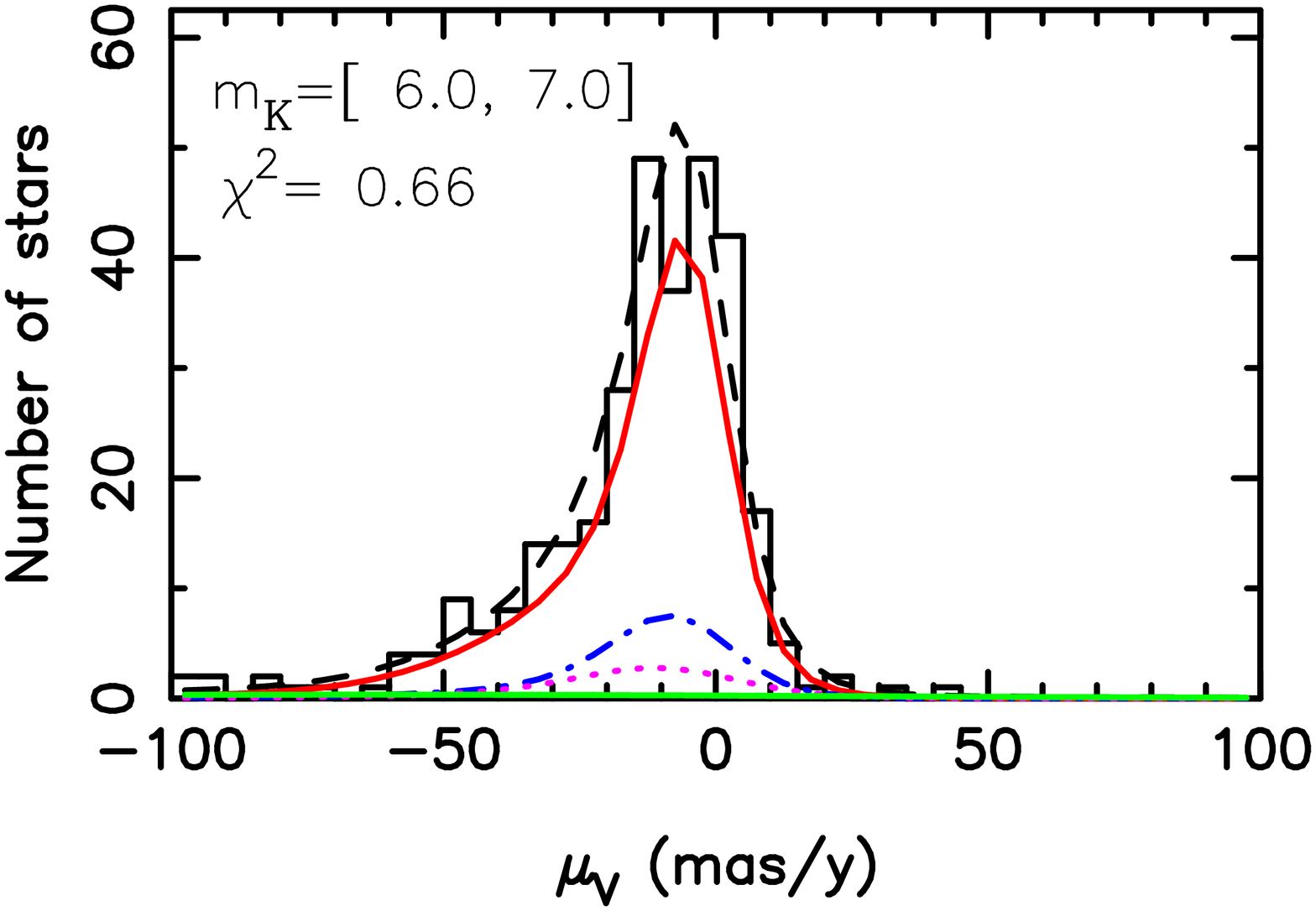}}
\scalebox{0.17}[0.21]{\includegraphics[angle=270]{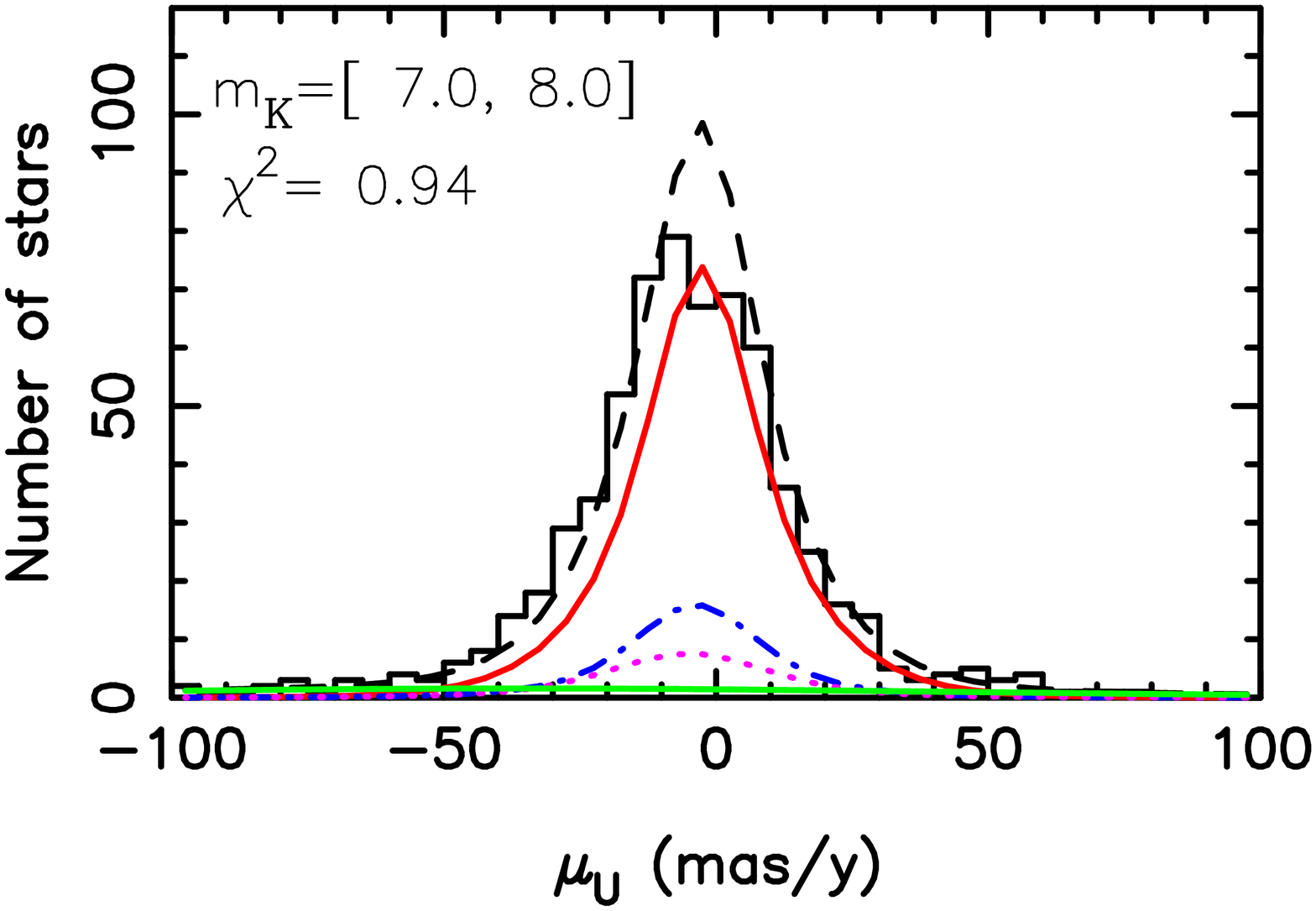}
	                      \includegraphics[angle=270]{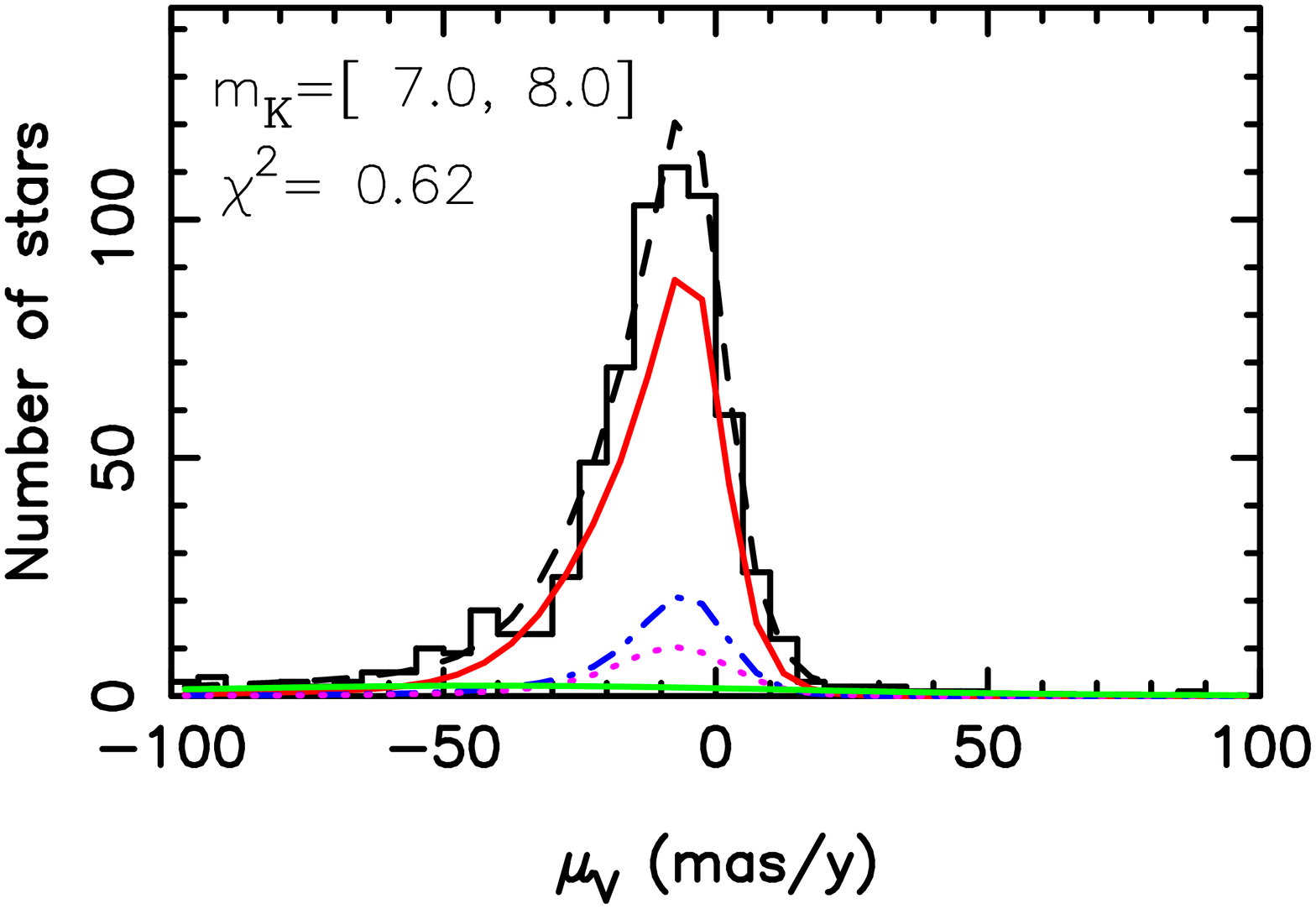}}
\scalebox{0.17}[0.21]{\includegraphics[angle=270]{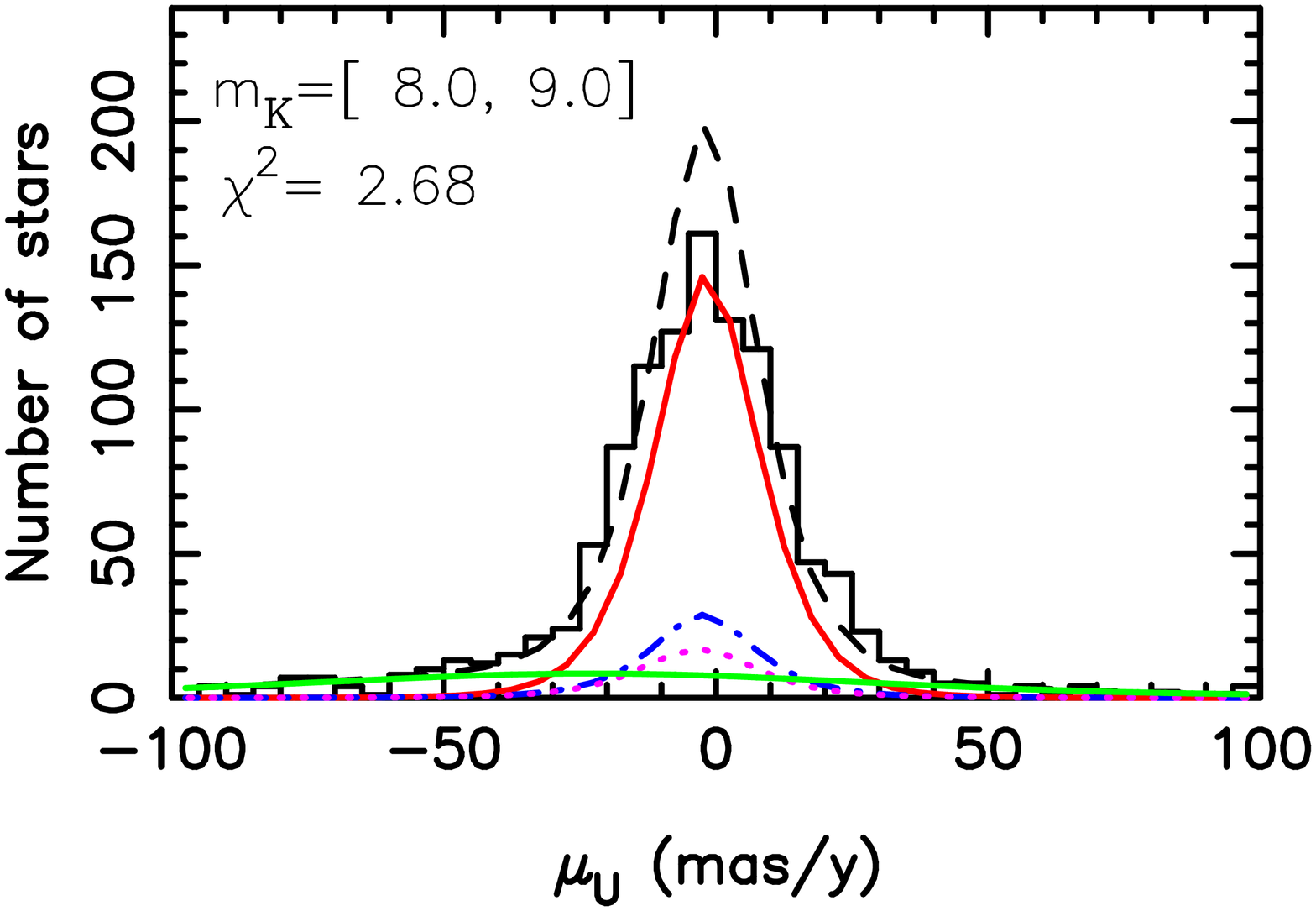}
	                      \includegraphics[angle=270]{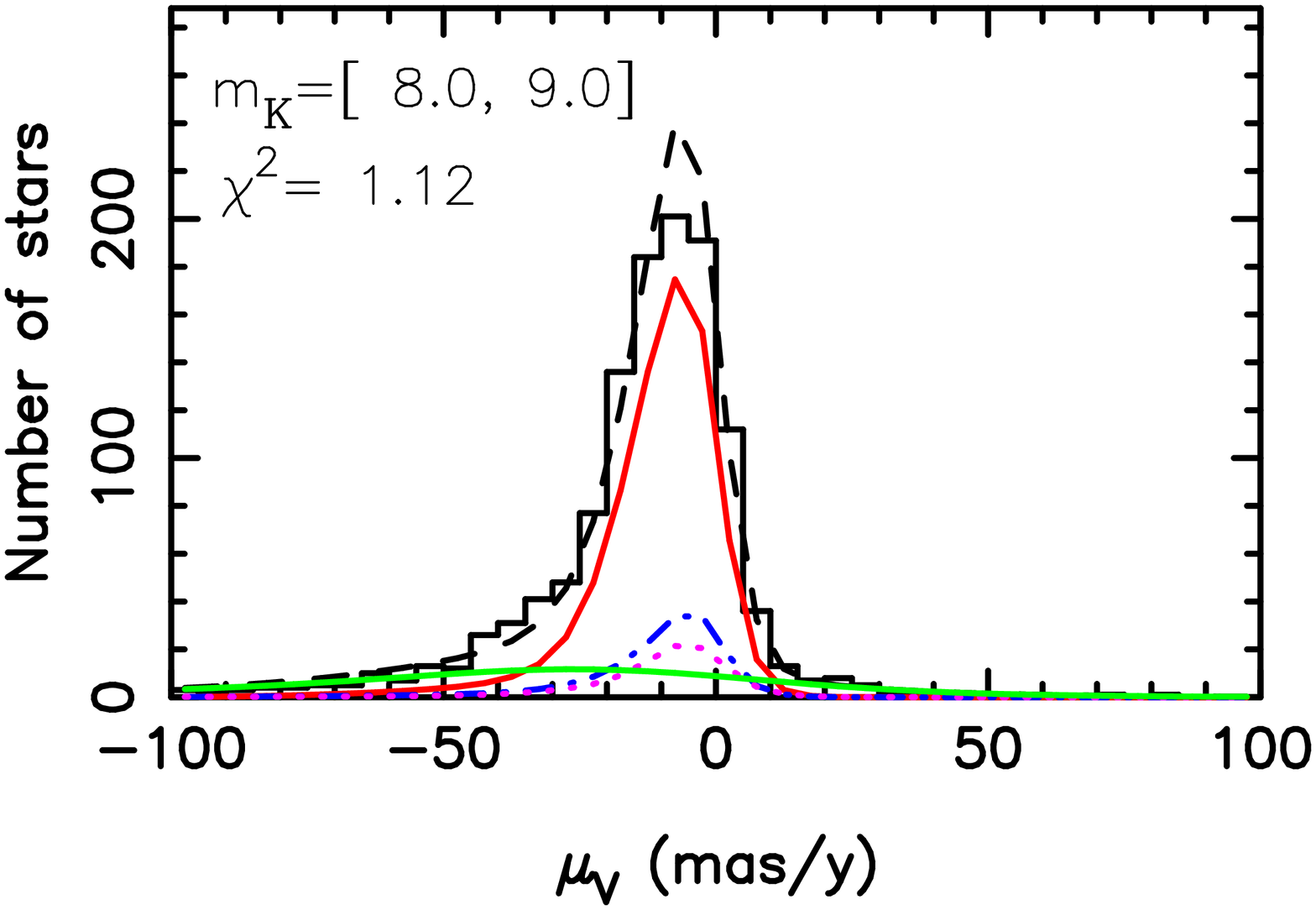}}
\scalebox{0.17}[0.21]{\includegraphics[angle=270]{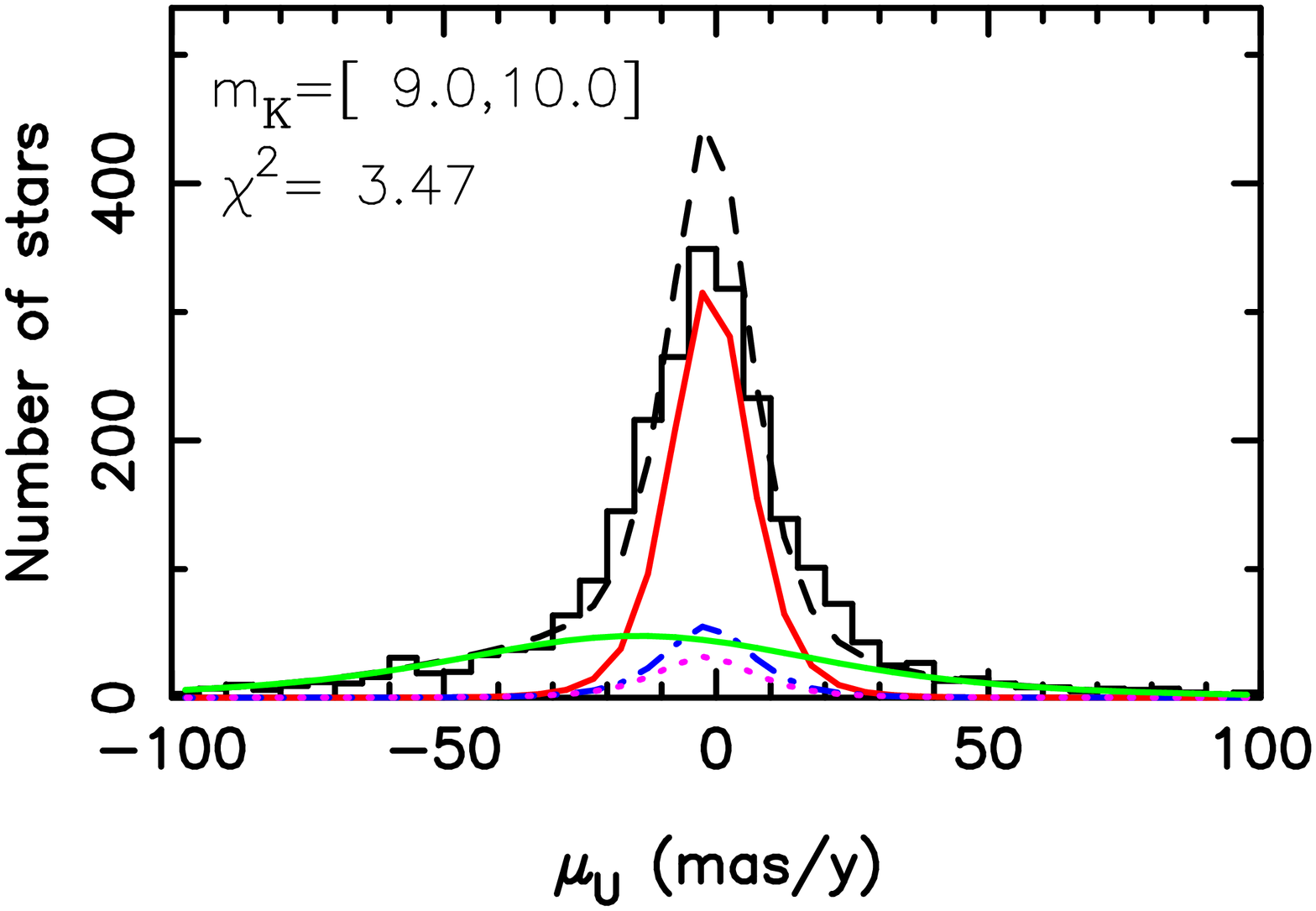}
	                      \includegraphics[angle=270]{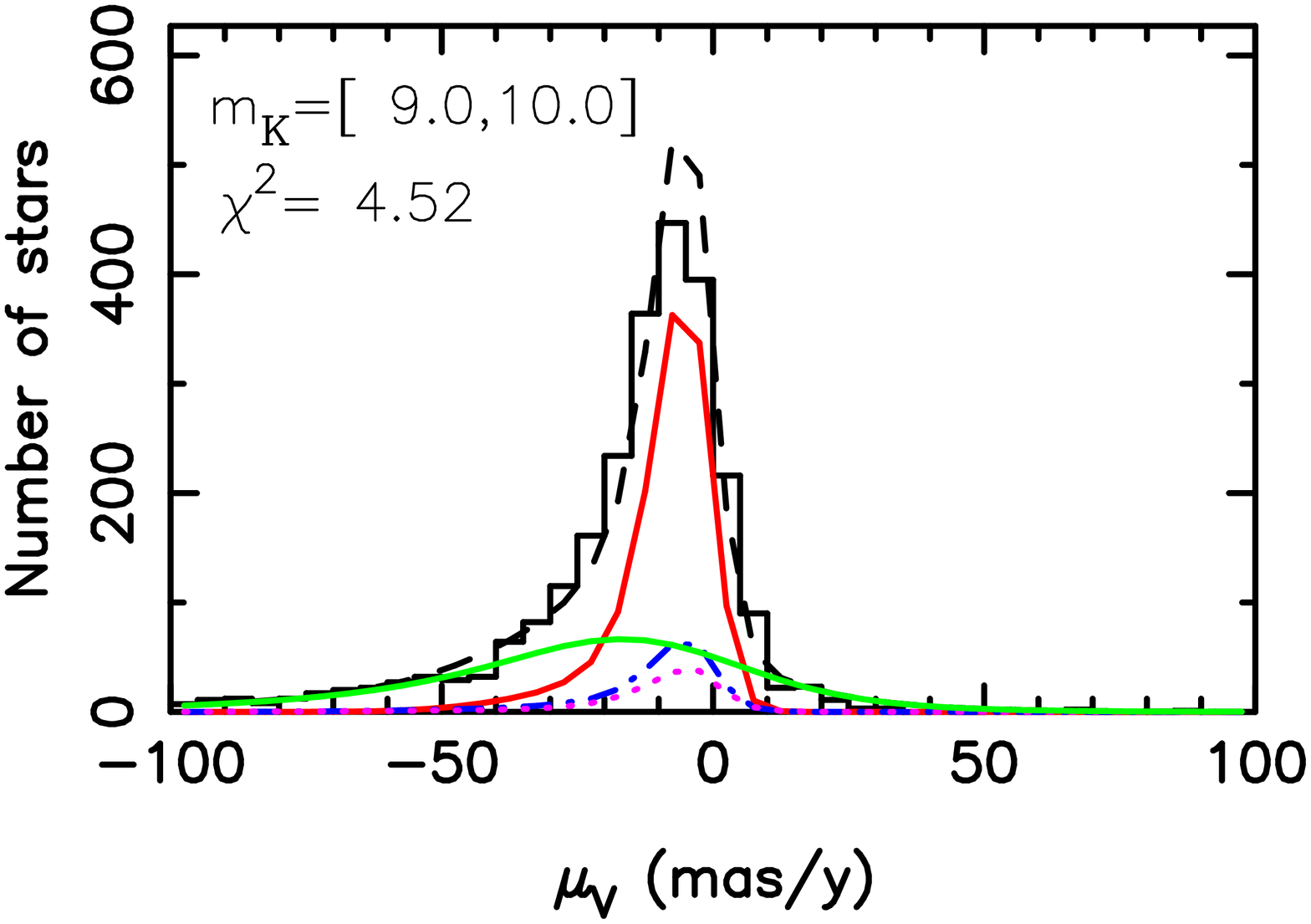}}
\end{minipage}
\caption{$\mu_U$  and $\mu_V$  histograms towards  the  North Galactic
Pole (right)  and the South Galactic  Pole (left) for  magnitudes 6 to
10: model (dashed  line) and contributions from the  different types of
stars:  giants (red  or dark  thin lines),  sub-giants  (dot-dashed and
dotted lines) and dwarfs (green or grey thick lines).}
\label{f:vitesse}
\end{figure*}

\begin{figure*}
\begin{minipage}{.48\textwidth}
\scalebox{0.17}[0.21]{\includegraphics[angle=270]{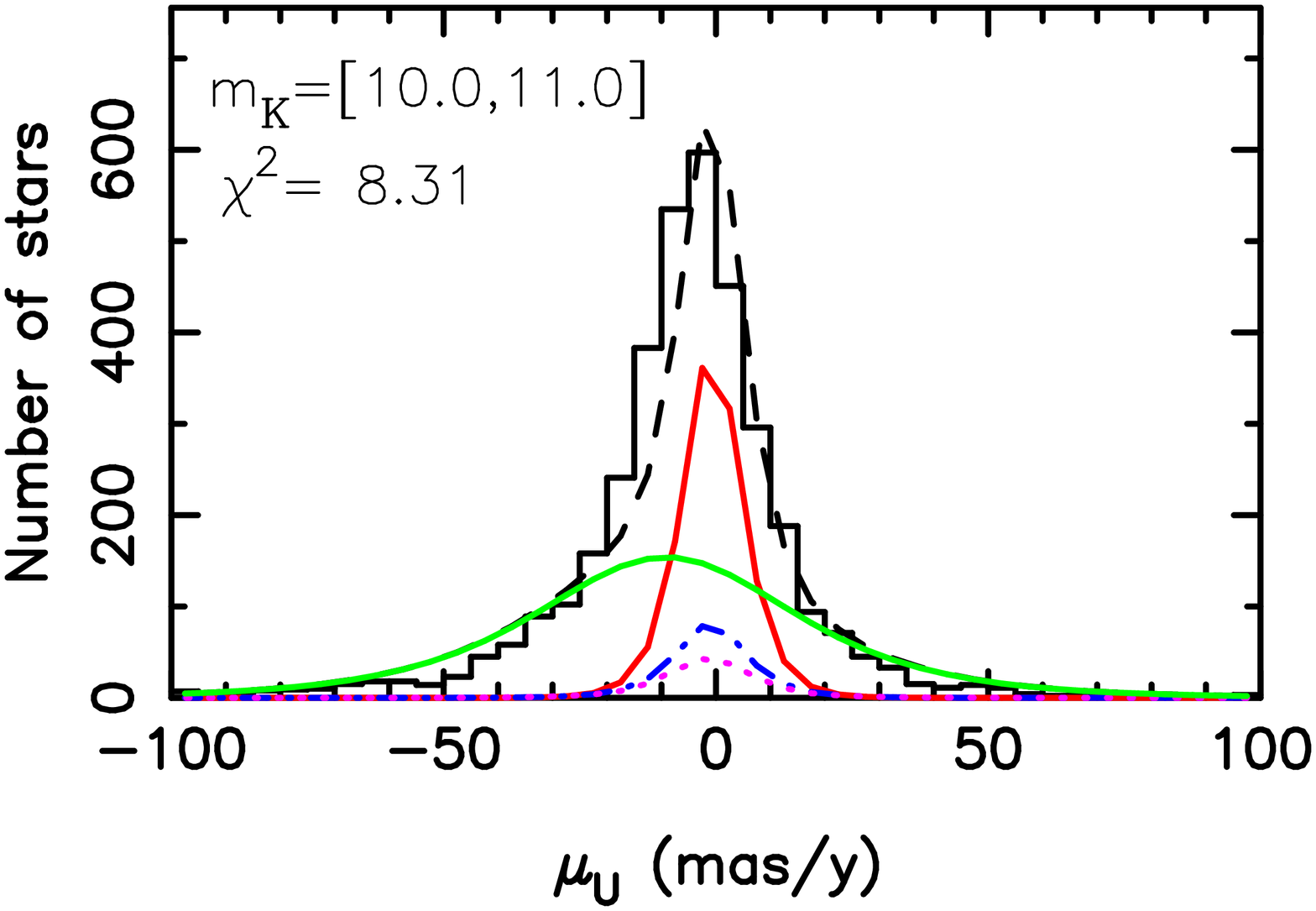}
	                      \includegraphics[angle=270]{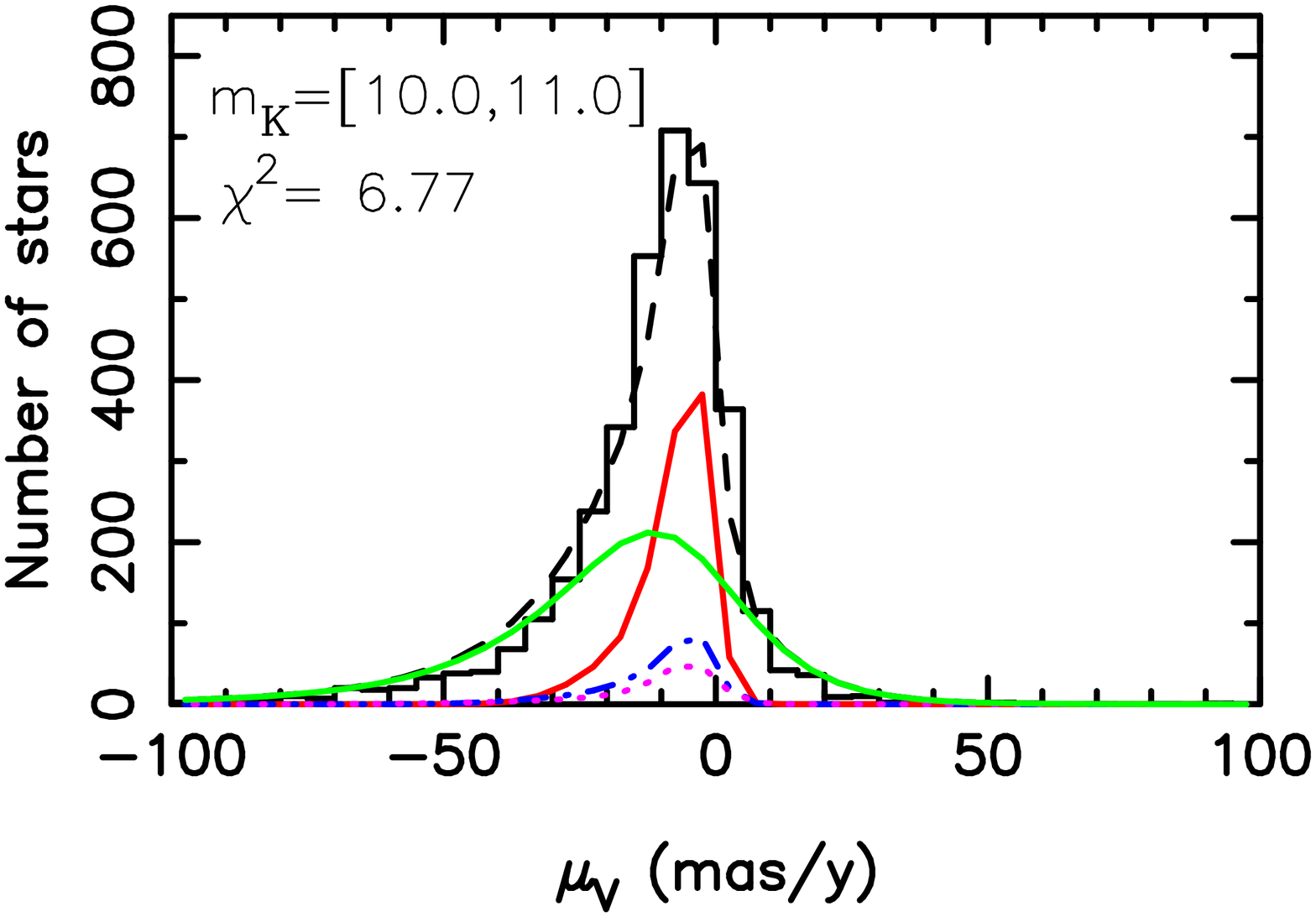}}
\scalebox{0.17}[0.21]{\includegraphics[angle=270]{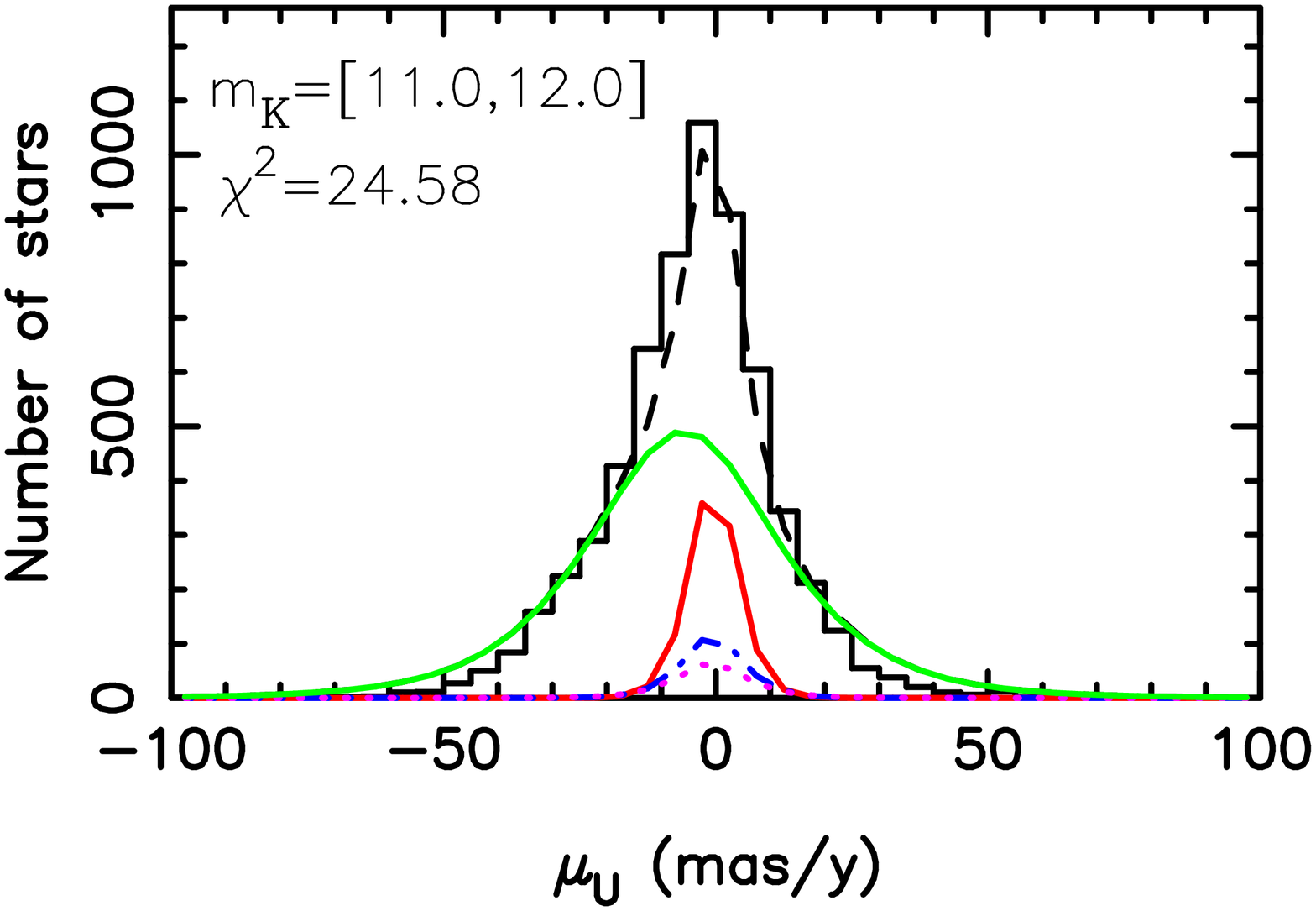}
	                      \includegraphics[angle=270]{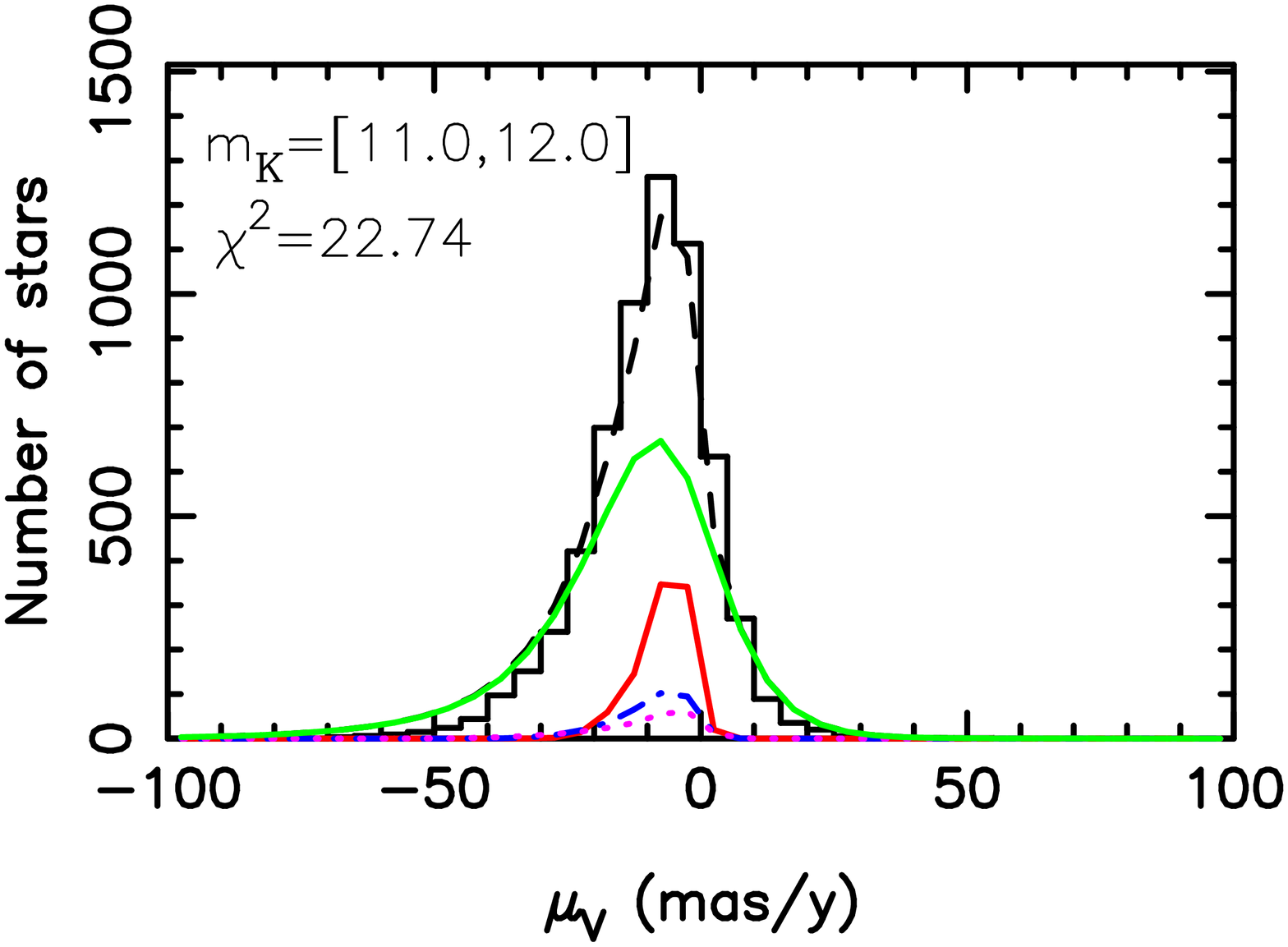}}
\scalebox{0.17}[0.21]{\includegraphics[angle=270]{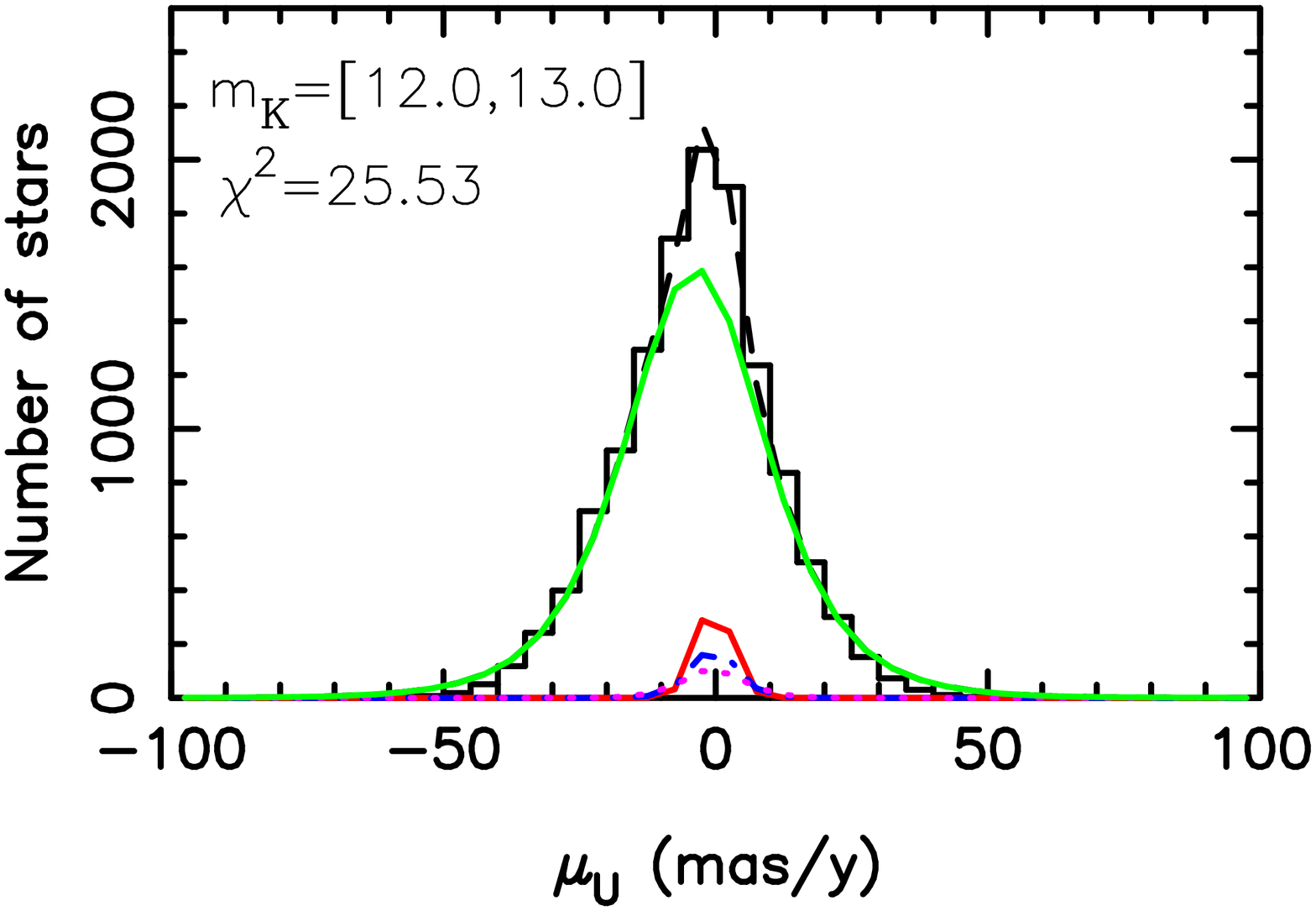}
	                      \includegraphics[angle=270]{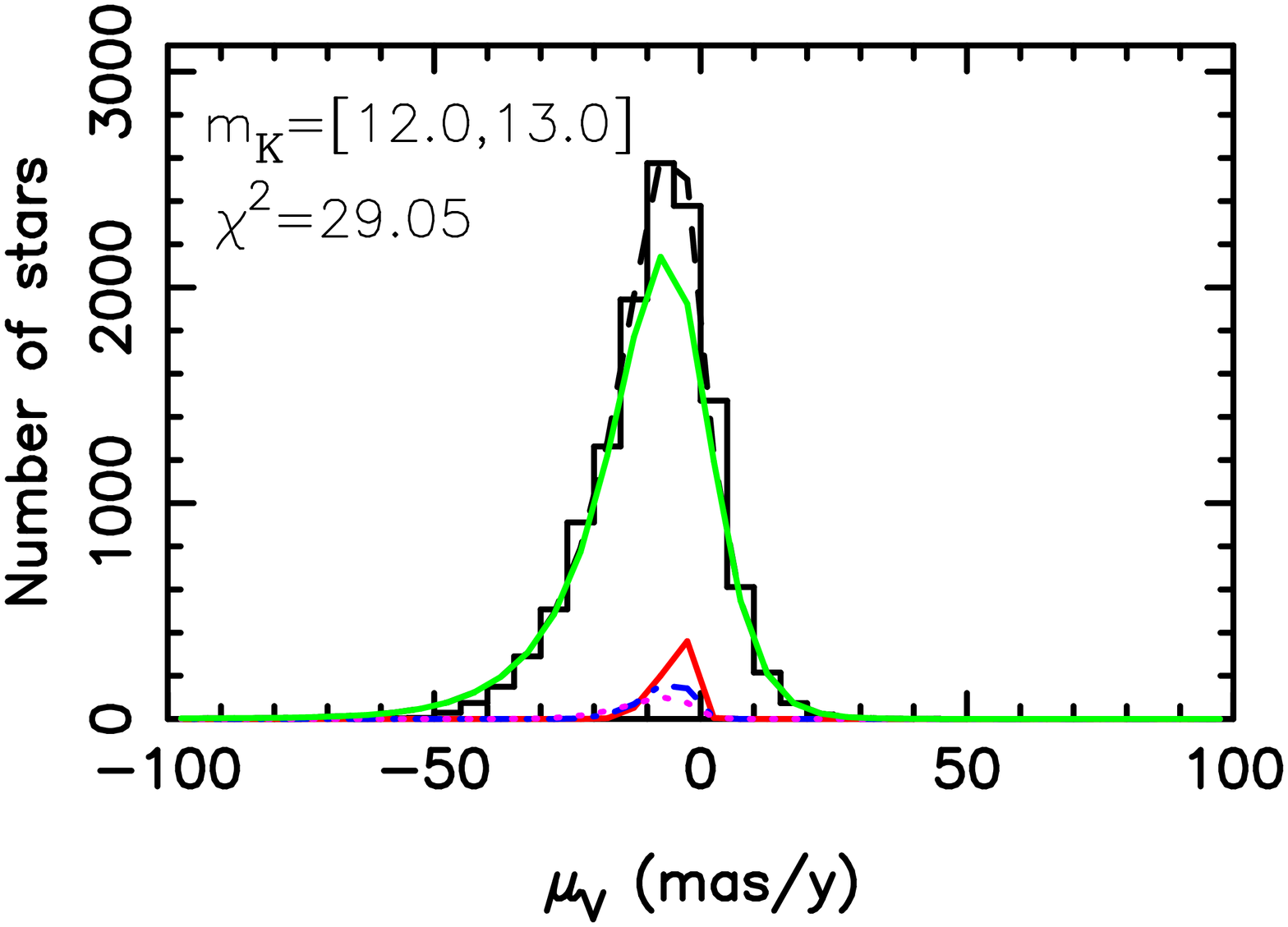}}
\scalebox{0.17}[0.21]{\includegraphics[angle=270]{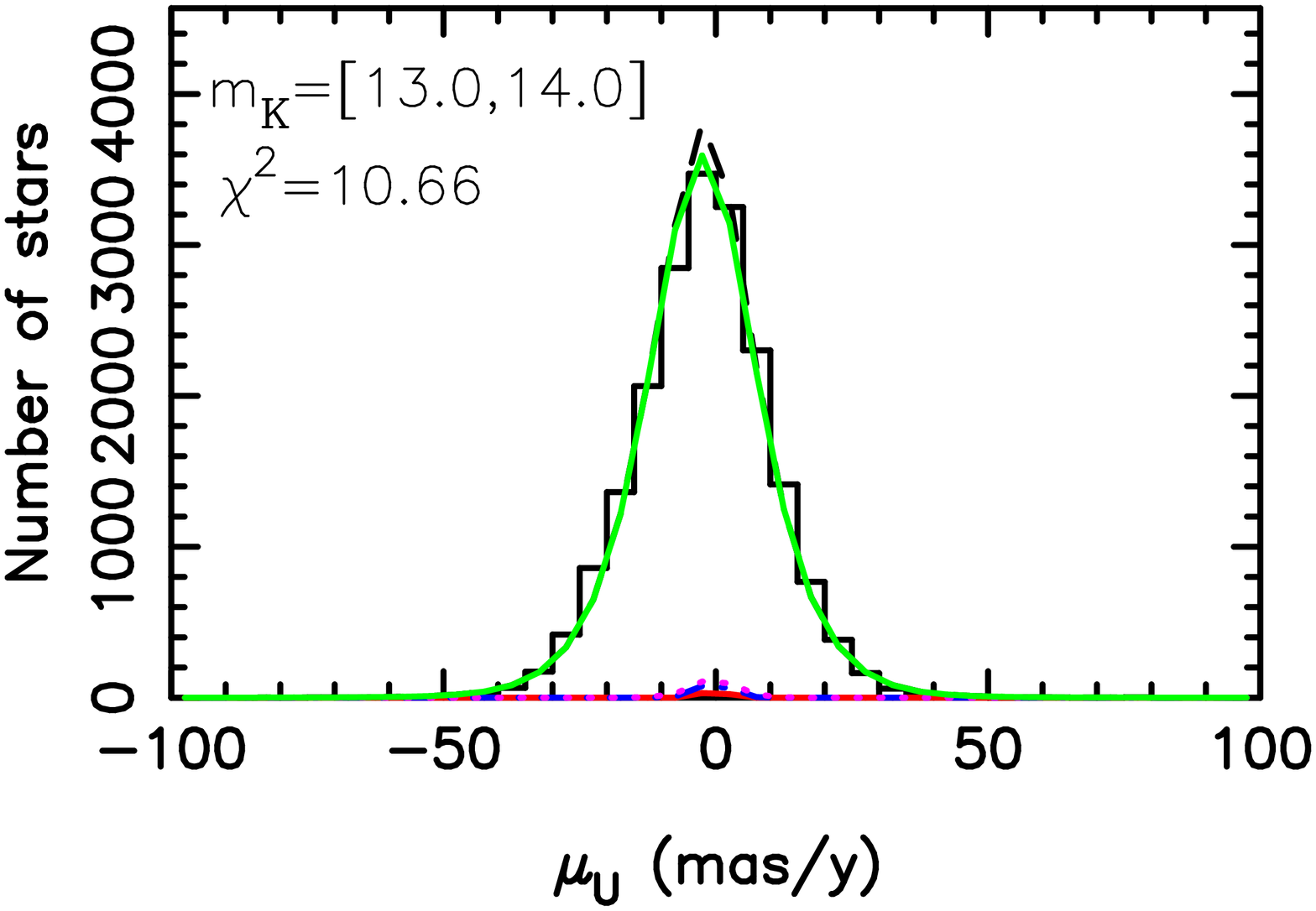}
	                      \includegraphics[angle=270]{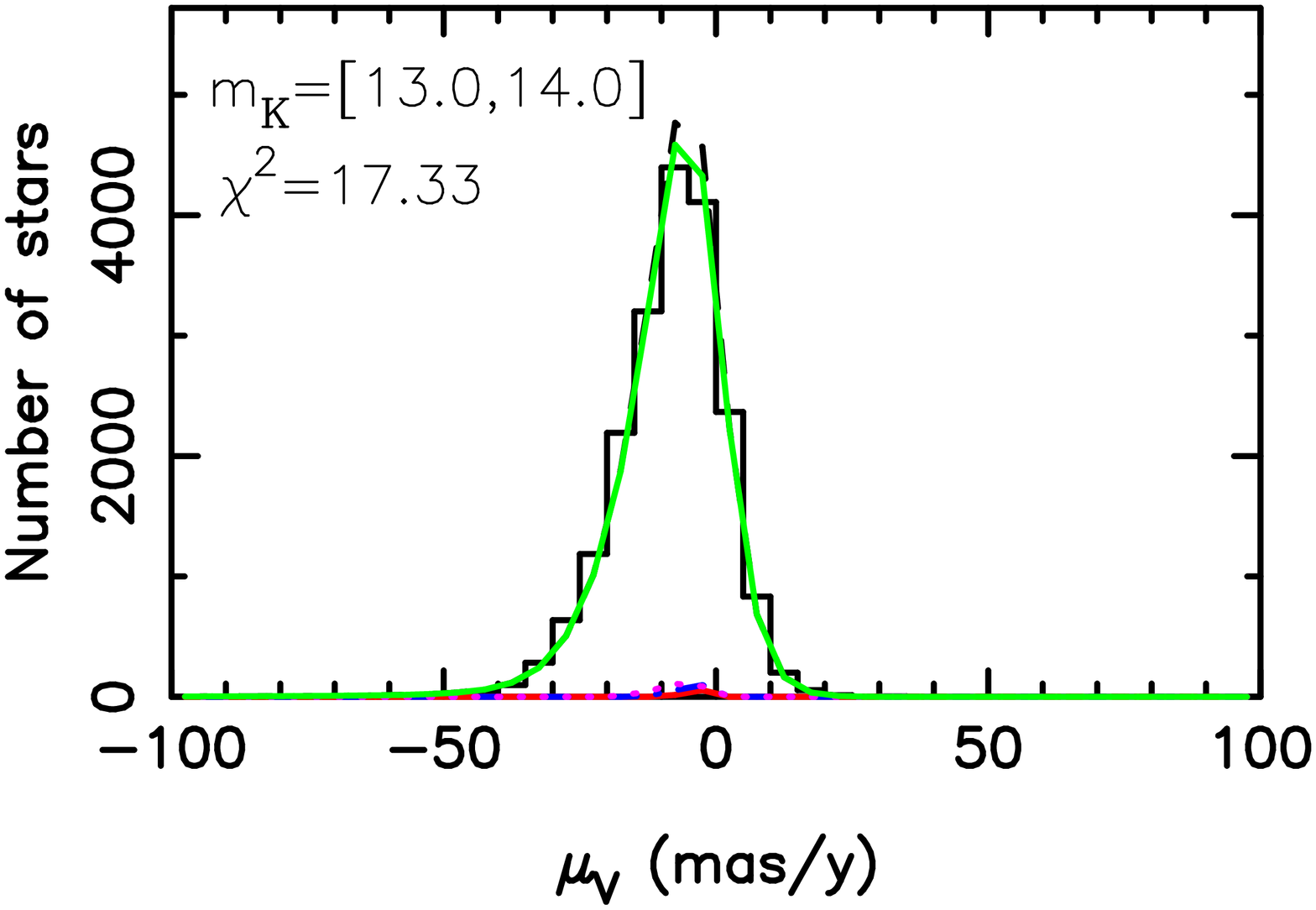}}
\end{minipage}
\hfill
\begin{minipage}{0.48\textwidth}
\scalebox{0.17}[0.21]{\includegraphics[angle=270]{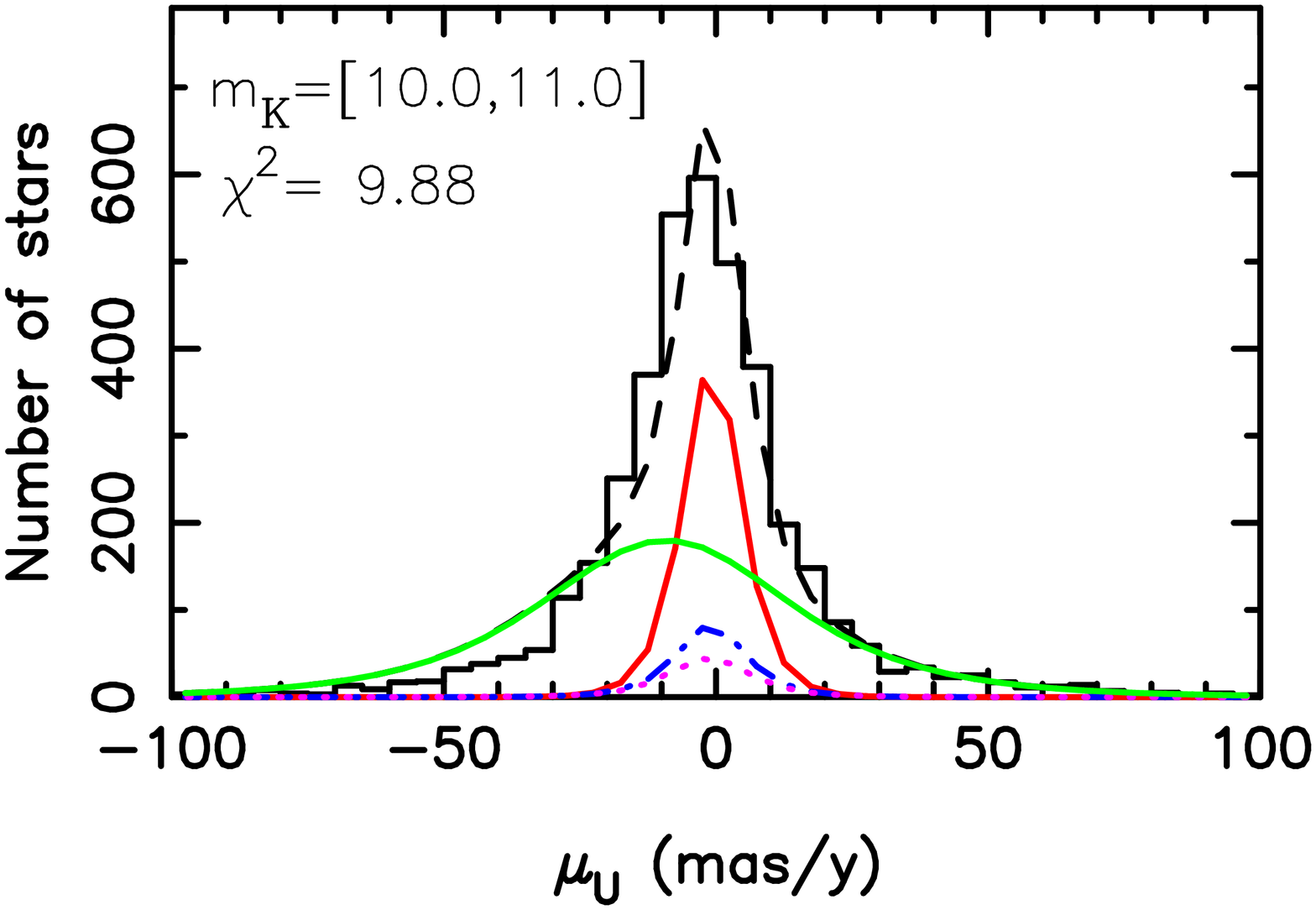}
	                      \includegraphics[angle=270]{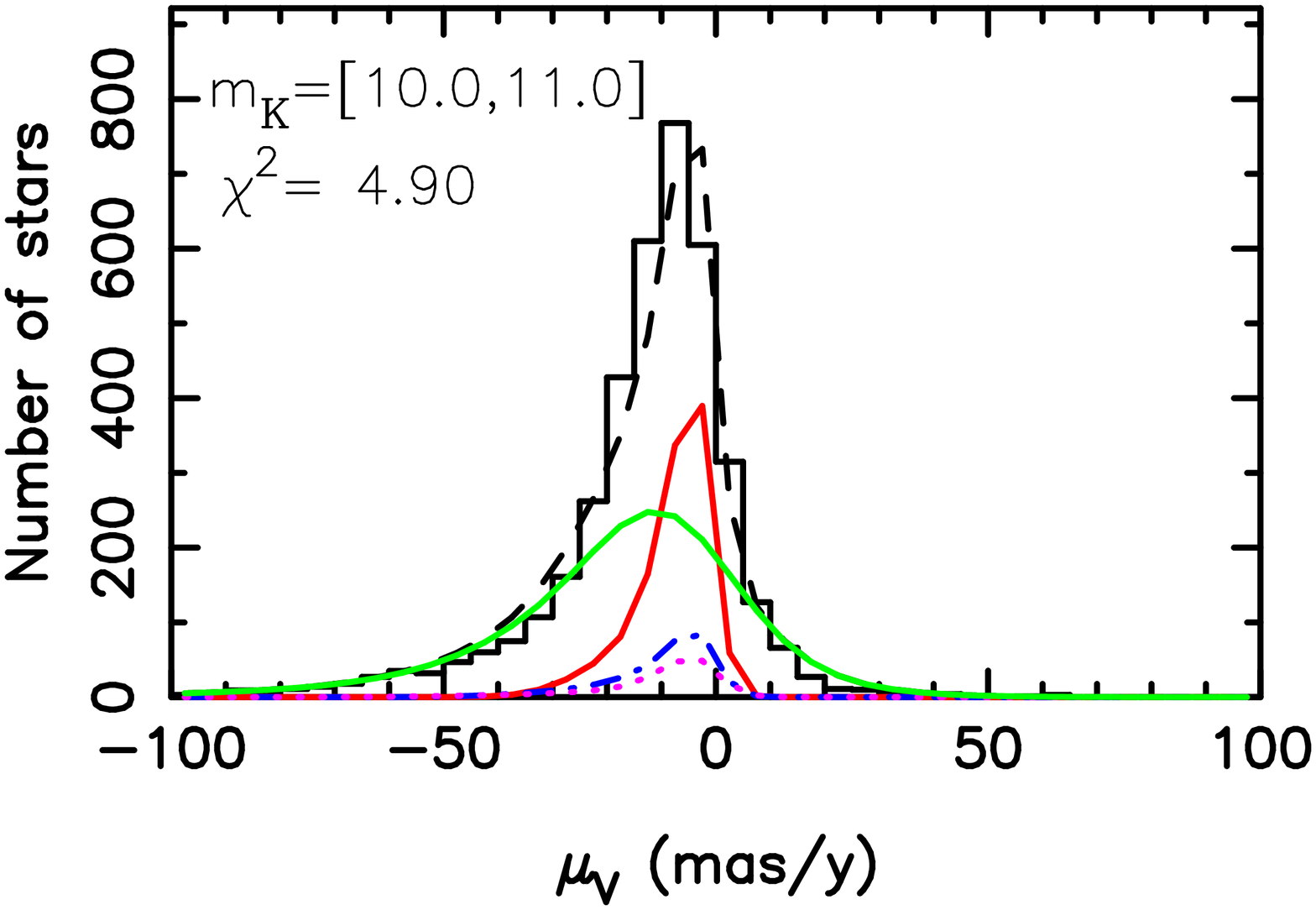}}
\scalebox{0.17}[0.21]{\includegraphics[angle=270]{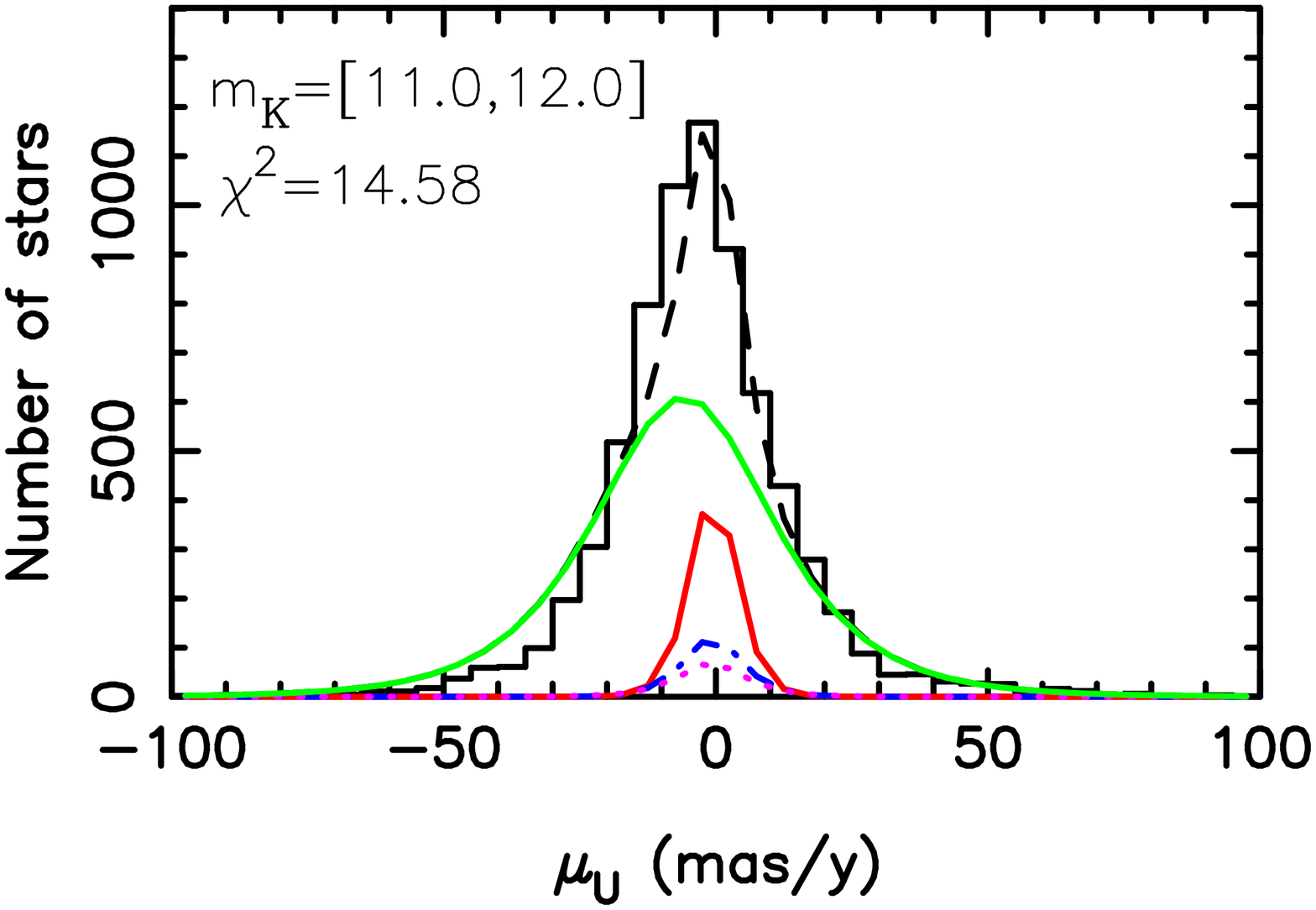}
	                      \includegraphics[angle=270]{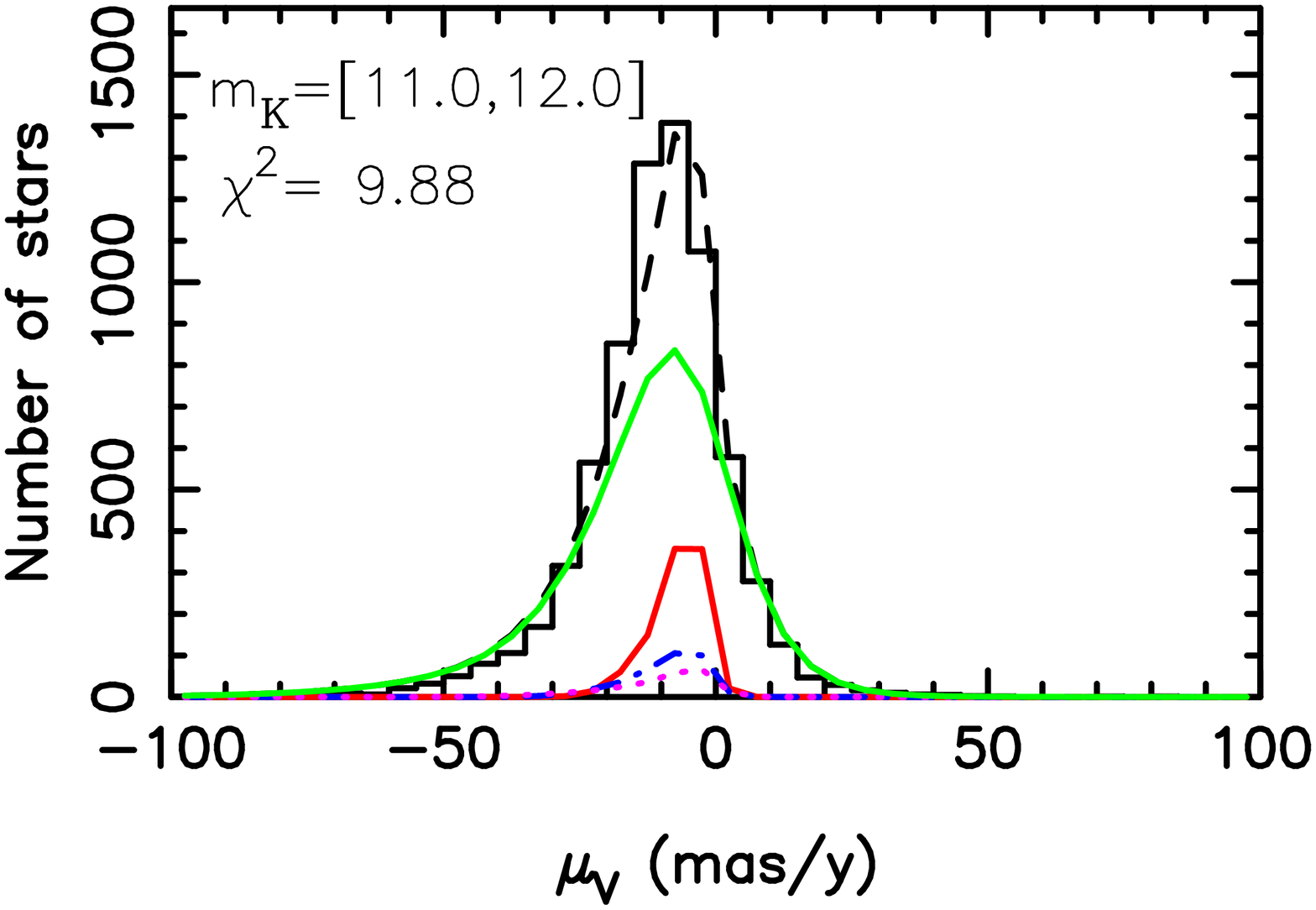}}
\scalebox{0.17}[0.21]{\includegraphics[angle=270]{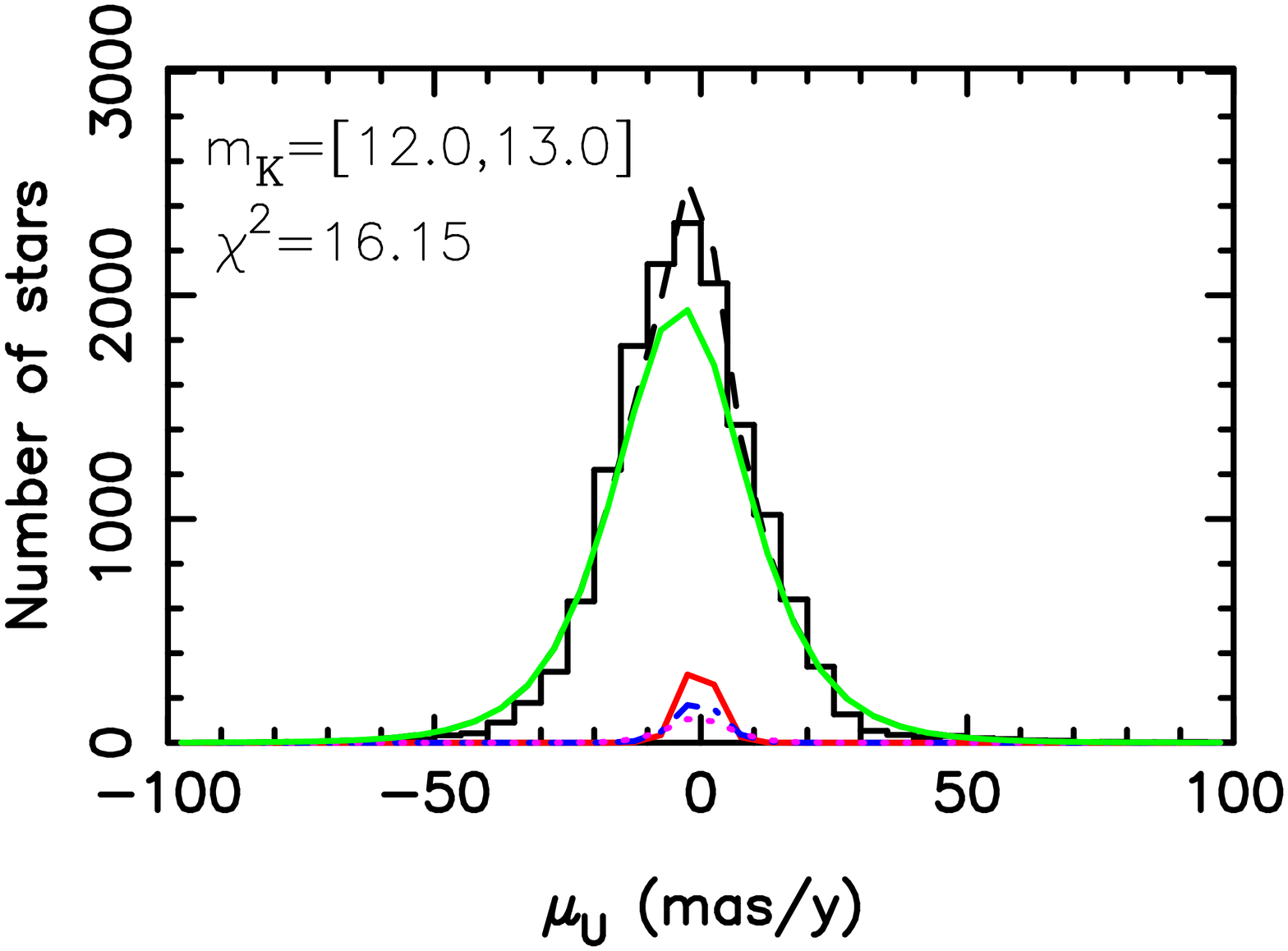}
	                      \includegraphics[angle=270]{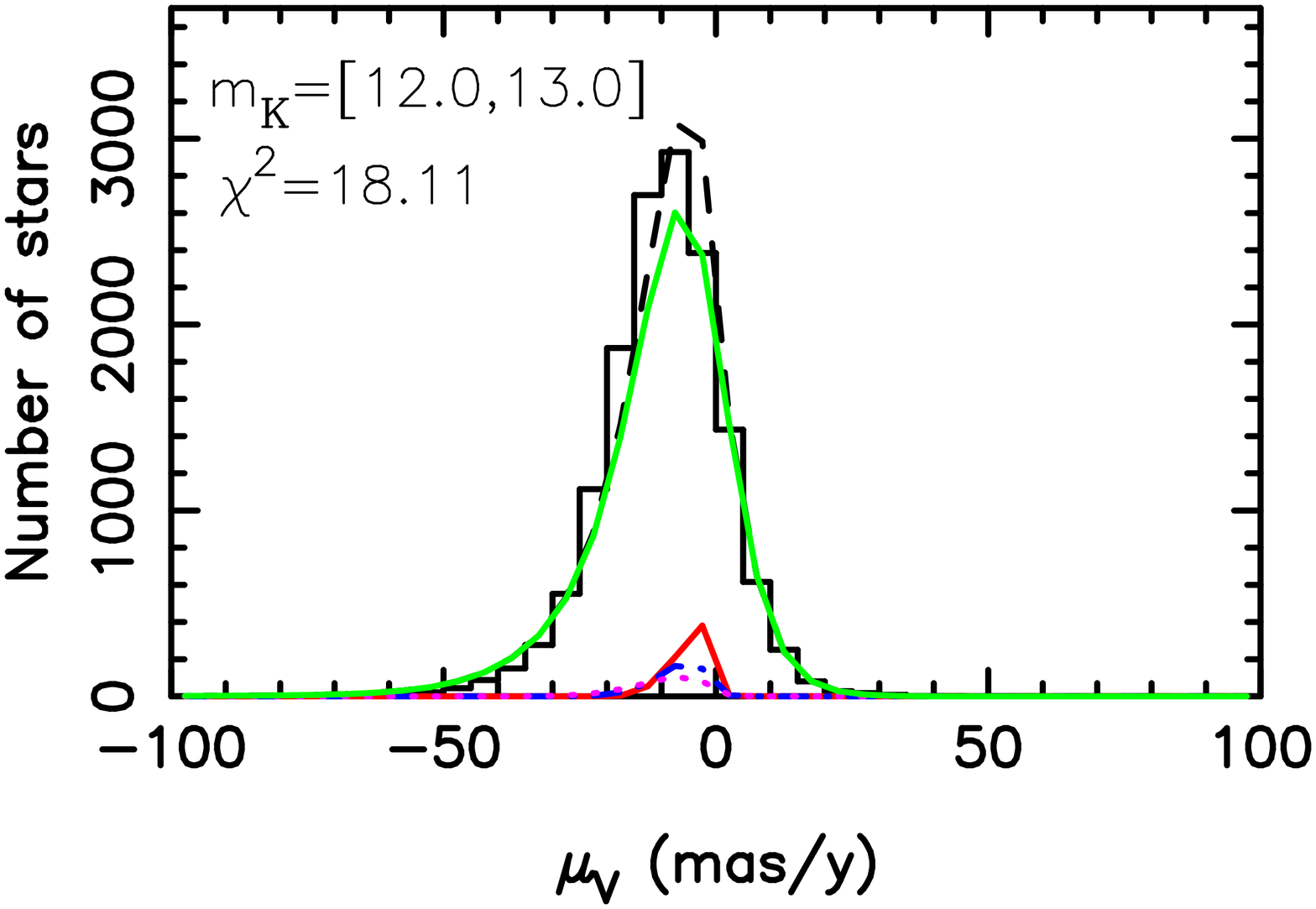}}
\scalebox{0.17}[0.21]{\includegraphics[angle=270]{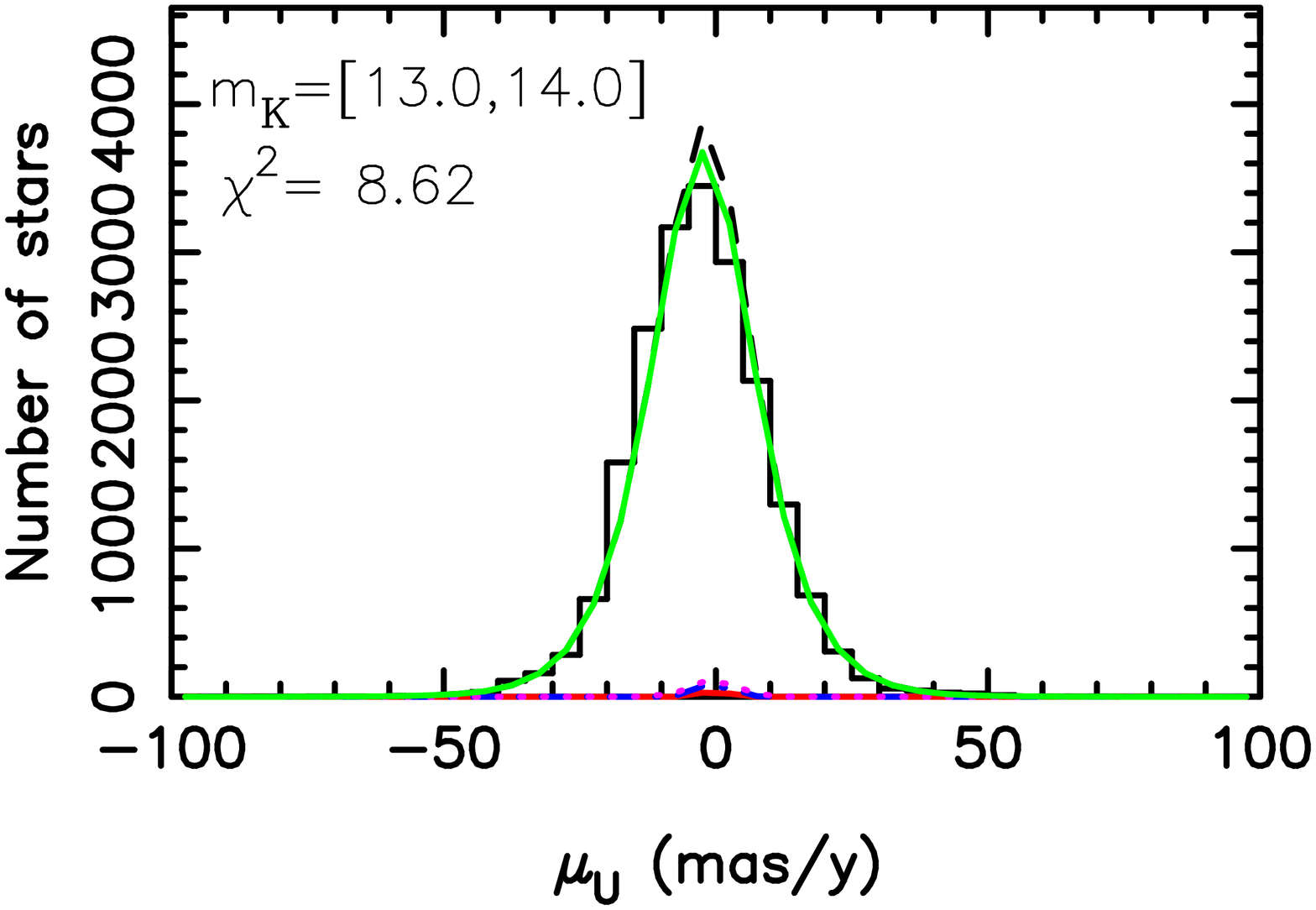}
	                      \includegraphics[angle=270]{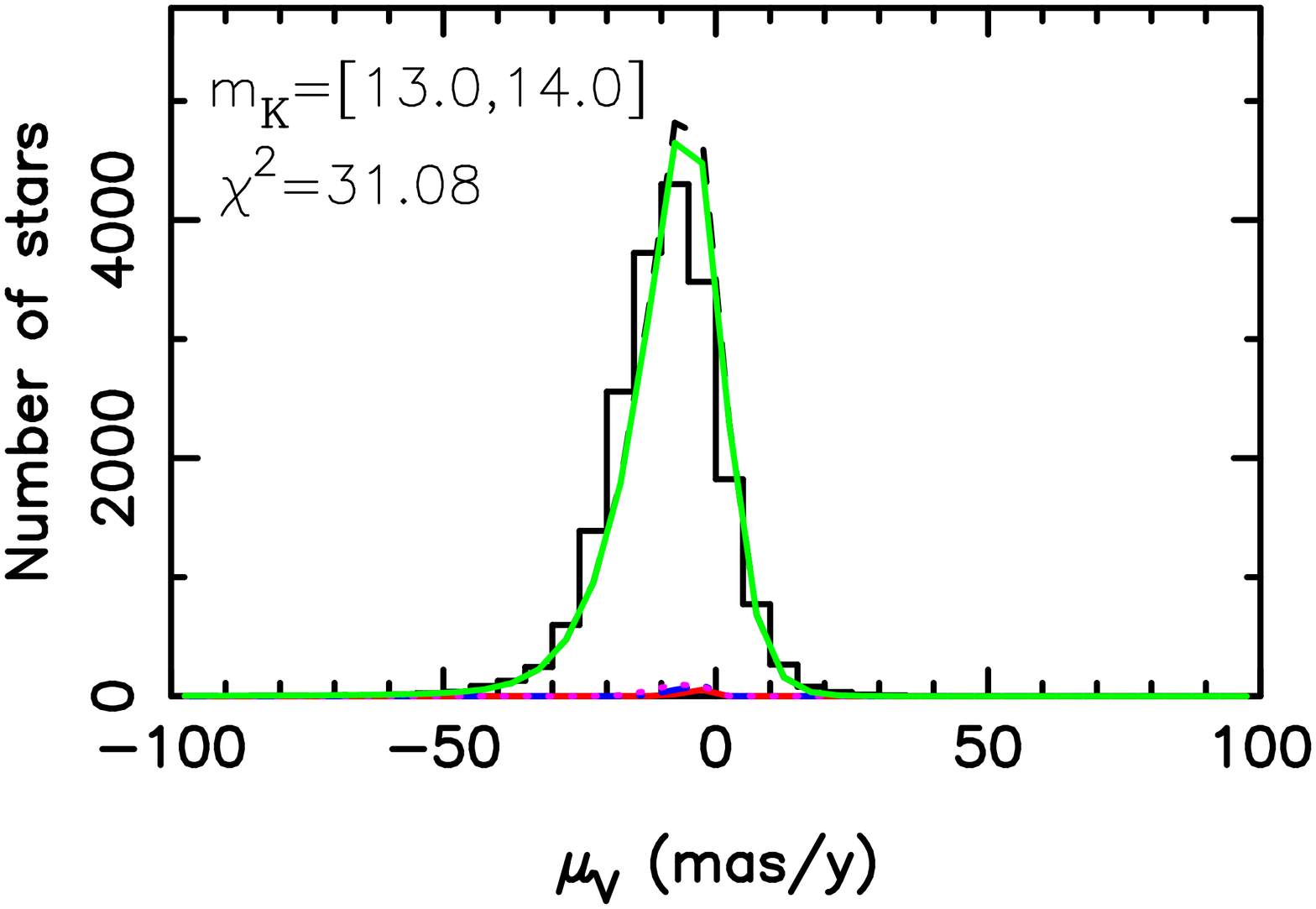}}
\end{minipage}
\caption{Same as Fig.~\ref{f:vitesse} for magnitudes 10 to 14.}
\label{f:vitesse2}
\end{figure*}

\begin{figure*}[!htbp]
\center   
\begin{tabular}{ccc}
\includegraphics[width=4.1cm,angle=270]{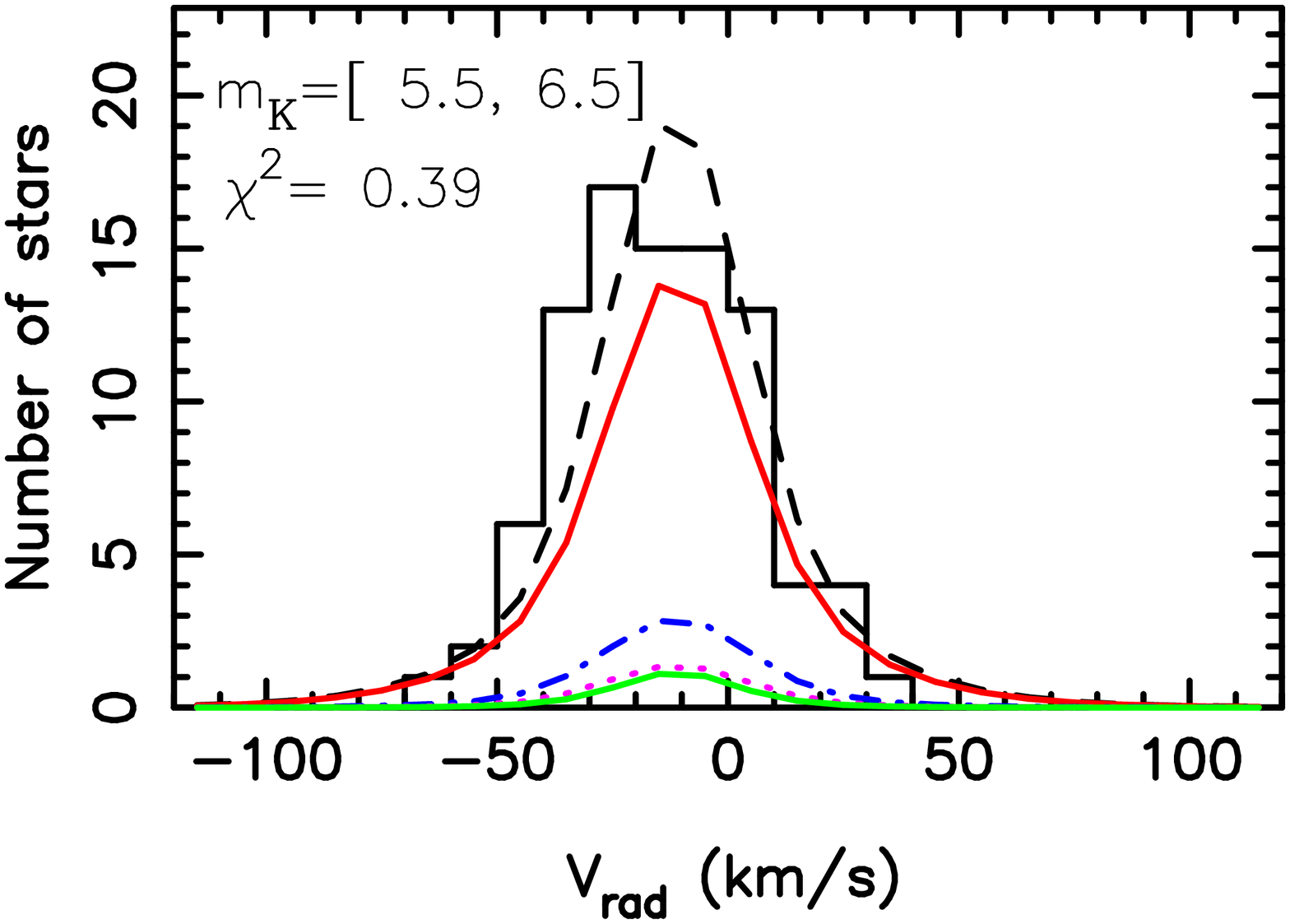} &
\includegraphics[width=4.1cm,angle=270]{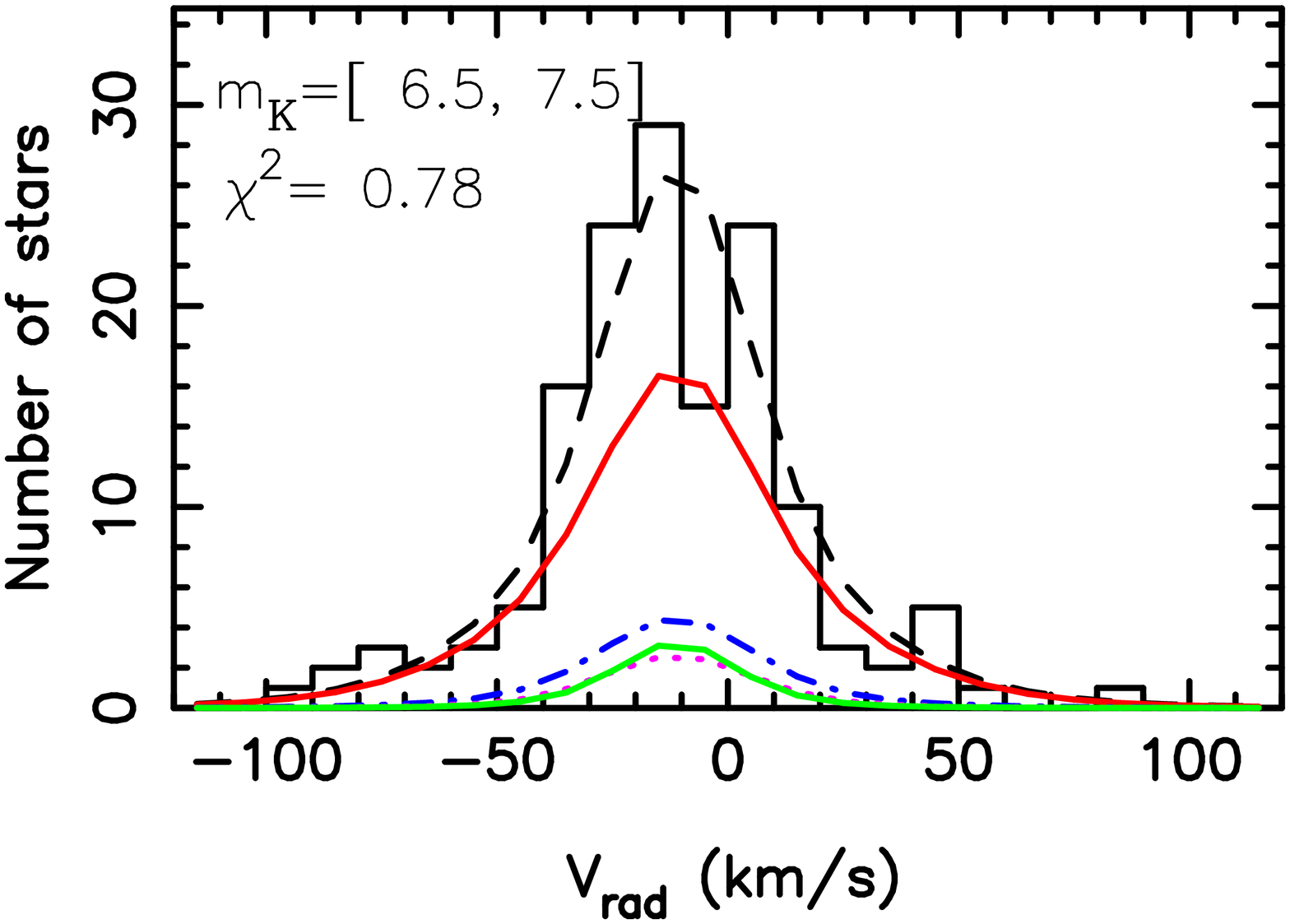} &
\includegraphics[width=4.1cm,angle=270]{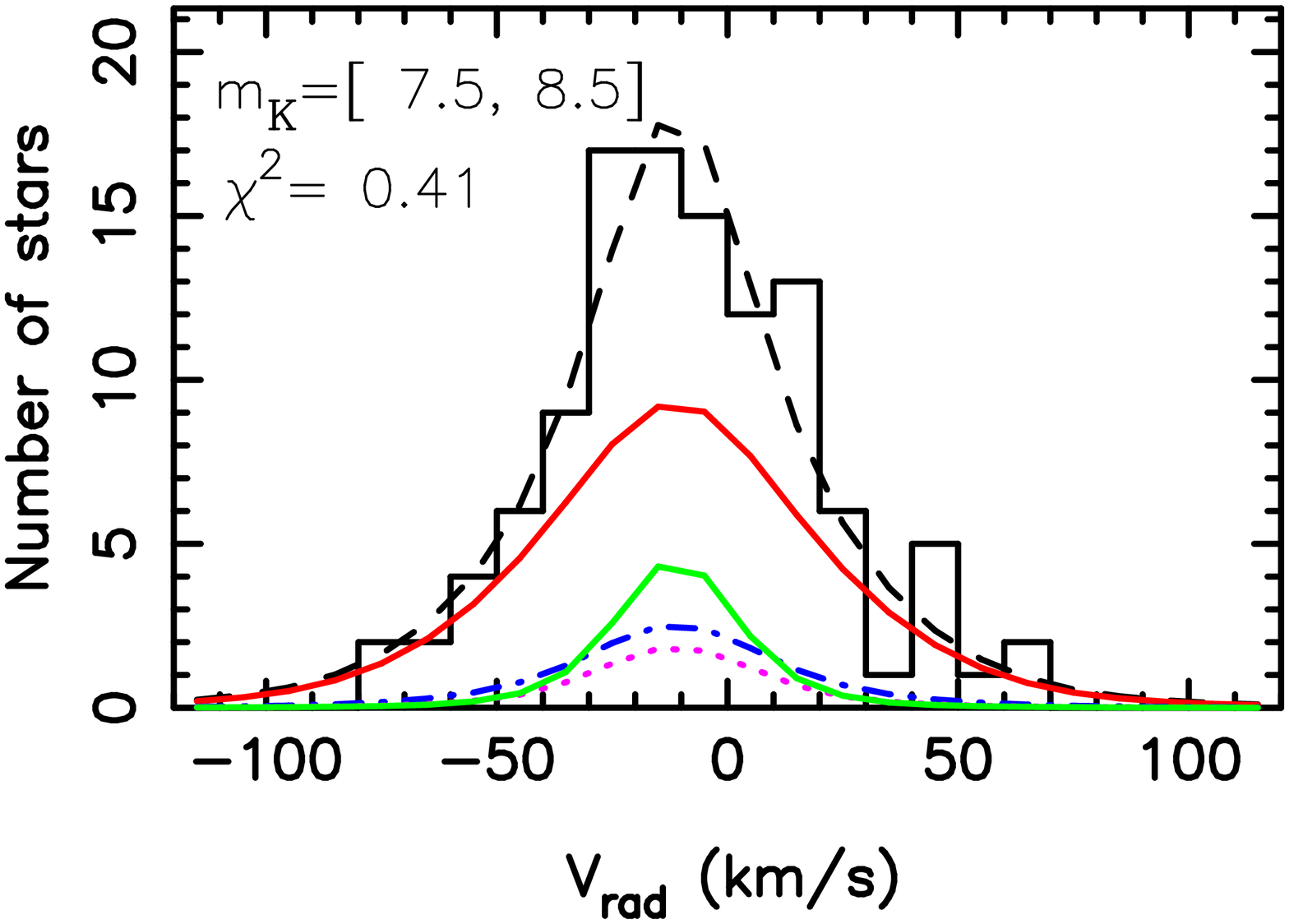} \\
\end{tabular}
\caption{Radial  velocity histograms towards  the North  Galactic Pole
for magnitudes  5.5 to  8.5 for ELODIE  data: model (dashed  line) and
contributions  of the  different type  of stars:  giants (red  or dark
lines), sub-giants  (dot-dashed and dotted)  and dwarfs (green  or grey
line).}
\label{f:elodie}
\end{figure*}

\begin{figure*}[!htbp]
\center   
\begin{tabular}{ccc}
\includegraphics[width=4.1cm,angle=270]{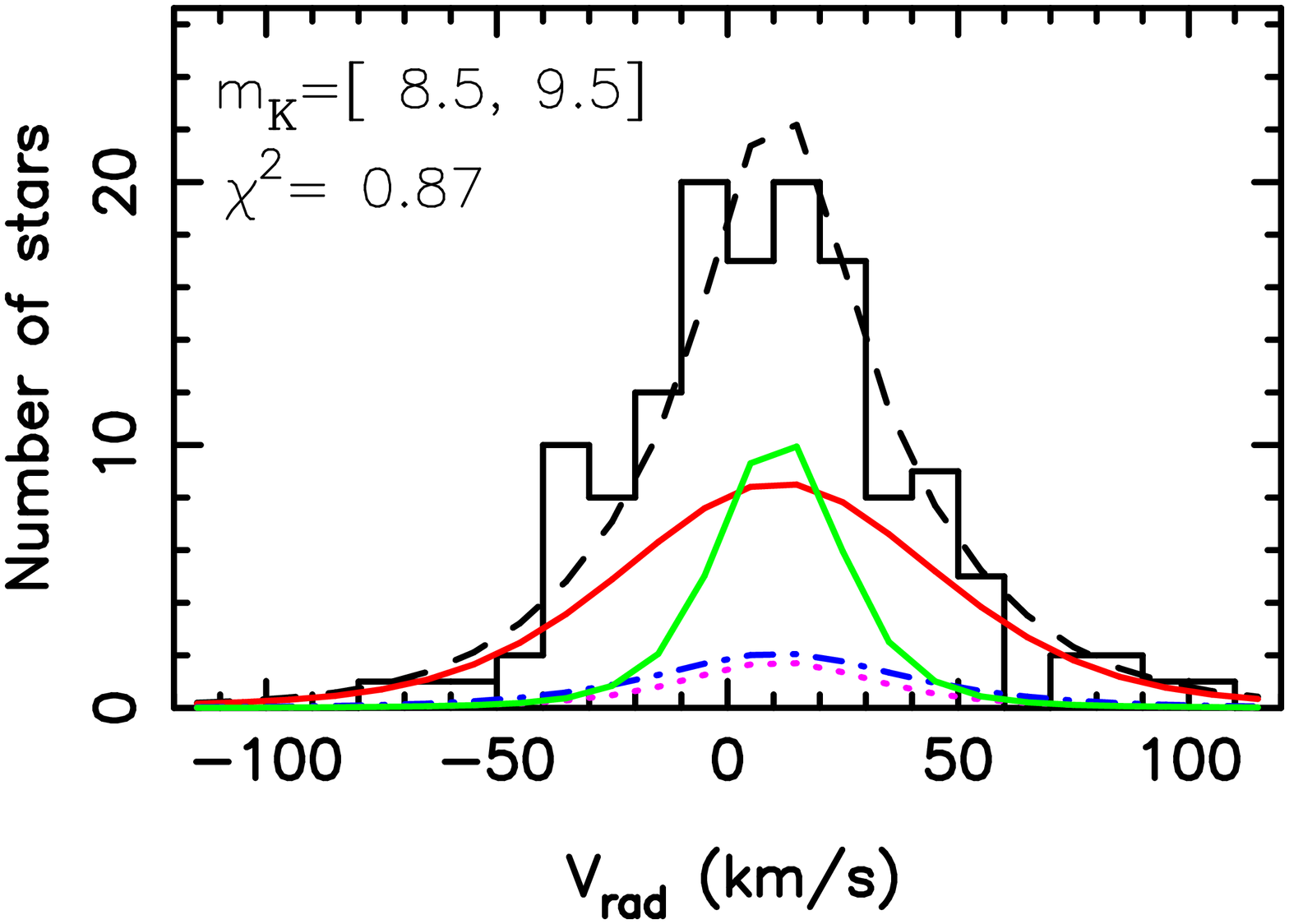} &
\includegraphics[width=4.1cm,angle=270]{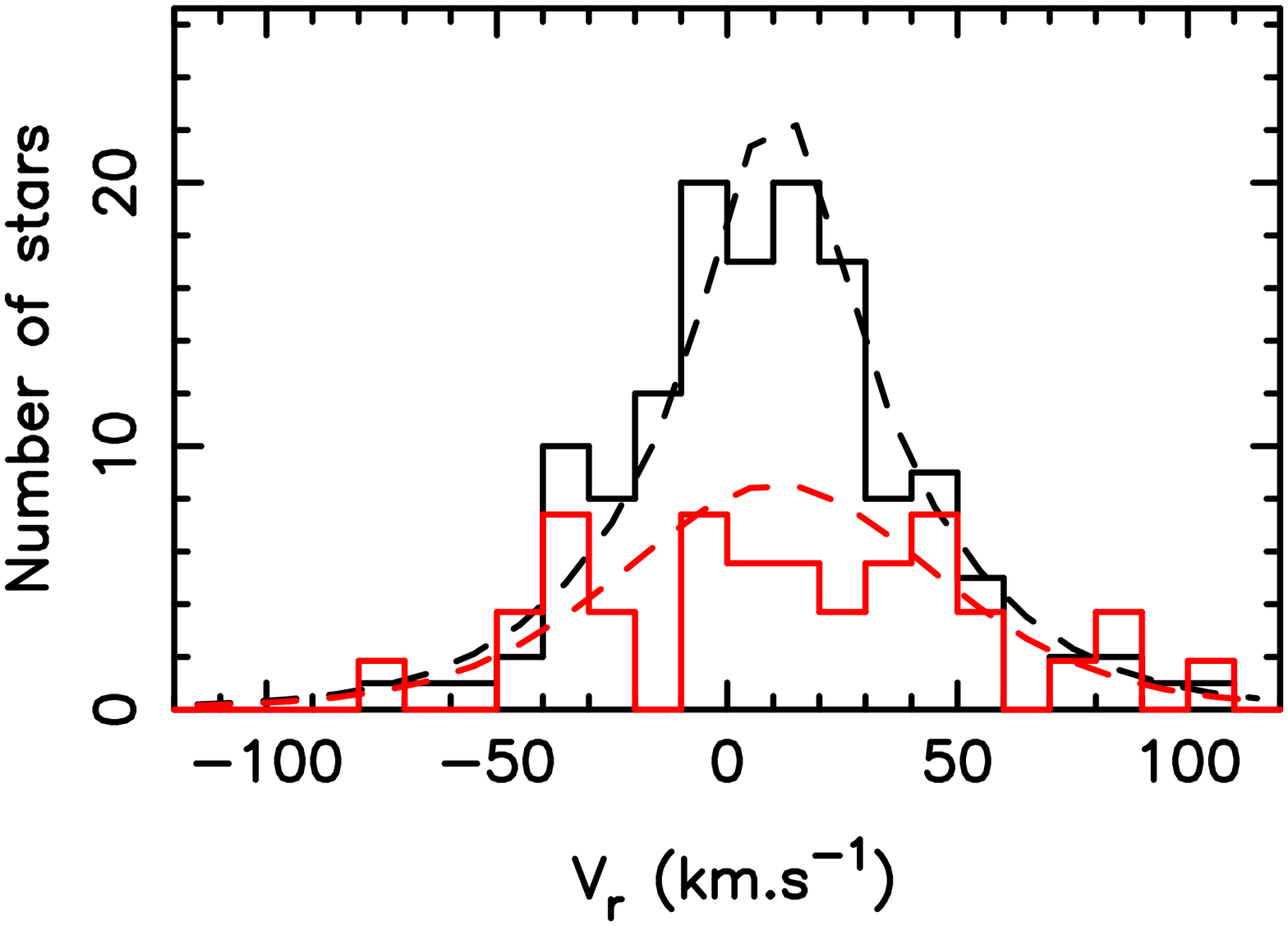} &
\includegraphics[width=4.1cm,angle=270]{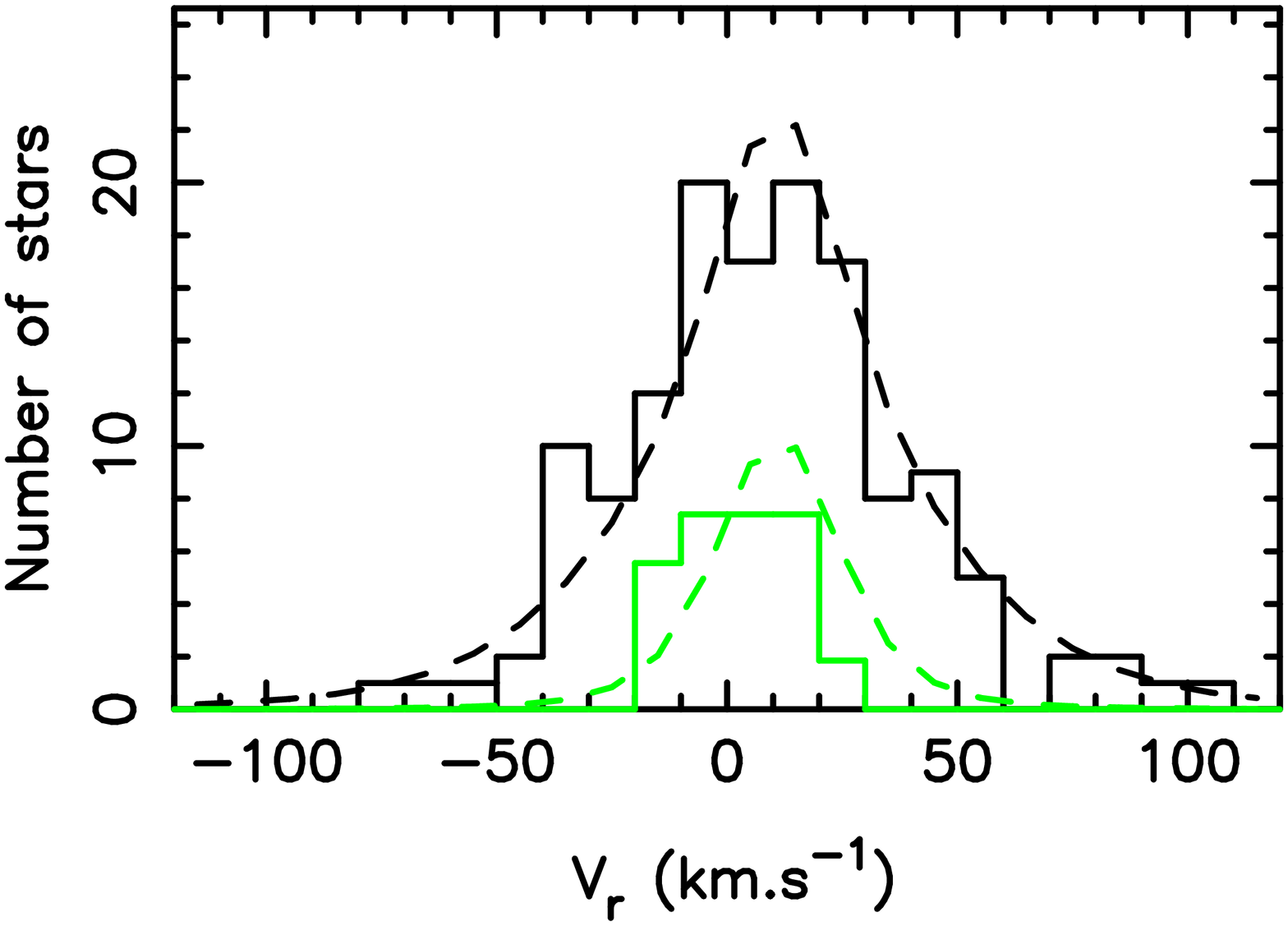} \\
\includegraphics[width=4.1cm,angle=270]{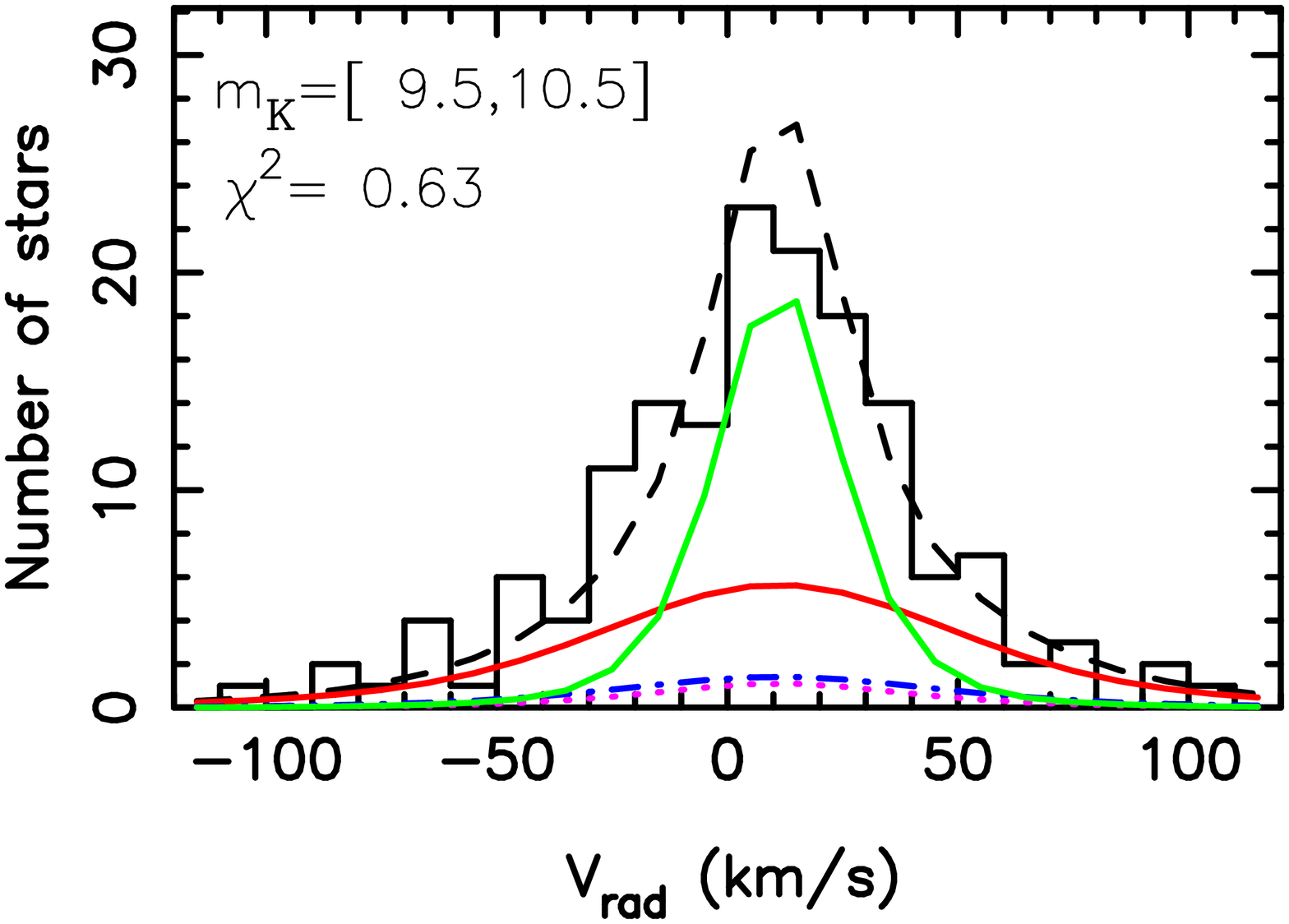} &
\includegraphics[width=4.1cm,angle=270]{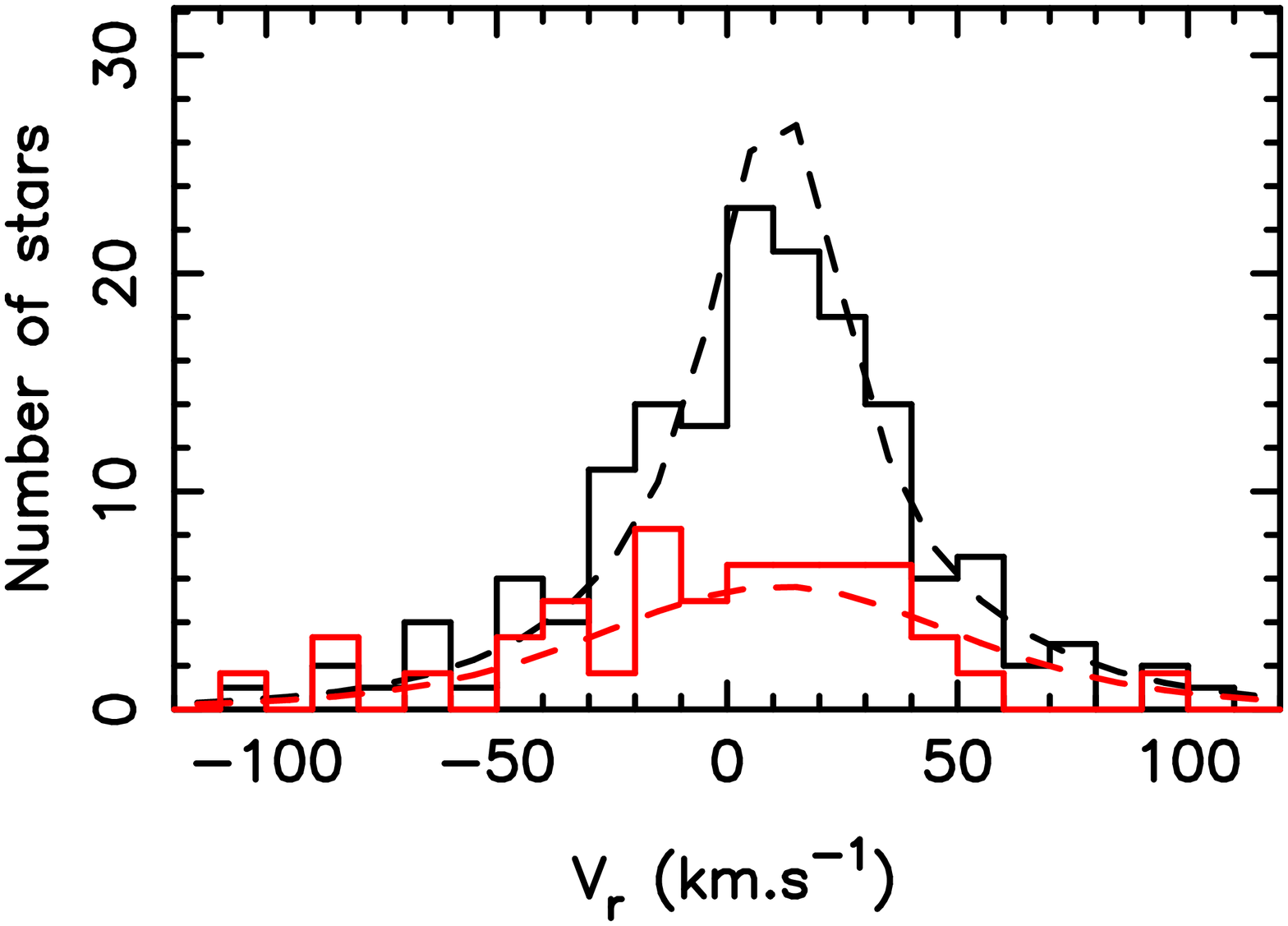} &
\includegraphics[width=4.1cm,angle=270]{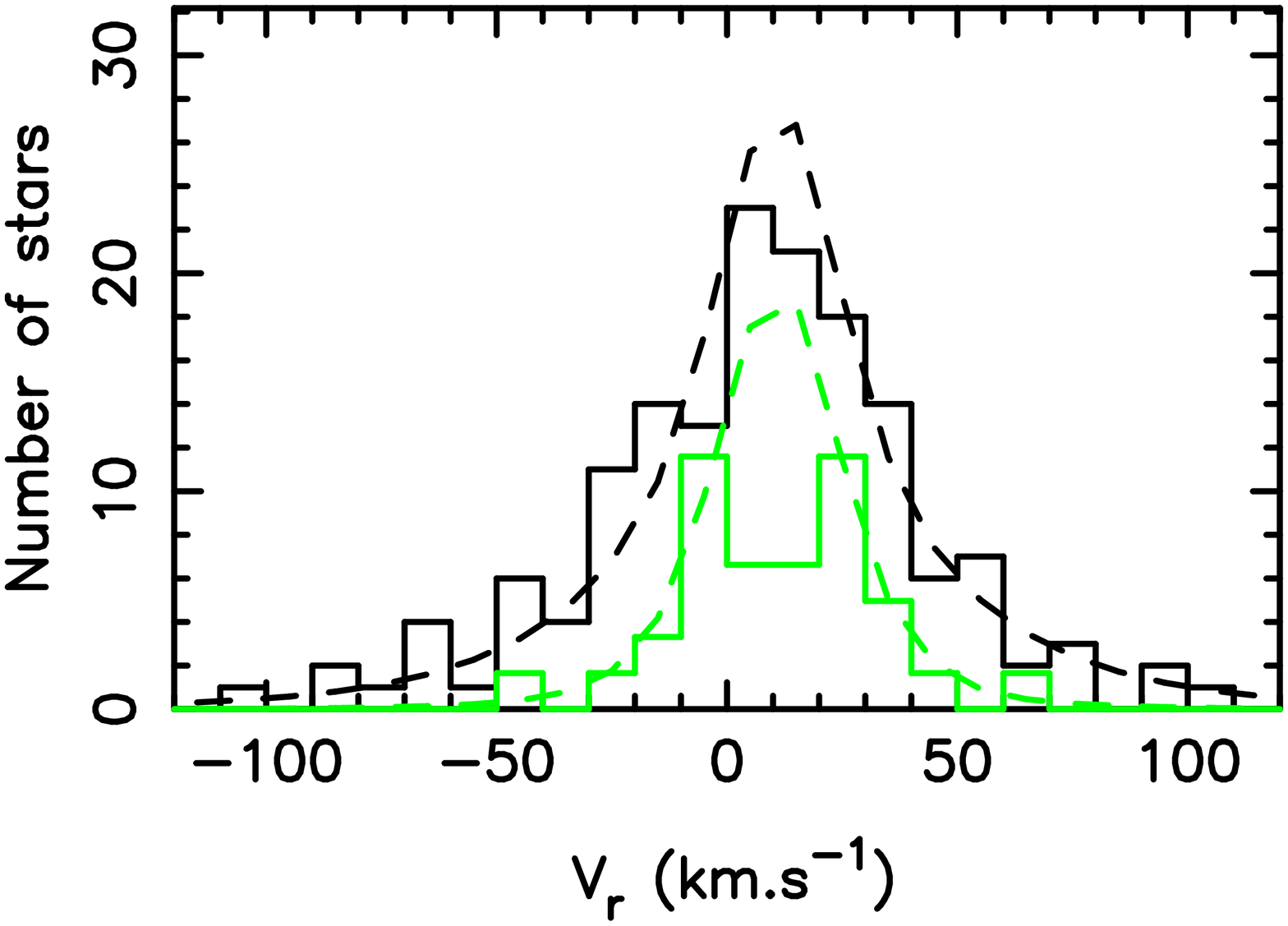} \\
\includegraphics[width=4.1cm,angle=270]{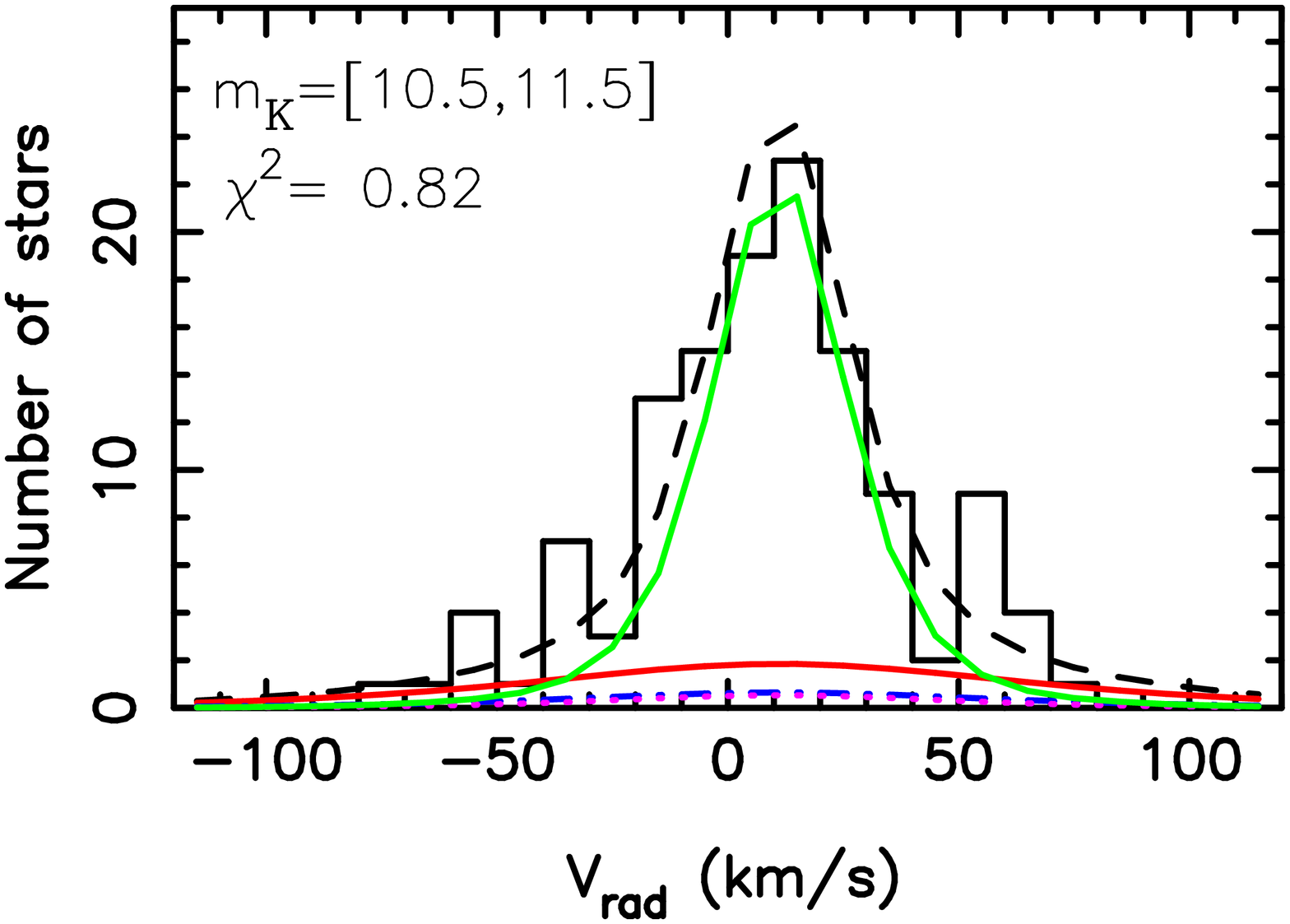} &
\includegraphics[width=4.1cm,angle=270]{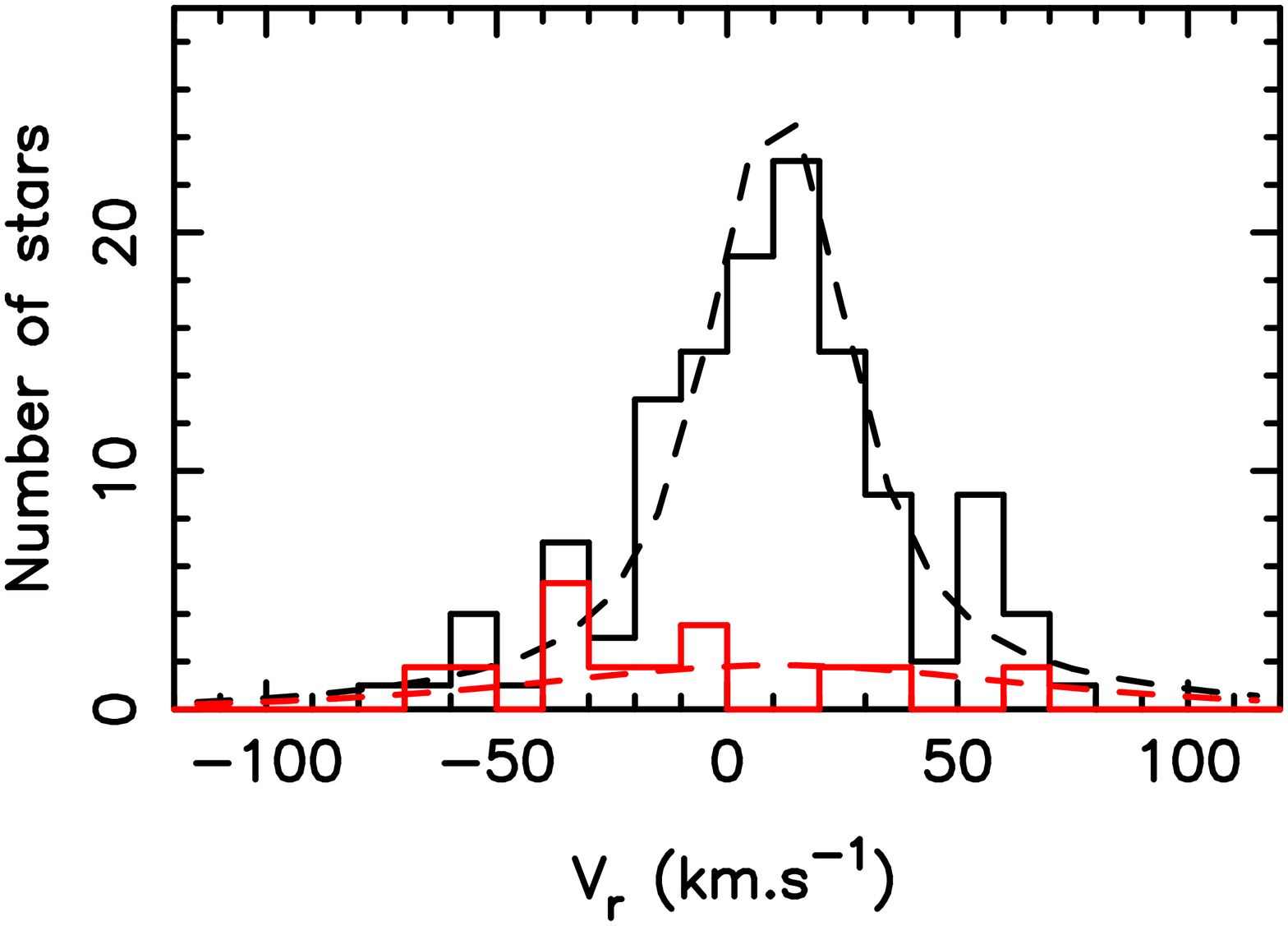} &
\includegraphics[width=4.1cm,angle=270]{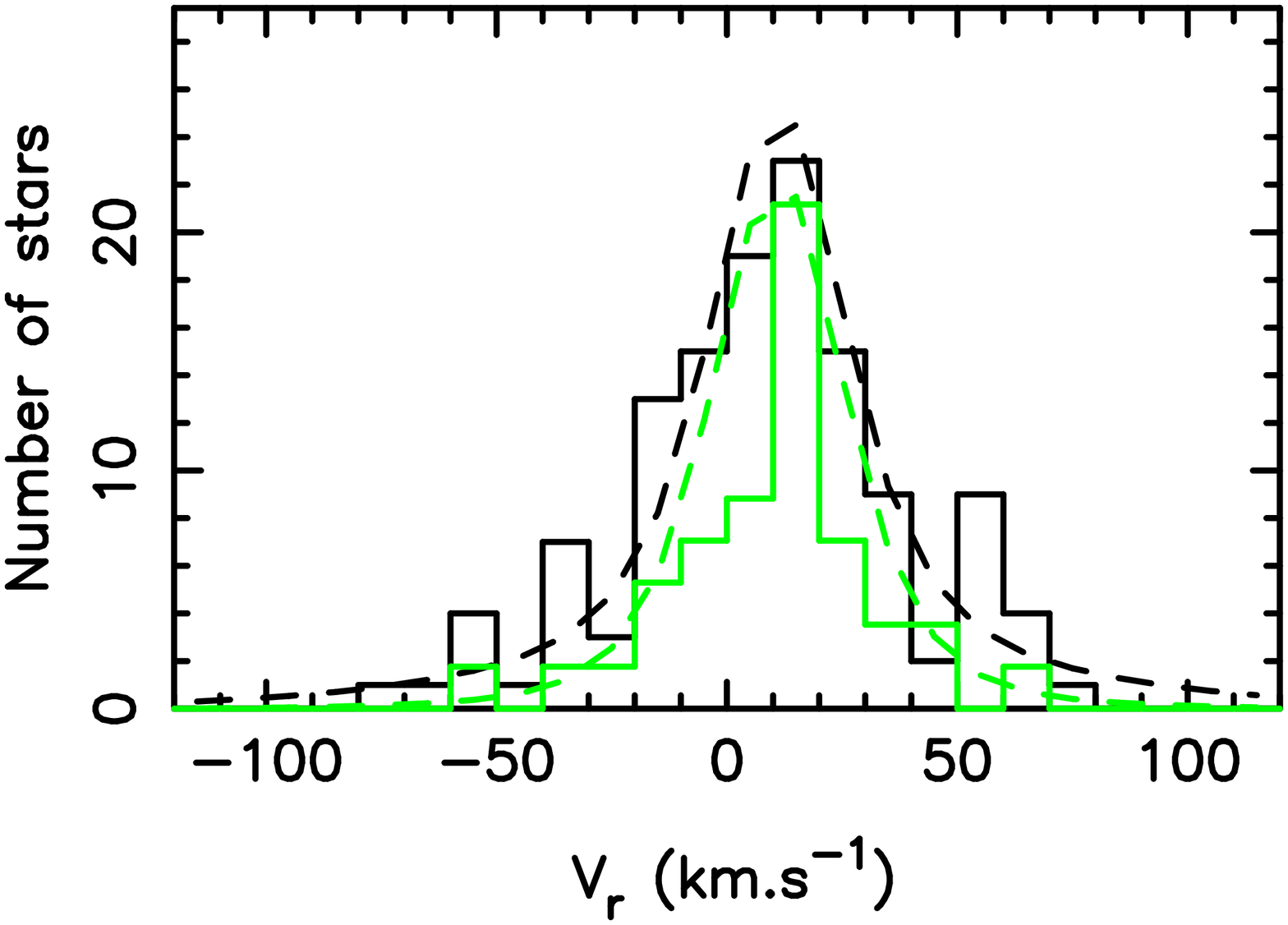} \\
\end{tabular}
\caption{Number of giants and dwarfs in RAVE data compared to model prediction. 
Left column: Radial  velocity histograms towards  the South  Galactic Pole for  magnitudes 8.5  to 11.5  for RAVE  data, model  (dashed  line) and contributions of the different type of stars: giants (red or dark 
lines), sub-giants  (dot-dashed and dotted lines) and  dwarfs (green or grey line). 
Center column: Radial  velocity histograms for all stars (black) and for giants (red or grey): model for all stars (black dashed line) and for giants (red or grey dashed line). 
Right column: Radial  velocity histograms for all stars (black) and for dwarfs (green or light grey): model for all stars (black dashed line) and for dwarfs (green or light grey dashed line). 
}
\label{f:rave}
\end{figure*}

The 181 free model  parameters are adjusted through simulations.  Each
simulation  is compared to  histograms of  counts, proper  motions and
radial velocities  (see Sect.~\ref{data}  for the description  of data
histograms                           and                           see
Figures~\ref{f:count},~\ref{f:vitesse},~\ref{f:vitesse2},~\ref{f:elodie},~\ref{f:rave})
for the comparison of the best fit model with data.  The adjustment is
done  by minimizing  a  $\chi^2$ function  using  the MINUIT  software
(James \cite{jam04}).  Equal weight is given to each of the four types
of  data (magnitude  counts,  $\mu_U$ proper  motions, $\mu_V$  proper
motions, and radial velocities).  This gives relatively more weight to
the radial velocity data whose contribution in number is two orders of
magnitude smaller than for the photometry and proper motions.

By  adjusting   our  Galactic  model,  we  derive the  respective
contributions of dwarfs  and giants, and of thin  and thick disks. One
noticeable result is the kinematic  gap between the thin and thick disk
components of our Galaxy. This discontinuity must be the consequence of
some specific process of formation for these Galactic components.

Fitting  a  multi-parameter  model  to  a large  data-set  raises  the question of the  uniqueness of the best fit  model, and the robustness of our solution  and conclusions.  For this purpose,  we have explored
the strength of the best Galactic model, by fitting various subsets of data, by modifying various  model parameters and adjusting the others. This  is a  simple, but  we expect  efficient, way  to  understand the
impact of parameter correlations and to see what is really constrained by model  {\bf or} by data.  A  summary of the main  outcomes is given below.

From  these explorations,  we choose  to fix  or bound  some important
Galactic model parameters which would otherwise be poorly constrained:
i) we fix  the vertical Galactic potential (adjusting  the $K_z$ force
does not give more accurate results than for instance in Bienaym\'e et
al., \cite{bie06}, since we only increase  by a factor 2 the number of
stars with  measured radial velocities), ii) the  asymmetric drifts of
all kinematic components are linked through a unique linear asymmetric
drift  relation  with just  one  free  parameter;  the solar  velocity
component  $V_\odot$  is  also  fixed,
iii)
the axis  ratio of the velocity  ellipsoids is bounded;  for thin disk
components      ($\sigma_W\le      25$\,km\,s$^{-1}$)      we      set
$\sigma_U/\sigma_W>1.5$, for thick disks ($\sigma_W>30$\,km\,s$^{-1}$,
$\sigma_U/\sigma_W>1.1$).

The  agreement between  our fitted  model and  the observed  counts is
illustrated  by  the  various  magnitude,  proper  motion  and  radial
velocity                                                  distributions
(Figs.~\ref{f:count},~\ref{f:vitesse},~\ref{f:vitesse2},~\ref{f:elodie},~\ref{f:rave}). We
can consider that globally the agreement is good, if we note the small
$\chi^2$  values  obtained. We  just  comment  the main  disagreements
visible within  these distributions.  They  can be compared  to recent
similar studies  (Girardi et al.  \cite{gir05}, Vallenari  et al.
\cite{val06}).

The agreement for the apparent magnitude distribution looks satisfying
in Fig.~\ref{f:count}.  

The  comparison  of  observed   and  modeled  $\mu_U$  proper  motion
distributions does not show satisfactory agreement close to the maxima
of histograms  at apparent magnitude $m_{\rm  K}<$10 ( NGP  or SGP, see
Fig.~\ref{f:vitesse}).  We have not been  able to determine if this is
due to  the inability of our  model to describe the  observed data, for
instance due  to simplifying assumptions (gaussianity  of the velocity
distribution,  asymmetric drift relation,  constant ratio  of velocity
dispersions, etc...).  We note  that this disagreement may just result
from an underestimate of the impact of the proper motion errors.

Some possible substructures are  seen in proper motion histograms for
the  brightest bins  ($ m_{\rm K}<$7,  Fig.~\ref{f:vitesse});  they are
close  to   the  level  of  Poissonian   fluctuations  and  marginally
significant. One  of the possible structures corresponds  to the known
Hercules    stream    ($\bar    U=-42\,$\,km\,s$^{-1}$    and    $\bar
V=-52$\,km\,s$^{-1}$, Famaey et al \cite{fam05b}).

For  faint  magnitude  ($ m_{\rm K}>$11)  bins  (Fig.~\ref{f:vitesse2}),
small shifts ($\sim$ 3-5\,mas\,yr$^{-1}$) of  $\mu_U$ explain most of the  differences between North
and South and the  larger $\chi^2$.

At  $  m_{\rm K}$ within  10-13  (Fig.~\ref{f:vitesse2}),  the wings  of
$\mu_U$ histograms look slightly different between North and South 
directions; it  apparently results from  shifts of  North histograms  
versus South ones.

A disagreement  of the model  versus observations also  appears within
  the  wings  of  $\mu_V$  distributions, ($  m_{\rm K}$  within  10-13,
  Fig.~\ref{f:vitesse2}).    This   may   introduce   some   doubt 
  concerning our  ability to  correctly recover the  asymmetric drift,
  because  the negative  proper motion  tail of  $\mu_V$ distributions
  directly  reflects   the  asymmetric  drift  of   the  $V$  velocity
  component.   However,  we estimate  that  our  determination of  the
  asymmetric drift  coefficient is  robust and marginally  correlated to
  the other model parameters.

These  comparisons  of observed  and  model  distributions suggest  new
directions to analyze data.  In the future, we plan to use the present
galactic model to simultaneously fit the RAVE radial velocity distribution 
in  all  available  galactic  directions.  This  result  will  be
compared to a fit of our model to proper motion distributions over all
galactic directions. This will give a better insight into the inconsistency
between radial velocity and proper  motion data, and also for possible
inconsistency in our galactic modeling.

\subsection{The transition from dwarfs to giants}

Within the  J$-$K=[0.5-0.7] interval, the proper motion  is an excellent distance indicator: there is a  factor of 14 between the proper motion of a  dwarf and the  proper motion of  a giant with the same apparent
magnitudes  and  velocities.  Combining  proper  motions and  apparent magnitudes, our best-fit  Galactic model  allows us  to  separate the contributions  of dwarfs and  giants (Fig.~\ref{f:count}).  

We deduce that, towards the Galactic poles, most of the bright stars are giants. At $m_{\rm K}=7.2$,  only 10\% are dwarfs and  at $m_{\rm K}=9.6$ only 50\% are giants. We have checked if the contribution of sub-giants with absolute magnitude $M_{\rm K}=[0.2-2]$ can change the contribution of dwarfs and giants. At  $m_{\rm K}<10$, the contribution of sub-gaint with $M_{\rm K}=[0.2-2]$ is at least one order of magnitude lower. So the ratio of giants and dwarfs is unchanged. Furthermore, the RAVE data confirm our model prediction. This is in contradiction with Cabrera-Lavers et al. (\cite{cab05}) statement based on the Wainscoat et al. (\cite{wai92}) model which estimates that, at magnitude $m_{\rm  K} < 10$, giants represent more than 90 \% of the stars. The Wainscoat model assumes only one disk with a scale height of 270\,pc for the giants and 325\,pc for the dwarfs. In our model, we find a scale height of 225\,pc both for the giants and the dwarfs. This explains why we find more dwarfs at bright magnitudes ( $m_{\rm  K} < 10$). 

Faint  stars are  mainly dwarfs,  80\%  at $m_{\rm  K}=11.6$ while  at $m_{\rm  K}=11.9$, only  10\%  are giants.   The 50\%-50\%  transition between  giants-sub-giants and  dwarfs occurs  at  $m_{\rm K}\sim10.1$. This is  a robust result from  our study that depends slightly on the absolute magnitude adopted for dwarf and giant stars. We have not tried to change our color range. If we take a broader color interval, the dispersion around the absolute magnitude of dwarfs will be larger, but our results are  not expected to change. For another color interval, we can expect this result to be different, since we would be looking at a different spectral type of star.     \\
A confirmation of the dwarf-giant separation between magnitudes $m_{\rm  K} = [5.5-11.5]$ comes from RAVE spectra. With the preliminary determination of the stellar parameters (T$_{eff}$, $\log(g)$ and [Fe/H]) of RAVE stars, we choose to define giant stars with $\log(g) < 3$ and dwarfs with $\log(g) > 4$. 
The comparison of the number of giants and dwarfs predicted by our best model 
to the observed one is in good agreement (see fig. \ref{f:rave}). 

\begin{figure}[!htbp]
\resizebox{\hsize}{!}{
\includegraphics[angle=270]{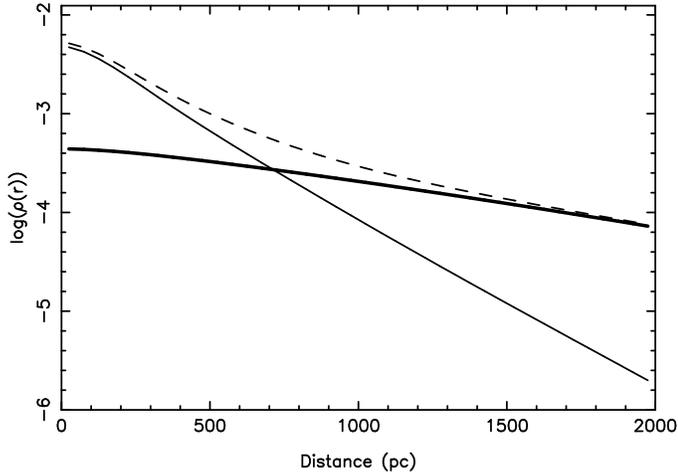}   } 
\caption{ Model  of  the vertical  stellar  density  $\rho(z)$
towards the  the North  Galactic Pole (dashed  line) and its  thin and
thick  disk decomposition  (respectively thin  and thick  lines).  The
thin   disk  includes   the  isothermal   kinematic   components  with
$\sigma_W\,<$25\,km\,s$^{-1}$, the thick disks include components with
$\sigma_W>$ 25\,km\,s$^{-1}$. }
\label{f:thin-thick}
\end{figure}

\subsection{The scale heights of stellar components} 

Our dynamical modeling  of star counts allows us to  recover the vertical density  distribution of each  kinematic component  $\rho_i(z)$, with the exact shapes depending on the adopted vertical potential $\Phi(z)$.  We recover the  well-known  double-exponential  shape  of  the  total vertical    number   density    distribution    $\rho_{\rm   tot}(z)$ (Fig.~\ref{f:thin-thick}).  Since  we  estimate that  the  kinematic  decomposition  in isothermal  components is  closer to  the idealized concept of stellar  populations and disks, we identify the thin  disk  as  the  components with  vertical  velocity  dispersions
 $\sigma_W$  smaller  than  25\,km\,s$^{-1}$  and the  thick  disk  with   $\sigma_W$    from   30   to   45.5\,km\,s$^{-1}$ (Fig.~\ref{f:KDF}).  Following this identification, we can  fit an exponential on the thin and thick disk  vertical density component (thin line and thick lines respectively of Fig.~\ref{f:thin-thick}). The scale  height of the thin disk is  225$\pm$10\,pc within 200-800\,pc.  For  the thick  disk, within 0.2-1.5\,kpc, the scale height is 1048$\pm$36\,pc. If we consider all the kinematic components without distinguishing between the thin and thick disk, we can fit a double exponential with a scale length of the thin disk  217$\pm$15\,pc and of the thick disk 1064$\pm$38\,pc. We calculate the error of the scale length from the error on the individual kinematic disk components $\phi_{kin,i}$ (see Tab. \ref{t:KDF}). We have performed a Monte-Carlo simulation on the value of the components and obtained the error bars for the scale length of the thin and thick disk both independently and together.

We note that our density distribution is not exponential for $z < 200$ pc: this mainly results  from the fact  that we do not   model    components    with   small    velocity    dispersions $\sigma_W<$\,8\,km\,s$^{-1}$.   Thus our  estimated density  at $z$=0 cannot be  directly compared, for instance, to  Cabrera-Lavers et al. (\cite{cab05}) results.  With this proviso, the star number density ratio of thick to thin disk stars at $z$=0\,pc is 8.7\% for the dwarfs.

One candidate to trace the thin and thick disk are the red clump giants. In fact, at $z$-distances larger than  $\sim$\,500\,pc (i.e.  $m_{\rm K}$ larger than $\sim$7.0, see Fig.~\ref{f:count}, there are more thick disk giants than thin disk giants. Cabrera-Lavers et al.  (\cite{cab05}) have analyzed them using 2MASS data. To do this, they select all stars with color J-K=[0.5-0.7] and magnitude $m_{\rm K} < 10$. But, beyond magnitude 9, the proportion of giants relative to sub-giants and dwarfs decreases quickly. At $m_{\rm K}$=9.6, giants just represent half of the stars, and  their distance  is about  1.7\,kpc.  Thus,  we must  be cautious when probing  the  thick disk  with  clump giants  and  we  have first  to determine the respective sub-giant and dwarf contributions.  However, Cabrera-Lavers et al.  (\cite{cab05}) obtained a  scale  height  of  267$\pm$13\,pc and 1062$\pm$52\,pc for the thin and thick disks which is in relatively good agreement with the values obtained from our model. 

For dwarfs that dominate the counts at faint  apparent  magnitudes  $m_{\rm  K} > 11$ (distances larger than $\sim$ 240\,pc), we use the  photometric distance:
\begin{equation}  
z_{phot}=10^{(m_{\rm   K}-M_{\rm  K}-5)/5}
\end{equation}
where $M_{\rm  K}$ is equal to 4.15 (the value for the dwarfs).

\begin{figure*}[!htbp]
\resizebox{\hsize}{!}{ 
\hspace{1.5cm}
\includegraphics{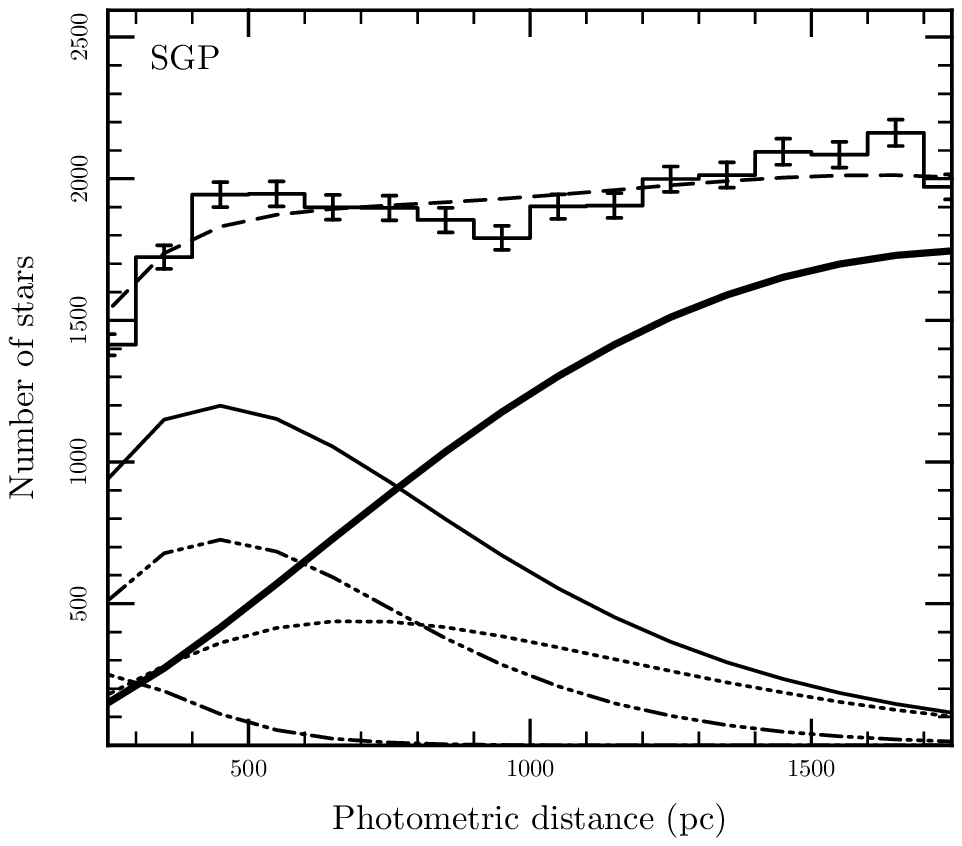} 
\hspace{3cm}
\includegraphics{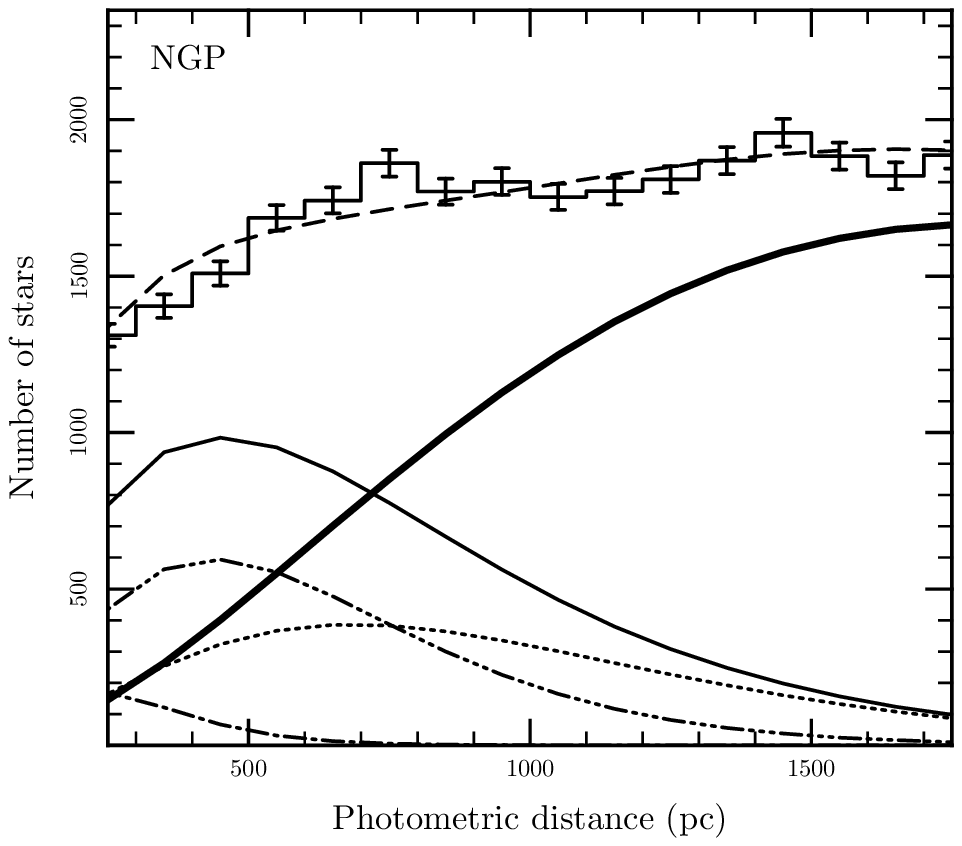}
\hspace{1.5cm}
}
\caption{Data (histogram with error bars) and model (dashed line) for the NGP  (left)
  and SGP  (right)  vertical density  distribution
using  photometric  distances  $n_{phot}$(z)  for  dwarf  stars.   The
transition between thin and thick  components is revealed by a minimum
at $z\sim$1\,kpc.   The main contributing components  are plotted, for
the thin  disk (thin continuous line) $\sigma_W$  = 10.5 (dot-dashed),
14 \& 17.5 (triple  dot-dashed), 21 \& 24.5\,km\,s$^{-1}$ (dotted) and
for  the  thick  disk   (thick  continuous  line)  $\sigma_W$  =  45.5
km\,s$^{-1}$.}
\label{f:nz}
\end{figure*}

Doing so, we obtain the number density $n(z_{phot})$ of  stars seen along the line of sight at the SGP and NGP (Fig.~\ref{f:nz}). These plots show  a well-defined first  maximum at  $z_{phot}$=500\,pc  (SGP) or  700\,pc (NGP)  related to the  distribution of thin disk  dwarfs.  At 0.9-1.1\,kpc, $n(z_{phot})$  has a minimum  and then rises  again at larger  distances, indicating the  thick disk  dwarf contribution. 

However, the use of photometric distances can introduce a systematic error for thick disk dwarfs that have lower metallicities . The mean metallicity of the thick disk population at 1 kpc is $\langle$[Fe/H]$\rangle \simeq$ -0.6 (Gilmore et al. \cite{gil95};  Carraro et al. \cite{car98}; Soubiran et al. \cite{sou03}). 

The metallicity variation from [Fe/H]=0.0 for the thin disk to [Fe/H]=-0.6 for the thick disk means that the absolute magnitude $M_{\rm  K}$ changes from 4.15 to 4.5.  So, we smoothly vary the absolute magnitude with the metallicity from the thin to the thick disk, in this way:  
\begin{equation}
M_{\rm  K}([Fe/H])=M_{{\rm  K},0} + 0.035 m_{\rm  K}
\end{equation}
where $M_{{\rm  K},0}$  is equal to 4.15.

\begin{figure*}[!htbp]
\resizebox{\hsize}{!}{ 
\hspace{1.5cm}
\includegraphics{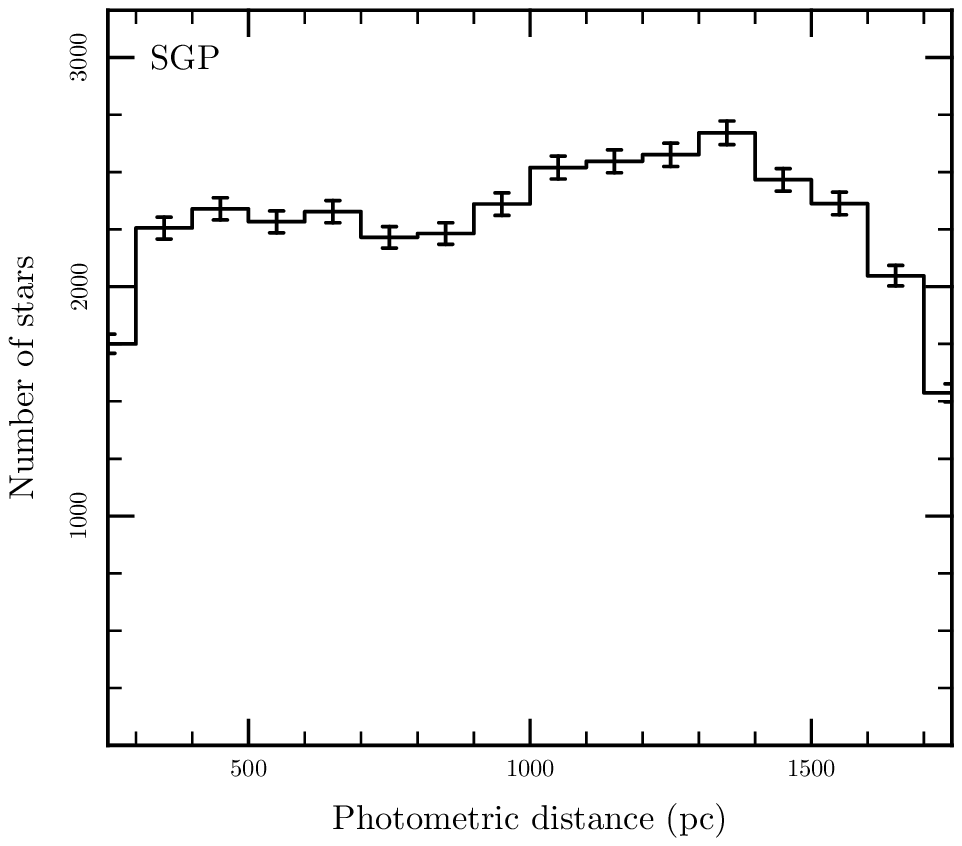} 
\hspace{3cm}
\includegraphics{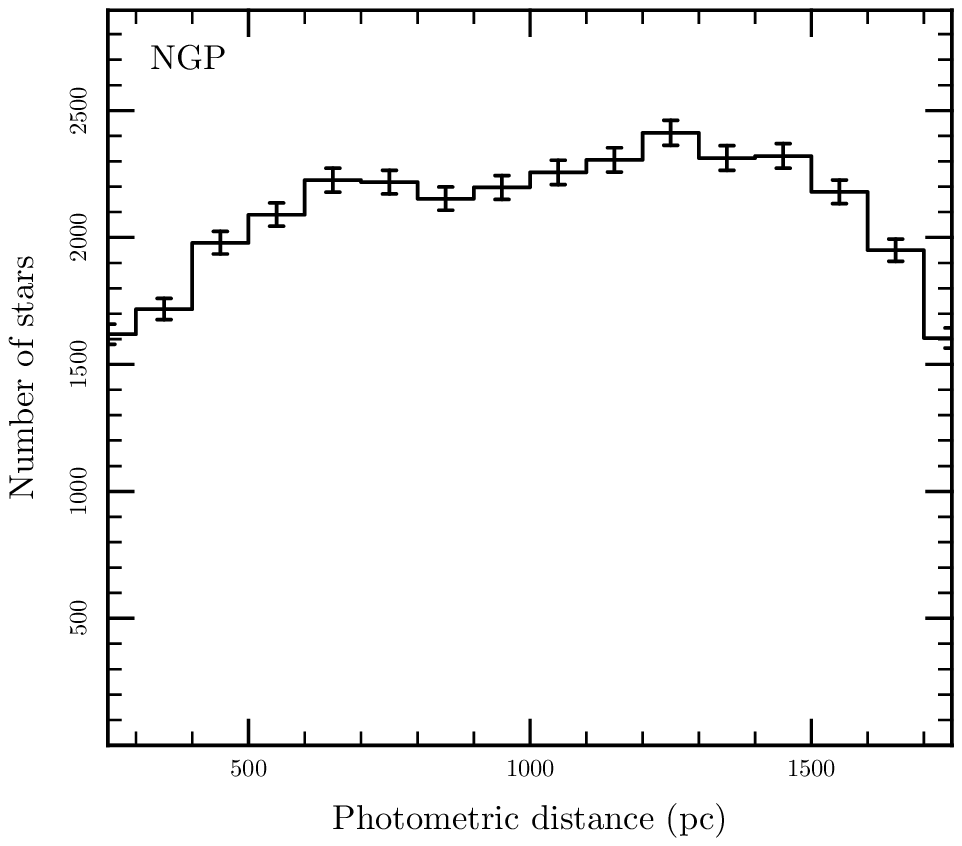}
\hspace{1.5cm}
}
\caption{Histograms of the vertical density  distribution  for the NGP (left) and SGP  (right) 
using  photometric  distances  $n_{phot}$(z)  for  dwarf  stars with a smooth variation in the [Fe/H] from the thin to the thick disk. }
\label{f:nz2}
\end{figure*}

\begin{figure*}[!htbp]
\resizebox{\hsize}{!}{ 
\hspace{1cm}
\includegraphics[angle=270]{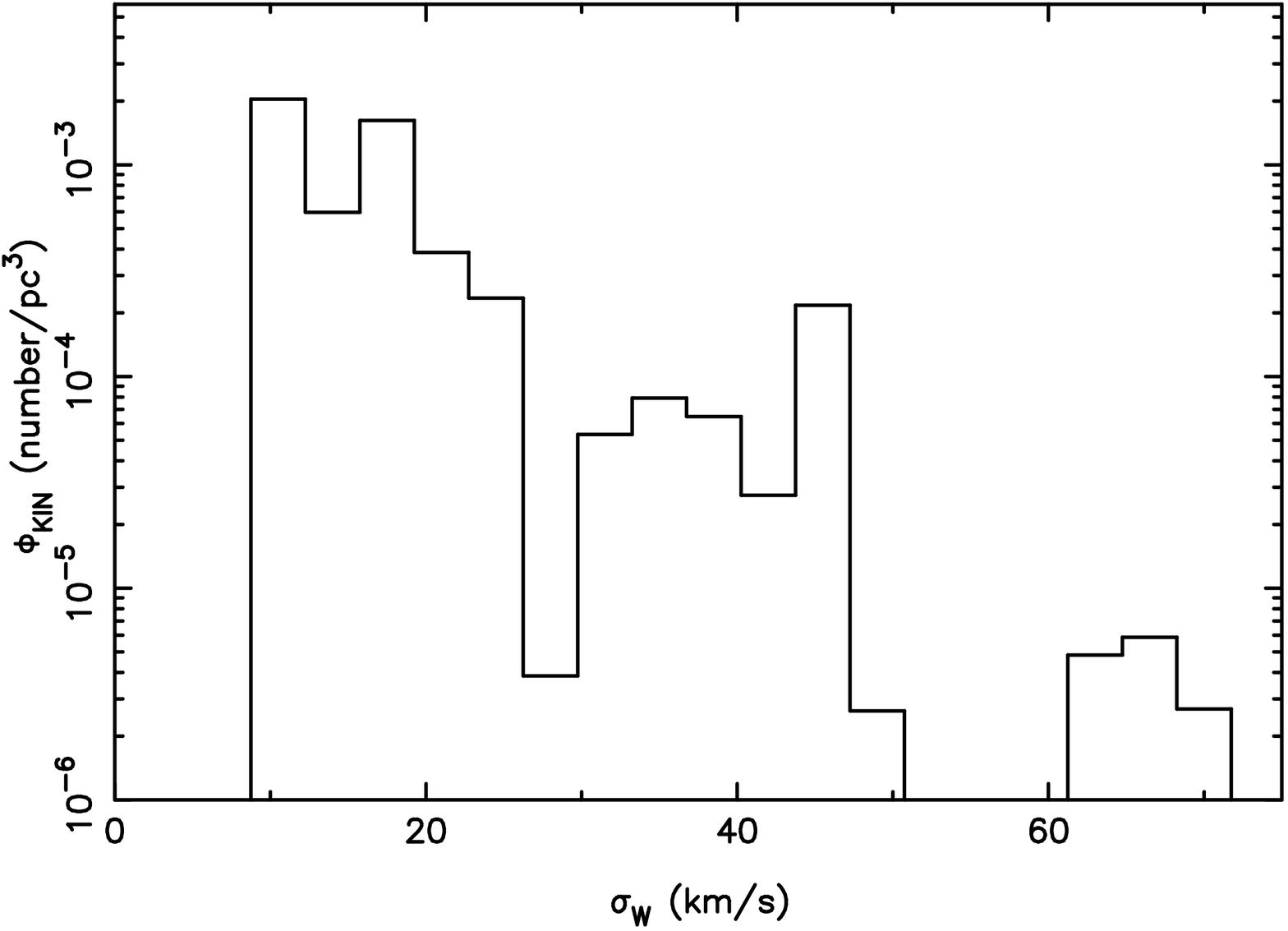}
\hspace{2cm}
\includegraphics[angle=-90]{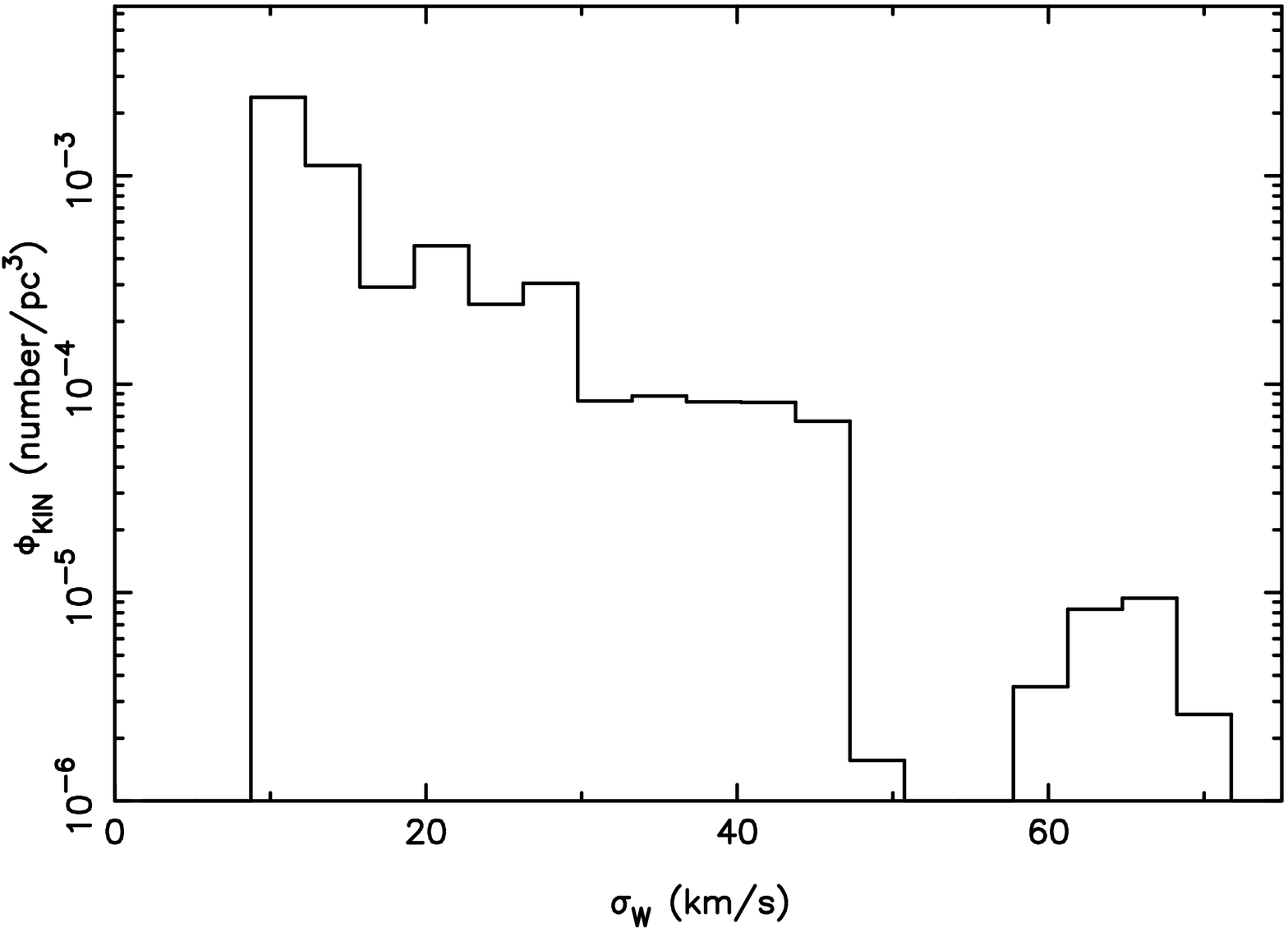}
\hspace{1cm}
}
\caption{ Left: The local  $\sigma_W$ kinematic
distribution function.  The contributing components to star counts can
be put together in a thin disk component ($\sigma_W<$25 km\,s$^{-1}$),
a thick disk (isothermal with $\sigma_W$=45.5 km\,s$^{-1}$) and a 
 hotter component with  $\sigma_W\sim$65 km\,s$^{-1}$.  The
two first components with $\sigma_W$=3.5 and 7 km\,s$^{-1}$ are set to
zero by construction. Right: A  Kinematic Distribution Function (KDF) that  tries to reproduce  the
magnitude star counts and the kinematic data:  this model has been obtained requiring  the  
continuity  of  the KDF  from  $\sigma_w$=10 to 48\,km\,s$^{-1}$. }
\label{f:KDF}
\end{figure*}

The counts continue to show two maxima (Fig.~\ref{f:nz2}), even if the minimum is less deep. The minimum delineates a discontinuous transition between the thin and thick components. \\

The superposition of the model on the number density $n(z_{phot})$ shows only approximate agreement (Fig.~\ref{f:nz}). We think that is due to non-isothermality of the real stellar components. Anyway, the fact that the model does not reproduce exactly the observation does not weaken the conclusion about the kinematic separation of the thin and thick disk. It reinforces the need for a clear kinematic separation between the two disks in the kinematic decomposition (Fig.~\ref{f:KDF}). \\ 
  
We  also  notice,   in  Fig.~\ref{f:nz},  the difference  in  counts  between   the  North  and  the  South.   This difference  allows  us to  determine the  distance of  the Sun  above the Galactic plane,  $z_\odot=+20.0\pm2.0$\,pc, assuming symmetry between North  and South.   We also  note that the  
 transition between thin and thick  disks is more visible towards the SGP than towards the NGP.

\subsection{The  thin--thick  disk   transition, and the kinematic distribution 
function}
\label{s:thin-thick}

The  minimum   at  $z$\,$\sim$\,1\,kpc  in   the  $n(z)$  distribution (Fig.~\ref{f:nz}) provides  very direct evidence  of the discontinuity between   stellar   components   with   small   velocity   dispersions
($\sigma_W$=10-25\,km\,s$^{-1}$)  and those with intermediate velocity   dispersions  ($\sigma_W\sim$\,45.5\,km\,s$^{-1}$) (left panel Fig. \ref{f:KDF}). 

\begin{table}
\begin{center}
\begin{tabular}{| r | r @{,} l |  r @{,} l |  r @{,} l | r @{,} l |}
\hline
n$^{o}$ & \multicolumn{2}{c|}{$\sigma_w$} & \multicolumn{2}{c|}{$\phi_{kin}$} &
 \multicolumn{2}{c|}{error} &  \multicolumn{2}{c|}{error} \\ 
 &  \multicolumn{2}{c|}{(km\,s$^{-1}$)} &  \multicolumn{2}{c|}{($\times 10^6$)} &  \multicolumn{2}{c|}{absolute} &   \multicolumn{2}{c|}{in \%} \\
\hline
1 &  3 &  5 &  0 &  00 &  \multicolumn{2}{c|}{--}  &  \multicolumn{2}{c|}{--}  \\ 
2 &  7 &  0 &  0 &  00 &  \multicolumn{2}{c|}{--}   &  \multicolumn{2}{c|}{--}  \\ 
3 &  10 &  5 &  2044 &  13 &  720 &  50 &  35 &  25 \\ 
4 &  14 &  0 &  596 &  69 &  493 &  81 &  82 &  76 \\ 
5 &  17 &  5 &  1618 &  79 &  169 &  57 &  10 &  48 \\ 
6 &  21 &  0 &  385 &  76 &  92 &  03 &  23 &  86 \\ 
7 &  24 &  5 &  234 &  53 &  54 &  72 &  23 &  33 \\ 
8 &  28 &  0 &  3 &  85 &  35 &  10 & \multicolumn{2}{l|}{ $>$100}  \\ 
9 &  31 &  5 &  53 &  21 &  33 &  09 &  62 &  19 \\ 
10 &  35 &  0 &  79 &  16 &  30 &  73 &  38 &  82 \\ 
11 &  38 &  5 &  64 &  71 &  63 &  76 &  98 &  53 \\ 
12 &  42 &  0 &  27 &  49 &  66 &  31 &  \multicolumn{2}{l|}{ $>$100}  \\ 
13 &  45 &  5 &  216 &  96 &  44 &  07 &  20 &  32 \\
14 &  49 &  0 &  2 &  63 &  39 &  19 & \multicolumn{2}{l|}{ $>$100}  \\ 
15 &  52 &  5 &  0 &  38 &  0 &  08 &  21 &  05 \\ 
16 &  56 &  0 &  0 &  04 &  0 &  04 &  100 &  00 \\ 
17 &  59 &  5 &  0 &  29 &  0 &  11 &  37 &  93 \\ 
18 &  63 &  0 &  4 &  83 &  31 &  72 & \multicolumn{2}{l|}{ $>$100}  \\ 
19 &  66 &  5 &  5 &  86 &  30 &  88 & \multicolumn{2}{l|}{ $>$100}  \\ 
20 &  70 &  0 &  2 &  69 &  0 &  05 &  1 &  86 \\ 
\hline
\end{tabular}
\end{center}
\caption{List of the values of the kinematic disk components $\phi_{kin,i}$ ( 10$^6 \times$ number of stars / pc$^3$ ) with the individual errors absolutes and relatives in percent.}
\label{t:KDF}
\end{table}

Another manifestation of this transition is well known from the $\log \rho(z)$ density distribution (Fig.~\ref{f:thin-thick}) which shows a change of slope at  $z$=500-700\,pc.  This  feature can be  successfully modeled with  two   (thin  and  thick)  components  (e.g.   Reid  and  Gilmore \cite{gil83}), which  is an indication of a  discontinuity between the thin  and thick disks  of our  Galaxy.  

It  is conclusive evidence,  only if we  show that we can not fit accurately the  star counts or vertical density  distributions with a continuous set  of kinematic components (without a gap  between the thin
and the thick disks). We  find that the constraint of a set of    kinematic    components    following    a    continuous    trend (right panel of Fig.~\ref{f:KDF})   raises   the   reduced  $\chi^2$,   in particular on  SGP magnitude  counts, from 1.59  to 3.40. This confirms  the robustness  of  our  result and  conclusion  on the  wide transition between thin and thick stellar disk components.

Adjusting  the  Galactic  model  to star counts,  tangential  and  radial velocities,  we  can recover  the  details  of  the kinematics of stellar  populations,  and   we  determine  the  local  $\sigma_W$ kinematic  distribution function  (left panel of Fig.~\ref{f:KDF} and Tab. \ref{t:KDF}).  This kinematic  distribution  function clearly shows  a large  step between the kinematic properties of  the thin and thick disks. We define the  thin disk as  the components with  $\sigma_W$ covering 10-25\,km\,s$^{-1}$, and the  thick  disk  as the components with $\sigma_W$ covering 30-45 km\,s$^{-1}$. The counts and radial velocities by themselves already show the kinematic transition that we obtain in the kinematic decomposition. The fit of proper motions confirms the conclusion from the star counts and radial velocities, even if a fraction of the proper motions $\mu_l$ and $\mu_b$ at magnitude m$_{\rm K}$ fainter than 13 have significant errors ($> 20$ km\,s$^{-1}$). The only consequence for the proper motion errors is that we obtained  an ellipsoid axis ratio $\sigma_U/\sigma_W$ different from the classical values (see Sec. \ref{kin}).

The last non-null components at approximately $\sigma_W\sim\,65\,$km\,s$^{-1}$ are necessary to fit  the faintest star counts at  $m_{\rm K}\sim15$. But, they do not result from the fit of proper motion histograms (since, unfortunately, they stop at $m_{\rm K}\sim$14). Thus their exact nature, a second thick disk or halo  (they would have very different  asymmetric drift) cannot  be solved in the context of our analysis.

\subsection{The  luminosity function of stellar components}
\label{lf}

\begin{figure}[!htbp]
\resizebox{\hsize}{!}{
\includegraphics{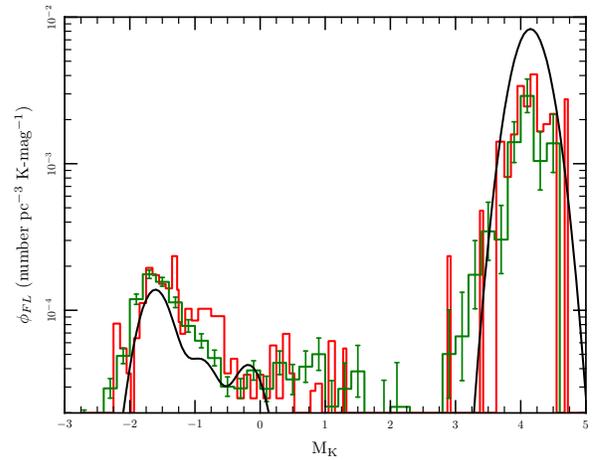}
}
\caption{ The local  Luminosity Function of K stars  from our modeling
of star counts towards the  Galactic poles (line) compared to the
LF function  from nearby  Hipparcos K stars  by Cabrera-Lavers  et al.
(\cite{cab05}) (red or black histogram) and our  own estimate of the local LF:
see text  (green or grey histogram  with error bars).  The  scale of 
Cabrera et al.'s LF has been arbitrarily shifted.}
\label{f:LF}
\end{figure}

Our distant star count and kinematic adjustment constrains the local
luminosity function (LF).  
We make the comparison with the local
LF  determined with  nearby stars.   However, the  brightest HIPPARCOS
stars needed to determine the local LF are saturated within
2MASS and  have less accurate photometry.   We can also  compare it to
the  LF determined by  Cabrera-Lavers et  al. (\cite{cab05})  who use
a cross-match  of  HIPPARCOS and  MSX  stars  and  estimate $m_{\rm  K}$
magnitudes  from MSX A band  magnitudes (hereafter [8.3]).  
However  we note from  our own
cross-match  of HIPPARCOS-MSX-2MASS (non saturated) stars  that their  
LF, for  stars
selected from V-[8.3], corresponds mainly to stars  with J--K colors
between 0.6-0.7 rather than between 0.5-0.7.  A second limitation for
a comparison of LFs is that  our modeling does not include the stellar
populations    with   small    velocity    dispersions   ($\sigma_W<8$
km\,s$^{-1}$).  For these reasons, we determine a rough local LF
based  on 2MASS-HIPPARCOS  cross-matches, keeping  stars with  V$<$7.3 or
distances $<$125\,pc,  and using the color selection  V--K between 2.0
and  2.6, that corresponds approximately to  J--K =  [0.5-0.7].
Using  V  and  K  magnitudes  minimizes  the  effects of the J--K
uncertainties.  Considering these limitations, there is reasonable
agreement between the  local LF obtained with our model using distant
stars  and  the  LF  obtained  from  nearby  Hipparcos 
stars  (see Fig.~\ref{f:LF}).

\subsection{The   stellar   kinematics}  
\label{kin}

Many of the  stellar disk kinematic properties obtained  with our best
fit Galactic  model are comparable with  previously published results.
We make the comparison with  the analysis of HIPPARCOS data (Dehnen \&
Binney  \cite{deh98};   Bienaym\'e  \cite{bie99};  N\"ordstrom   et  al.
\cite{nor04};   Cubarsi  \&  Alcob\'e   \cite{cub04};  Famaey   et  al.
\cite{fam05}),  and also  with results  published from  remote
stellar samples using a wide variety of processes to identify thin and
thick kinematic components (Barta\u{s}i\={u}t\.{e} \cite{bar94}; Flynn
\&  Morrel \cite{fly97};  Soubiran  et al.  \cite{sou03};  Pauli et  al.
\cite{pau05}).

We   obtain    for   the   Sun    motion   relative   to    the   LSR,
$u_{\odot}$=8.5$\pm$0.3\,km\,s$^{-1}$                               and
$w_{\odot}$=11.1$\pm$1.0\,km\,s$^{-1}$.   We find  for  the asymmetric
drift coefficient,  $k_{a}$=76$\pm$4\,km\,s$^{-1}$,  compared to
80$\pm$5\,km\,s$^{-1}$  for nearby HIPPARCOS  stars (Dehnen  \& Binney
\cite{deh98}) and the thick disk lag is $V_{\rm lag}= \sigma_R^2 /k_{a}=33
\pm$2\,km\,s$^{-1}$ relative to the LSR.  We note that this value of
the thick disk lag is close to the value of Chiba \& Beers (\cite{chi00}) and other 
estimates prior to this. It is less in agreement with the often-mentioned values  
of 50-100\,km\,s$^{-1}$ from pencil-beam
samples. These  may be more affected  by Arcturus group  stars which are
more dominant at higher $z$-values.

Our determination of
the asymmetric drift coefficient  is highly correlated to $V_{\odot}$.
The  reason  is that  we  do not  fit  populations  with low  velocity
dispersions and small  $V_{\rm lag}$ since we do not  fit star counts with
$m_{\rm K}<6$: as  a consequence the slope of  the relation, $V_{\rm lag}$
versus  $\sigma_U$, is  less well constrained.  To  improve the  $k_{a}$
determination, we adopt $V_{\odot}=5.2$\,km\,s$^{-1}$ (Dehnen \& Binney
\cite{deh98};     Bienaym\'e      \cite{bie99}).      The     adjusted
$\sigma_U/\sigma_V$  velocity  dispersion  ratio,  taken to be the  
same  for  all components,  is  $1.44 \pm 0.02$.   
We obtain  $\sigma_U/\sigma_W$  ratios
significantly  smaller than  those  published using  nearby samples  of
stars.  For the thin disk components, we find $\sigma_U/\sigma_W$=1.50
to 1.62 (compared to published values  $\sim$2 by authors using
HIPPARCOS    stars).     For     the    thick    disk,    we    obtain
$\sigma_U/\sigma_W$=1.1,  instead of  $\sim 1.5-1.7$  typically obtained
with  nearby thick  disk stars  by other  authors.

While   there    is   no   dynamical   reason   preventing   the variation of
$\sigma_U/\sigma_W$ with $z$,  we suspect that our low
$\sigma_U/\sigma_W$ ratio at large $z$ for the thick disk results from
a bias  within our model due to the outer  part of the wings  of some
proper motion histograms not being accurately adjusted.
This  may  be  the  consequence  of  an  incorrect adopted vertical
potential or, as  we think, more likely the non-isothermality of
the real  velocity distributions.   This suspicion is reinforced since
fitting each proper  motion histograms separately with a set of gaussians
gives us larger values for $\sigma_U/\sigma_W$. 

   Our results can be directly  compared with the very recent analysis
by Vallenari et al.  (\cite{val06}) of stellar populations towards the
NGP   using   BVR   photometry   and   proper   motions   (Spagna   et
al. \cite{spa96}).  Their model is dynamically consistent but based on
quite  different hypotheses  from ours;  for each  stellar population,
they assume that in the Galactic plane $\sigma_{zz}^2$ is proportional
to the  stellar density $\rho$  (Kruit \& Searle  \cite{vdk82}).  They
also  assume   that  both  velocity   dispersions,  $\sigma_{zz}^2$  and
$\sigma_{RR}^2$,  follow  exponential  laws with  the  same  scale
exponential  profile as  the surface  mass density  (Lewis  \& Freeman
\cite{lew89}).   Vallenari  et al.   (\cite{val06})  found thick  disk
properties (see their  table 6) quite similar to  the ones obtained in
this    paper.     They    obtain:    $\sigma_W$=38$\pm$7km\,s$^{-1}$,
$\sigma_U/\sigma_V$=1.48,
$V_{\rm lag}=42\pm7$ km\,s$^{-1}$, and for  the thick
disk  scale height:  900\,pc.  However, they find $\sigma_U/\sigma_W$=1.9.
 They also  claim  that "no  significant
velocity gradient is found in the thick disk", implying that the thick
disk must be an isothermal component.

\subsubsection{Radial velocities}

The number  of RAVE and ELODIE stars  used in this analysis  is a tiny fraction  of the total  number of  stars used  from 2MASS  or UCAC2  catalogues.   However they play  a key  role in  constraining  Galactic model parameters: the magnitude  coverage of RAVE stars towards the  SGP, from $m_{\rm K}=8.5$ to 11.5, can be used to discriminate between the  respective   contributions  from  each   type  of   star,  dwarfs, sub-giants,  giants.   A future RAVE data release (Zwitter et al. submitted) will  include  gravities, allowing for easier identification of dwarfs and  red clump giants; it will  also include element abundances allowing for better description of  stellar disk  populations and  new insights  into the  process of their formation.

\section{Conclusion}

We revisit  the thin-thick disk transition using  star counts
and kinematic data towards  the Galactic poles.  Our Galactic modeling
of star count, proper motion and radial velocity allows us to recover the
LF, their kinematic distribution  function, their vertical
density distribution, the relative distribution of giants, sub-giants
and  dwarfs, the  relative contribution from  thin  and thick  disk
components, the  asymmetric drift  coefficient and the  solar velocity
relative to the LSR.

The double exponential fitting of the vertical disk stellar density distribution is not sufficient to fully characterize the thin and thick disks. A more complete description of the stellar disk is given by its  kinematical decomposition. 

From  the star  counts, we see a  sharp transition  between the
thick and  thin components.  Combining star counts with kinematic data,
and applying a  model with 20 kinematic components,  we discover a gap
between the vertical velocity dispersions of thin disk components with
$\sigma_W$  less  than  21  km\,s$^{-1}$  and a  dominant  thick  disk
component at $\sigma_W$=45.5km\,s$^{-1}$.  The thick disk scale height
is found to be 1048$\pm$36\,pc.
We identify this thick  disk with the intermediate metallicity ([Fe/H]
$\sim$\,--0.6  to  --0.25)  thick  disk described,  for  instance,  by
Soubiran et  al. (\cite{sou03}).  This  thick disk is also  similar to
the thick disk measured by Vallenari et al (\cite{val06}) who find "no
significant velocity  gradient" for  this stellar component.   We note
that star  counts at  $m_{\rm K}\sim15$  suggest a  second 
thick disk or halo component with  $\sigma_W\sim$65km\,s$^{-1}$.

Due to the separation of the thin and thick components, clearly 
identified with stars counts and visible within the kinematics, 
the thick disk measured in this paper cannot be the result of 
dynamical heating of the thin disk by massive molecular clouds 
or by spiral arms. We would expect otherwise a continuous kinematic 
distribution function with significant kinematic components covering without 
discontinuity the range of   $\sigma_W$ from
10 to 45 km\,s$^{-1}$.

 We find that, at the solar position, the surface mass density of the thick disk 
 is 27\% of the surface mass density of the thin disk. The thick disk has velocity dispersions
$\sigma_U=50\,$km\,s$^{-1}$,     $\sigma_W$=45.5\,km\,s$^{-1}$,    and
asymmetric drift $V_{\rm lag}=33 \pm 2$\,km\,s$^{-1}$.
Although clearly  separated from the  thin disk, this  thick component
remains a  relatively `cold' thick  disk and has  characteristics that
are close to the thin disk properties.
This  `cold'  and  rapidly  rotating  thick disk  is  similar  to  the
component identified by many kinematic  studies of the thick disk (see
Chiba \& Beers  \cite{chi00} for a summary). Its  kinematics appear to
be  different  from  the  thick  disk stars  studied  at  intermediate
latitudes  in pencil  beam surveys  (eg Gilmore  et  al \cite{gil02}),
which  appear to be  significantly affected  by a  substantial stellar
stream with  a large lag  velocity. They interpret  this stellar
stream  as  the possible  debris  of  an  accreted satellite  (Gilmore
\cite{gil02};    Wyse    et al. \cite{wys06}). Maybe some  connections exist with streams identified
in  the solar  neighborhood as  the  Arcturus stream  (Navarro et  al
\cite{nav04}).

 Some  mechanisms of  formation connecting  a thin  and a   thick
components are compatible with our  findings.  It may be, for instance
a `puffed-up' thick disk, i.e. an earlier thin disk puffed up by the accretion
of  a  satellite (Quinn  et  al.  \cite{qui93}).  Another  possibility,
within the monolithic  collapse scenario, is a thick  disk formed from
gas with  a large vertical scale  height before the  final collapse of
the gas  in a thin disk,  i.e.  a `created  on the spot' thick disk.   We also
notice the Samland (\cite{sam04})  scenario: a chemodynamical model of
formation of  a disk galaxy within  a growing dark  halo that provides
both a `cold'  thick disk and a metal-poor `hot' thick disk.

A  popular scenario  is  the  `accreted' thick  disk  formed from  the
accretion of satellites.  If the thick disk results from the accretion
of just a  single satellite, with a fifth of the  mass of the Galactic
disk, this has been certainly a major event in the history of the
Galaxy,  and it  is hard  to  believe that  the thin  disk could  have
survived this upheaval.

Finally,  from the thick  disk properties  identified in
this paper, we  can reject the most improbable  scenario of formation:
the  one of type  `heated' thick disk  (by molecular  clouds or
spiral arms).

\begin{acknowledgements}

Funding   for  RAVE   has  been   provided  by   the  Anglo-Australian
Observatory, by the Astrophysical Institute Potsdam, by the Australian
Research Council,  by the German Research foundation,  by the National
Institute for Astrophysics at Padova, by The Johns Hopkins University,
by  the Netherlands  Research  School for  Astronomy,  by the  Natural
Sciences and Engineering Research  Council of Canada, by the Slovenian
Research  Agency, by  the Swiss  National Science  Foundation,  by the
National  Science   Foundation  of  the  USA   (AST-0508996),  by  the
Netherlands  Organisation  for Scientific  Research,  by the  Particle
Physics  and Astronomy  Research Council  of  the UK,  by Opticon,  by
Strasbourg Observatory,  and by the Universities  of Basel, Cambridge,
and Groningen. The RAVE web site is at www.rave-survey.org.

Data  verification is  partially based  on observations  taken  at the
Observatoire de  Haute Provence (OHP, France), operated  by the French
CNRS.

This publication makes  use of data products of the  2MASS, which is a
joint  project of  the University  of Massachusetts  and  the Infrared
Processing and Analysis Center, funded by the NASA and NSF

It is  a pleasure to thank  the UCAC team  who supplied a copy  of the
UCAC CD-ROMs in July 2003.

This  research  has made  use  of  the  SIMBAD and  VIZIER  databases,
operated at CDS, Strasbourg,  France.

This paper  is based  on data from  the ESA {\it  HIPPARCOS} satellite
(HIPPARCOS and TYCHO-II catalogues).

\end{acknowledgements}

%

\end{document}